\documentclass[10pt]{iopart}

\usepackage[english]{babel}

\usepackage{subfigure}
\usepackage{amssymb}
\usepackage{bm}
\usepackage{graphics}
\usepackage{graphicx}
\usepackage{iopams} 
\usepackage{float}

\usepackage[utf8x]{inputenc}
\usepackage[T1]{fontenc}
\usepackage{ae,aecompl}

\usepackage[usenames,dvipsnames]{xcolor}

\usepackage{color}
\definecolor{darkgreen}{rgb}{0,0.6,0}
\definecolor{darkblue}{rgb}{0,0,0.6}
\definecolor{darkred}{rgb}{0.6,0,0}
\definecolor{darkpurple}{rgb}{0.5,0,0.5}

\usepackage{hyperref}

\hypersetup{
bookmarksopen=true,
pdftitle=Power countings versus physical scalings in 1D DES,
pdfauthor=E Agoritsas and V Lecomte, 
pdftoolbar=false, % toolbar hidden
pdfstartview={FitH},		% fits the width of the page to the window
pdfmenubar=true,			%menubar shown
pdfhighlight=/O,			%effect of clicking on a link
colorlinks=true,			%couleurs sur les liens hypertextes
urlcolor=darkblue,
citecolor=darkblue,		%couleur des liens pour les citations
linkcolor=darkpurple	%darkred %couleur des liens hypertextes internes
}

\newcommand{\argc}[1]{\left[#1\right]}
\newcommand{\arga}[1]{\left\lbrace #1\right\rbrace }
\newcommand{\argp}[1]{\left(#1\right)}
\newcommand{\valabs}[1]{\vert #1\vert}
\newcommand{\moy}[1]{\left\langle  #1 \right\rangle }
\newcommand{\moydes}[1]{\overline{#1}}

\newcommand{\specialcell}[2][c]{\begin{tabular}[#1]{@{}c@{}}#2\end{tabular}}

\newcommand{\tf}{t_{\rm{f}}}

\newcommand{\cc}{\rm{c}}

\newcommand{\FF}{\rm{F}}
\newcommand{\KPZ}{\rm{KPZ}}
\newcommand{\var}{\rm{var}}

\newcommand{\HHh}{\hat {\mathcal H}}
\newcommand{\HHt}{\widetilde {\mathcal H}}

\renewcommand{\phi}{\varphi}

\newcommand{\dbar}{d\mkern-6mu\mathchar'26\!}

\newcommand\ronron[1]{%
  {% make an Ord atom
   \mathop{\kern0pt #1}\limits^{% set a box over the variable
     \vbox to-1.85ex{
       \kern-2ex % lower the ring accents
       \hbox to 0pt{\hss\normalfont\kern.1em \r{}\kern-.3em \r{}\hss}%
       \vss % fill
     }% end of \vbox
   }% end of the superscript
  }% end of \mathop
}

\newcommand\titilde[1]{%
  {% make an Ord atom
   \mathop{\mathop{\kern0pt #1}\limits^{% set a box over the variable
     \vbox to-1.85ex{
       \kern-1.9ex % lower the ring accents
       \hbox to 0pt{\hss\normalfont\kern.1em \textasciitilde\kern-.01em\hss}%
       \vss % fill
     }% end of \vbox
   }% end of the superscript
  }\limits^{%
     \vbox to-1.85ex{
       \kern-1.6ex % lower the ring accents
       \hbox to 0pt{\hss\normalfont\kern.1em \textasciitilde\kern-.01em\hss}%
       \vss % fill
  }}
  }% end of \mathop
}

\providecommand{\tfrac}[2]{\textnormal{$\frac{#1}{#2}$}}
\providecommand{\eqref}[1]{\eref{#1}}
\providecommand{\text}[1]{{\rm{#1}}}

\begin{document}

%-------------------------

\title{%
Power countings \textit{versus} physical scalings
in disordered elastic systems
\\
{\large --~Case study of the one-dimensional interface}
}

%-------------------------

\author{Elisabeth Agoritsas$^{1,2}$ and Vivien Lecomte$^3$}

\address{$^1$
Universit\'e Grenoble Alpes, LIPhy, F-38000 Grenoble, France
}
\address{$^2$
CNRS, LIPhy, F-38000 Grenoble, France
}
\address{$^3$
Laboratoire Probabilit\'e  et Mod\`eles Al\'eatoires, CNRS UMR 7599, Universit\'e Paris Diderot, Paris Cit\'e Sorbonne, B\^atiment Sophie Germain, Avenue de France, F-75013 Paris, France}

\ead{\\elisabeth.agoritsas@univ-grenoble-alpes.fr\\ vivien.lecomte@math.univ-paris-diderot.fr}
%	elisabeth.agoritsas@univ-grenoble-alpes.fr
%	vivien.lecomte@univ-paris-diderot.fr

\vspace{10pt}

%_____________________________________________________________________________________________________

\begin{abstract}

We study the scaling properties of a one-dimensional interface at equilibrium,
at finite temperature and in a disordered environment with a finite disorder correlation length.
We focus our approach on the scalings of its geometrical fluctuations as a function of its length.
At large lengthscales, the roughness of the interface, defined as the variance of its endpoint fluctuations, follows a power-law behaviour whose exponent characterises its superdiffusive behaviour.
In 1+1 dimensions, the roughness exponent is known to be the characteristic $2/3$ exponent of the Kardar-Parisi-Zhang (KPZ)
universality class.
An important feature of the model description is that its Flory exponent, obtained by a power counting argument on its Hamiltonian, is equal to $3/5$ and thus does not yield the correct KPZ
roughness exponent.
In this work, we review the available power-counting options, and relate the physical validity of the exponent values that they predict, to the existence (or not) of well-defined optimal trajectories in a large-size or low-temperature asymptotics.
We identify the crucial role of the `cut-off' lengths of the problem (the disorder correlation length and the system size), which one has to carefully follow throughout the scaling analysis.
To complement the latter, we device a novel Gaussian Variational Method (GVM) scheme to compute the roughness, taking into account the effect of a large but finite interface length.
Interestingly, such a procedure yields the correct KPZ roughness exponent, instead of the Flory exponent usually obtained through the GVM approach for an infinite interface.
We explain the physical origin of this improvement of the GVM procedure and discuss possible extensions of this work to other disordered systems.

%---------------------------------

\vspace{2pc}

\end{abstract}

%_____________________________________________________________________________________________________

\newpage
\begin{small}
	\tableofcontents
\end{small}


%Uncomment for keywords


%\maketitle
 

%_____________________________________________________________________________________________________
%_____________________________________________________________________________________________________
\newpage
\section{Introduction}
\label{sec:introduction}

%-------------------------
%% PRAISE BUT BEWARE THE SCALINGS

Scaling analysis is a very powerful tool in theoretical physics, as it often allows to take remarkably fast shortcuts of otherwise long and cumbersome computations.
Indeed, in many cases, the dominant physical behaviour predicted by a model can be pointed out by literally back-of-the-envelope calculations based on such scaling analyses.
Nevertheless, at the same time, scaling arguments are usually presented as untrustworthy beforehand, but rather as \textit{a posteriori} explanations of a given problem, constructed by reasoning on the interplay between its `typical' scales.
They allow in fact to recover some physical intuition on the model predictions, disregarding the level of technical difficulties associated to their derivation.
There is thus a great temptation to perform straightforward power countings on a Hamiltonian or an equation of motion, as in the so-called `Imry-Ma'~\cite{imry-ma_1975_PhysRevLett35_1399,nattermann_1987_EPL4_1241} or `Flory' constructions~\cite{gennes_scaling_1979,halpin-healy_1989_PhysRevLett62_442}, and then to assume that the corresponding scaling behaviours are physically meaningful.
But in order to assess this, one must validate independently the implicit assumptions on which such scaling arguments rely.
In particular, when dealing with averages of observables in field theories,
a natural approach consists in examining the explicit implementation of such scaling procedures directly on the corresponding path integrals.

%-------------------------
%% STANDARD CLASSICAL STATPHYS PROBLEM: ROUGHNESS OF 1+1 DES

From that perspective, one problem of particular interest is the characterisation of the geometrical fluctuations of a one-dimensional (1D) interface, with a short-range elasticity and at equilibrium at finite temperature in a quenched random-bond Gaussian disorder.
As a standard %statistical-physics
problem in classical disordered systems, it has been extensively studied over the last decades --~analytically, numerically, and also in relation with experiments \cite{agoritsas_2012_ECRYS2011,TG_DES_Springer}.
It can moreover be exactly mapped on the characterisation of a directed polymer (DP) end-point fluctuations in a disordered 2D plane, whose free energy evolves according to the 1D Kardar-Parisi-Zhang (KPZ) equation
\cite{kardar_1986_originalKPZ_PhysRevLett56_889,huse_henley_fisher_1985_PhysRevLett55_2924}
with `sharp-wedge' initial conditions.
This specific problem hence relates more broadly to the 1D KPZ universality class \cite{corwin_2011_arXiv:1106.1596,halpin-healy_takeuchi_2015_JStatPhys160_794},
%
%% \cite{huse_henley_fisher_1985_PhysRevLett55_2924}
%%		Huse, David A. and Henley, Christopher L. and Fisher, Daniel S.
%%		"Huse, Henley, and Fisher respond"
%% \cite{kardar_1986_originalKPZ_PhysRevLett56_889}
%%		Kardar, Mehran and Parisi, Giorgio and Zhang, Yi-Cheng
%%		"Dynamic Scaling of Growing Interfaces"
%% \cite{corwin_2011_arXiv:1106.1596}
%%		Ivan Corwin
%%		"The Kardar-Parisi-Zhang equation and universality class"
%% \cite{halpin-healy_takeuchi_2015_JStatPhys160_794}
%%		Halpin-Healy, Timothy and Takeuchi, Kazumasa A.
%%		"A KPZ Cocktail-Shaken, not Stirred..."
and as we will see, it provides an interesting illustration of and a useful insight into some of the issues that may arise when invoking scaling arguments.

%-------------------------
%% STILL AN OPEN ISSUE: EXACT COMPUTATION OF THE ROUGHNESS FUNCTION

One observable that still has to be computed exactly, even after all these decades of studies, is the complete roughness function ${B(t)}$ of the  static 1D interface, \textit{i.e.}~the variance of its relative displacements as a function of the lengthscale $t$, with a finite disorder correlation length ${\xi}$ and at finite temperature ${T}$.
For an uncorrelated %$\delta$-correlated
disorder (${\xi=0}$), the scaling of the asymptotic roughness at large lengthscales has been known for a long time to be
${B(t) \sim T^{-2/3} t^{4/3}}$ at ${t \to \infty}$, \textit{i.e.}~with the KPZ roughness exponent ${\zeta_{\rm{KPZ}}=2/3}$ \cite{halpin-healy_takeuchi_2015_JStatPhys160_794},
although its numerical prefactor itself has only recently been exactly computed \cite{schehr_2012_JStatPhys149_385}.
%% \cite{schehr_2012_JStatPhys149_385}
%%	Gregory Schehr
%%	"{Extremes of N Vicious Walkers for Large N: Application to the Directed Polymer and KPZ Interfaces}"
%
However, at ${\xi>0}$, the only available analytical predictions for the complete roughness function have been obtained in a Gaussian-Variational-Method (GVM) approximation scheme
\cite{agoritsas_2010_PhysRevB_82_184207,agoritsas-2012-FHHpenta,phdthesis_Agoritsas2013}.
Such GVM approximations provide precious analytical shortcuts,
but the predicted roughness is not exact %and most importantly not all of its features are to be trusted.
so it is crucial to identify which of its features can be trusted (or not).
For instance, the GVM procedure starting from the Hamiltonian of an infinite 1D interface predicts the wrong asymptotic scaling
${B(t) \sim T^0 t^{6/5}}$ at ${t \to \infty}$,
\textit{i.e.}~with the so-called Flory exponent ${\zeta_{\rm{F}}=3/5}$.
This scaling can otherwise be obtained by a straightforward power counting on the Hamiltonian \cite{agoritsas_2010_PhysRevB_82_184207,phdthesis_Agoritsas2013}.
This is not a coincidence, since these GVM scaling predictions are intimately related to the power countings on the Hamiltonian, and in turn, their identification as the `true' physical scalings corresponds to specific assumptions in the saddle-point analysis of the underlying path integrals.

%-------------------------
%% AIM OF THE STUDY AND OUTLINE

The aim of the present study is precisely to investigate when and why do simple power countings help to correctly predict the asymptotic roughness ${B(t)}$, and to understand how they relate either to the existence of path-integral saddle points or to GVM predictions.
The outline of this paper is thus the following.
We first recall in \sref{sec:def-models} the definitions of the model and observable that we are considering.
Then in \sref{sec:powercounting-Flory} we review different options of rescalings based on straightforward power countings, either on the 1D interface Hamiltonian or on the DP endpoint free energy,  without or with replicas.
In \sref{sec:saddlepts-optimaltrajectories} we identify how the physically meaningful power countings can be derived from a saddle point analysis of path integrals, \textit{i.e.}~from the existence of optimal trajectories either at zero temperature or at asymptotically large timescales.
Similarly, we discuss why the Flory scaling on the Hamiltonian fails to predict the physical roughness exponent at large lengthscales.
%
%% TEASER
A key ingredient will be to keep simultaneously track of a finite disorder correlation length $\xi$, a finite temperature $T$, and a finite total length $\tf$ of the interface, in order to properly rescale the observable averages.
Combining these power-counting and saddle-point considerations,
we revisit in \sref{sec:GVM} the GVM approximation schemes for the computation of the roughness,
recalling on the one hand the main assumptions and predictions of previous GVM computations
\cite{agoritsas_2010_PhysRevB_82_184207,agoritsas-2012-FHHpenta,phdthesis_Agoritsas2013}.
On the other hand, we show how it is possible to recover by GVM the correct KPZ asymptotic scaling of the roughness (with a roughness exponent ${\zeta=2/3}$), in other words how to avoid the usual Flory pitfall leading to ${\zeta_{\rm{F}}=3/5}$.
At last, we conclude in \sref{sec:conclusion}
%by summarising the physical picture we have obtained, and presenting some perspective to this work.
%by presenting a synthetic summary of
on the physical picture we have obtained --~that can \textit{a priori} be applied to other disordered systems~-- and present some perspectives to this work.
Note that we have gathered in appendix~\ref{sec:details-gvm-comp} and appendix~\ref{sec:app-iter-num_GVM} the details of a GVM computation and of a numerical procedure used to analyse GVM variational equations.

%_____________________________________________________________________________________________________
%_____________________________________________________________________________________________________

\section{Static 1D interface and growing 1+1 directed polymer (DP)}
\label{sec:def-models}
%% "N-ième variation sur un même terme"  avec N --> {0,\infty}, comme usuellement dans la méthode des répliques

%-------------------------

We start by defining the model of the 1D interface that we are considering (in \sref{sec:def-models-Hamiltonian-1Dinterface}),
and the observable we want to compute (in \sref{sec:def-models-geom-fluct-roughness})
--~its static roughness, which quantifies the variance of its geometrical fluctuations at thermal equilibrium.
A broader context and more details to this model are given for instance in \cite{agoritsas_2010_PhysRevB_82_184207,agoritsas_2012_FHHtri-analytics,phdthesis_Agoritsas2013}.
We then briefly recall how this specific problem can be mapped on studying the endpoint fluctuations of a random walk or a growing directed `polymer' (in \sref{sec:def-models-mapping-DP-KPZ}),
%whose free-energy evolves according to a KPZ equation.
%
along with the main features of these fluctuations that are exactly known (in \sref{sec:def-models-previous-results}).
% of the geometrical and free-energy fluctuations of the DP endpoint.
%
So at the end we will have the explicit expressions of the quantities on which, in the next section, we will perform power countings and scaling arguments, which are either path integrals or simple integrals, either before or after using the so-called `replica trick'. %in order to average over disorder.
The reader already familiar to this method could thus go directly to \sref{sec:powercounting-Flory}.

%-------------------------
\subsection{Hamiltonian of the 1D interface}
\label{sec:def-models-Hamiltonian-1Dinterface}

%% 1) Hamiltonian of the 1D interface with short-range elasticity and random-bond disorder with a finite correlation length ${\xi>0}$

\begin{figure}
\begin{center}
\includegraphics[width=0.8\textwidth]{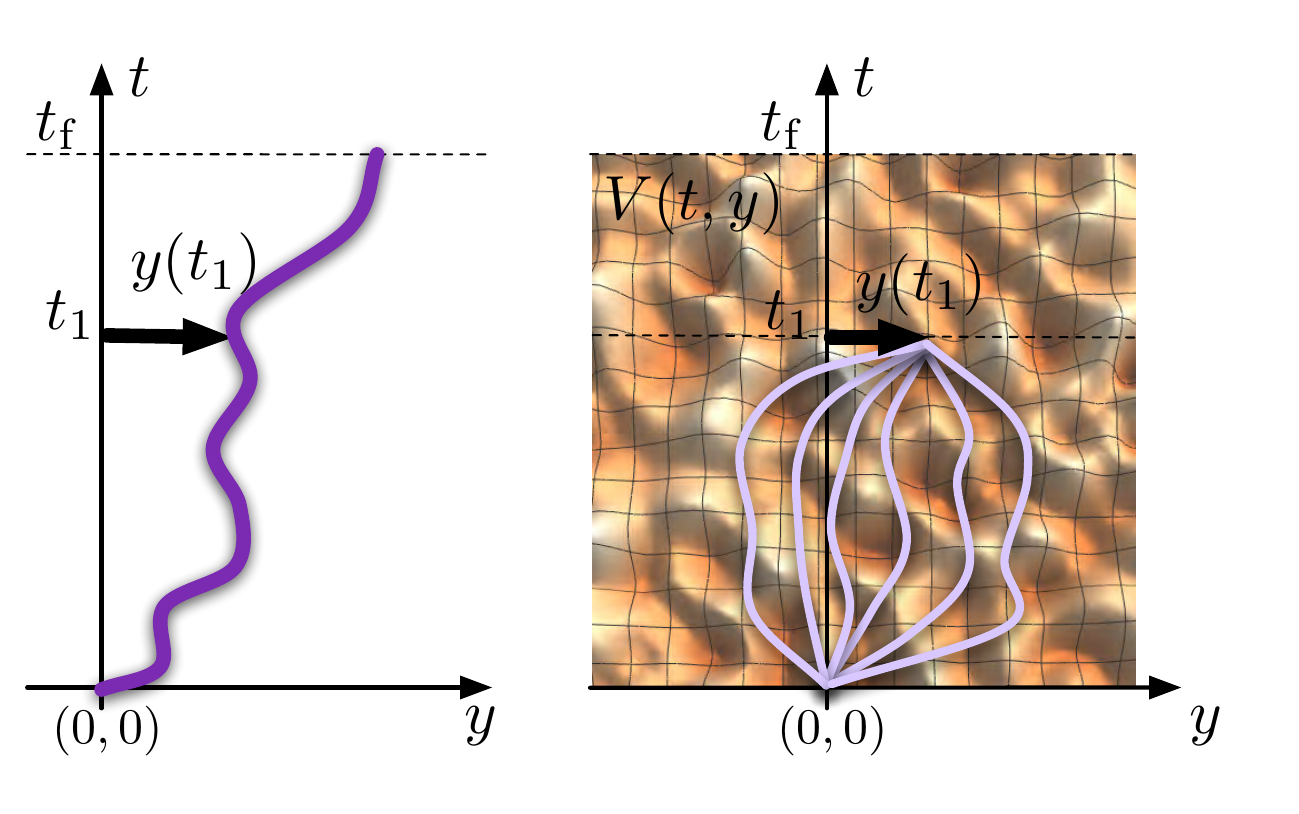}
\caption{
\label{fig-mapping-1Dinterface-DP}
Illustration of the mapping between the static 1D interface and the growing 1+1 DP.
(\textbf{Left})~Configuration of a 1D interface, of finite size $\tf$, parametrized by the `displacement field' ${y(t)}$, with respect  to a flat reference axis.
(\textbf{Right})~DP representation of a finite segment of the interface: the DP endpoint encodes all the possible paths from ${t=0}$ to ${t=t_1}$, in the disordered energy landscape ${V(t,y)}$, into its free energy ${F_V(t_1,y)}$.
}
%% a) 1D interface, axes $(t,y)$ (internal versus transverse directions), $\tf$, $y(t)$
%% b) 1+1 DP endpoint: distribution ${\mathcal{P}_V(t,y)=e^{-\frac{1}{T} F_V(t,y)}}$
%% with underlying quenched random potential ${V(t,y)}$.
\end{center}
\end{figure}

We consider a 1D interface with a short-range elasticity, embedded in a quenched random-bond (\textit{i.e.}~short-range) Gaussian disorder with a finite correlation length $\xi$.
We restrict ourselves to the case %where it has no bubbles no overhangs, meaning that
where its position can be parametrized by a univalued function with respect to a flat reference axis,
the `displacement field' ${y(t) \in \mathbb{R}}$ at a position ${t \in \mathbb{R}}$,
as illustrated in \fref{fig-mapping-1Dinterface-DP}.
We thus restrict ourselves to the case of an interface without bubbles nor overhangs.

Assuming that the interface is only weakly distorted, in other words that it can be described in the `elastic limit', the energy associated with a configuration ${y(t)}$ is given by the following Hamiltonian:
\begin{eqnarray}
\label{eq-Hamiltonian-original}
& \mathcal{H} \argc{y(t),V;\tf}
	= \int_0^{\tf} dt \, \argc{\frac{c}{2} \argp{\partial_t y(t)}^2 + V(t,y(t))}
\end{eqnarray}
where $\tf$ is the total length of the interface,
the elastic constant $c$ is the elastic energy per unit of length,
and ${V}$ is a quenched random potential. %in which the elastic interface is embedded.
Characterised by its distribution ${\bar{\mathcal{P}}\argc{V}}$, the disorder is assumed to be 
%translation invariant in distribution and
Gaussian, hence fully described by its zero mean and its two-point correlation function:
\begin{eqnarray}
\label{eq-disorder-mean}
\overline{V(t,y)}=0
\\
\label{eq-disorder-variance}
\overline{V(t,y)V(t',y')}= D \, \delta(t-t') \, R_{\xi}(y-y')
\end{eqnarray}
with $\overline{\vphantom{|}\cdots}$ denoting the statistical average over disorder.
Accounting for a random-bond disorder, the correlator ${R_{\xi}(y)}$ is assumed to decay exponentially fast at large $y$, %\lim_{\valabs{y} \to \infty} R_1(y)/y^n \to 0 \quad \forall n \in \mathbb{N}
and to scale as a Gaussian function would typically do:
\begin{eqnarray}
\label{eq-disorder-correlator-scaling}
R_{\xi}(y) = \xi^{-1} R_{1}(y/\xi)
\qquad {\rm{\textit{e.g.}}} \qquad
R_{\xi}^{\rm{G}}(y) = \frac{e^{-y^2/(4\xi^2)}}{\sqrt{4 \pi} \xi}
\end{eqnarray}

The Hamiltonian \eref{eq-Hamiltonian-original} has a quadratic deterministic part, which reduces the problem to Gaussian path integrals in absence of disorder, and a linear additive stochastic part, which allows to compute the averages over disorder.
This is thus a standard `textbook' Hamiltonian.
Moreover, we emphasise that it keeps explicitly track of the \emph{finite total length} $\tf$ of the interface, which has to be taken care of in the course of rescaling procedures, and will actually play a crucial role for the new GVM computation presented in \sref{sec:GVM-upgraded-finite-length}.

%-------------------------
\subsection{Geometrical fluctuations and roughness of the static interface}
\label{sec:def-models-geom-fluct-roughness}

%% 2) Study of the geometrical fluctuations at equilibrium at fixed lengthscale, with or without replicas, and focus on the roughness function.

We want to characterise the geometrical fluctuations as a function of the lengthscale ${t \leq \tf}$,
when the interface is equilibrated at finite temperature $T$.
Denoting by $\moy{\cdots}$ the statistical average over thermal fluctuations at fixed disorder,
this amounts to fix one point of the interface (${y(0)=0}$) and to consider observables depending solely on ${y(\tf)}$:
\begin{eqnarray}
\label{eq-pathintegral-geom-fluct-v1}
\fl
%% Def of the average over thermal fluctuations and quenched disorder
\moydes{\moy{\mathcal{O}\argc{y(\tf)}}}
%% wrt to the Hamiltonian
 &= \int \mathcal{D}V \, \bar{\mathcal{P}}\argc{V}
 		\frac{
 		\int_{y(0)=0} \mathcal{D}y(t) \, \mathcal{O}\argc{y(\tf)} \, e^{-\frac{1}{T} \mathcal{H} \argc{y(t),V;\tf}}
 		}{
 		\int_{y(0)=0} \mathcal{D}y(t) \, e^{-\frac{1}{T} \mathcal{H} \argc{y(t),V;\tf}}}
\\
%% wrt to the replicated Hamiltonian of the 1D interface
\label{eq-pathintegral-geom-fluct-v2}
\fl
&= \lim_{n \to 0} \int_{y_1(0)=0} \!\!\!\!\!\!\!\!\!\!\!\! \mathcal{D}y_1(t) \, \argp{\dots} \int_{y_n(0)=0} \!\!\!\!\!\!\!\!\!\!\!\! \mathcal{D}y_n(t) \, \mathcal{O}\argc{y_1(\tf)} \, e^{-\frac{1}{T} \widetilde{\mathcal{H}} \argc{y_1(t), \dots, y_n(t);\tf}}
\end{eqnarray}
where we have transformed \eref{eq-pathintegral-geom-fluct-v1} into \eref{eq-pathintegral-geom-fluct-v2} by using the `replica trick', %in order to average over disorder,
as detailed for instance in \cite{agoritsas_2010_PhysRevB_82_184207,book_beyond-MezardParisi,castellani-cavagna_2005_JStatMechP05012}.
%
%% \cite{castellani-cavagna_2005_JStatMechP05012}
%%	Tommaso Castellani AND Andrea Cavagna
%%	"Spin-Glass Theory for Pedestrians"
%
The `roughness' function is then simply defined as the variance of the geometrical fluctuations:
${B(\tf) \equiv \moydes{\moy{y(\tf)^2}}}$.

Introducing $n$ replicas of the system and performing the disorder average, we can directly define the replicated Hamiltonian  ${\widetilde{\mathcal{H}} \argc{y_1(t), \dots ,y_n(t);\tf}}$ as:
\begin{eqnarray}
\label{eq-Hamiltonian-replicated-definition}
\fl
\exp \argp{-\frac{1}{T} \widetilde{\mathcal{H}} \argc{y_1(t), \dots ,y_n(t);\tf}}
\equiv
\overline{\ \exp \argp{
-\frac{1}{T} \sum_{a=1}^{n} \mathcal{H} \argc{y(t),V;\tf}
}\ }
\end{eqnarray}
and thanks to the additive and Gaussian nature of the disorder and its zero mean, we have:
\begin{eqnarray}
\label{eq-Hamiltonian-replicated-explicit}
\fl
\widetilde{\mathcal{H}} \argc{y_1(t), \dots ,y_n(t);\tf}
=\int_0^{\tf} dt \, \argc{\frac{c}{2} \sum_{a=1}^{n} \argp{\partial_t y_a(t)}^2 - \frac{D}{T} \sum_{a,b=1}^{n} R_{\xi}(y_a(t)-y_b(t))}
\end{eqnarray}
We emphasise that this is an exact expression, although physically meaningful in the peculiar ${n \to 0}$ limit of \eref{eq-pathintegral-geom-fluct-v1}.

%-------------------------
\subsection{Free energy of the growing 1+1 DP endpoint}
\label{sec:def-models-mapping-DP-KPZ}

%% 3) Mapping on the growing 1+1 directed polymer (DP): endpoint fluctuations at fixed growing time,
%% 4) Study of the free-energy fluctuations via the KPZ equation / Mention the STS for focus on $\bar{F}_V(t,y)$

Since we are focusing on observables depending solely on ${y(\tf)}$, given that ${y(0)=0}$,
this problem can be mapped exactly to the study of a DP endpoint fluctuations in dimension ${1+1}$ (\textit{i.e.}~internal and transverse dimensions are both equal to 1),
as initially stated for instance in \cite{huse_henley_fisher_1985_PhysRevLett55_2924} and illustrated in \fref{fig-mapping-1Dinterface-DP}.

In this language, the statistical average ${\moydes{\moy{\mathcal{O}\argc{y(\tf)}}}}$ becomes straightforwardly ${\moydes{\moy{\mathcal{O}(y)}}_{\tf}}$, in other words the average of an observable depending on the DP endpoint~$y$ after a fixed growing `time' $\tf$, as for instance the roughness function can be obtained as
${B(\tf) = \moydes{\moy{y^2}}_{\tf}}$. %\equiv \moydes{\moy{y(\tf)^2}}
The path integral \eref{eq-pathintegral-geom-fluct-v1} can then be rewritten, before and after the replica trick, as:
\begin{eqnarray}
\label{eq-pathintegral-geom-fluct-DP-v1}
\fl
%% Def of the average over thermal fluctuations and quenched disorder
\moydes{\moy{\mathcal{O}(y)}}_{\tf}
%% wrt to the free-energy of the 1+1 DP
 &= \int \mathcal{D}\bar{F}_V \, \bar{\mathcal{P}}\argc{\bar{F}_V,\tf}
 		\frac{
 		\int_{\mathbb{R}} dy \, \mathcal{O}(y) \, e^{-\frac{1}{T} \argc{F_{V=0} (\tf,y) + \bar{F}_V (\tf,y)}}
 		}{
 		\int_{\mathbb{R}} dy \, e^{-\frac{1}{T} \argc{F_{V=0} (\tf,y) + \bar{F}_V (\tf,y)}}
 		}
\\
\label{eq-pathintegral-geom-fluct-DP-v2}
\fl
%% wrt to the replicated free-energy of the 1+1 DP
 &= \lim_{n \to 0} \int_{\mathbb{R}} dy_1 \, \argp{\dots} \int_{\mathbb{R}} dy_n \, \mathcal{O}(y_1) \, e^{-\frac{1}{T} \widetilde{F} (\tf, y_1, \dots, y_n)}
\end{eqnarray}
with the total free energy ${F_V(t,y)}$ of the DP endpoint at fixed disorder having two contributions:
\begin{eqnarray}
\label{eq-free-energy-original}
& F_V(t,y) = F_{V=0}(t,y) + \bar{F}_V(t,y)
\end{eqnarray}
On the one hand, ${F_{V=0}(t,y)}$ is the free energy in absence of disorder, and it corresponds to a Boltzmann weight given by a normalised Gaussian function of variance ${B_{\rm{th}}(t) = T t/c}$:
\begin{eqnarray}
\label{eq-thermal-free-energy}
& F_{V=0}(t,y)
	= -T \ln \arga{\frac{\exp \argc{- \frac{y^2}{2 B_{\rm{th}}(t)}}}{\sqrt{2 \pi B_{\rm{th}}(t)}}}
	= \frac{c y^2}{2t} + \frac{T}{2}\ln \frac{2 \pi T t}{c}
\end{eqnarray}
On the other hand, the disorder free energy ${\bar{F}_V(t,y)}$
--~strictly zero in absence of disorder~--
is in fact translation invariant in distribution in the $y$ direction, at fixed $\tf$:
\begin{eqnarray}
\bar{\mathcal{P}}\argc{\bar{F}_V(\tf,\cdot+Y),\tf}
=\bar{\mathcal{P}}\argc{\bar{F}_V(\tf,\cdot),\tf}
\end{eqnarray}
as is also the underlying random potential $V(t,y)$ \eref{eq-disorder-variance},
thanks to the Statistical Tilt Symmetry (STS), as detailed in Appendix~B of \cite{agoritsas_2012_FHHtri-analytics}.

Similarly to \eref{eq-pathintegral-geom-fluct-v1}-\eref{eq-pathintegral-geom-fluct-v2}, introducing $n$ replicas of the system and performing the disorder average, we can directly define the replicated free energy as
\begin{eqnarray}
\label{eq-free-energy-replicated-definition}
\exp \argp{-\frac{1}{T} \widetilde{F} (\tf, y_1, \dots ,y_n)}
\equiv
\overline{\exp \argp{
-\frac{1}{T} \sum_{a=1}^{n} F_V (\tf,y_a)
}}
\end{eqnarray}
Its distribution is however not Gaussian, except for in the steady state ($\tf=\infty$) for the ${\xi=0}$ case.
Hence, the replicated free energy includes \textit{a priori} all the cumulants of ${\bar{\mathcal{P}}\argc{\bar{F}_V(t,y),\tf}}$, along with a non-zero mean:
\begin{eqnarray}
\label{eq-free-energy-replicated-explicit}
\fl
\widetilde{F} (\tf, y_1, \dots, y_n)
	= & \sum_{a=1}^{n} F_{V =0} (\tf, y_a)
		+ \frac{1}{4T} \sum_{a,b=1}^{n} \bar{C}(\tf,y_a-y_b)
\\		
\nonumber
\fl
	& + n \overline{\bar{F}_V(\tf,0)}
		- \frac{n^2}{2T} \overline{\bar{F}_V(\tf,0)^2}^c + \textnormal{higher cumulants}
\end{eqnarray}
with the two-point correlator
${\bar{C}(\tf,y_a-y_b) = \overline{\argp{\bar{F}_V(\tf,y_a)-\bar{F}_V(\tf,y_b)}^2}}$.

At last, in order to close this definitions section, we recall that the total free energy ${F_V(t,y)}$ evolves according to a 1D KPZ equation with `sharp wedge' initial condition at ${t=0}$~\cite{huse_henley_fisher_1985_PhysRevLett55_2924,halpin_zhang_1995_PhysRep254}:
\begin{eqnarray}
\label{eq-free-energy-KPZ}
\fl
& \partial_t F_V(t,y)
 = \frac{T}{2c} \partial_y^2 F_V(t,y)
 	-\frac{1}{2c} \argc{\partial_y F_V(t,y)}^2
 	+ V(t,y)
\\
\label{eq-free-energy-KPZ-sharpwedge}
\fl
& e^{-F_V(0,y)/T} = \delta(y)
\end{eqnarray}
and the disorder free energy itself follows a `tilted' KPZ equation with flat initial condition at ${t=0}$:%~\cite{agoritsas_2012_FHHtri-analytics}:
\begin{eqnarray}
\label{eq-free-energy-tilted-KPZ}
\fl
& \partial_t \bar{F}_V(t,y) + \frac{y}{t} \partial_y \bar{F}_V(t,y)
 = \frac{T}{2c} \partial_y^2 \bar{F}_V(t,y)
 	-\frac{1}{2c} \argc{\partial_y \bar{F}_V(t,y)}^2
 	+ V(t,y)
\\
\label{eq-free-energy-tilted-KPZ-flat}
\fl
& \bar{F}_V(0,y) = 0
\end{eqnarray}
as we have presented in \cite{agoritsas_2012_FHHtri-analytics},
and is furthermore detailed in the section~3.3 of \cite{phdthesis_Agoritsas2013}.

%-------------------------
\subsection{Exactly known scalings ${\xi=0}$ and predictions at ${\xi>0}$}
\label{sec:def-models-previous-results}

%% 5) Summary of the results which are known ($t_{\rm{sat}}$, $L_{\rm{c}}$, \dots)

%-------------------------

One of the advantages of the mapping from the 1D interface to the 1+1 DP is that it transforms the path integrals ${\int \mathcal{D}y(t)}$ %${\moydes{\moy{\mathcal{O}\argc{y(\tf)}}}}$
of \eref{eq-pathintegral-geom-fluct-v1} into
simple integrals ${\int dy}$ %${\moydes{\moy{\mathcal{O}(y)}}_{\tf}}$
of \eref{eq-pathintegral-geom-fluct-DP-v1}.
%%
%% \cite{corwin_2011_arXiv:1106.1596}
%%	Ivan Corwin
%%	"The Kardar-Parisi-Zhang equation and universality class"
%% \cite{quastel_lecture-notes-arizona_2012}
%%	Quastel, Jeremy
%%	"Introduction to {KPZ}"
%% \cite{halpin-healy_takeuchi_2015_JStatPhys160_794}
%%	Halpin-Healy, Timothy and Takeuchi, Kazumasa A.
%%	"A KPZ Cocktail-Shaken, not Stirred..."
%
From the point of view of the static interface, it amounts to considering an effective description at fixed lengthscale or `time' $\tf$,
in which the Boltzmann weights of possible trajectories ${y(t)}$ at intermediate `times' ${t \in [0,\tf]}$, at fixed disorder $V$, are encoded in a normalised spirit into the evolution of the free energy.

In order to do so, the price to pay is that the corresponding free energy ${F_V(t,y)}$ follows a non-linear stochastic  partial differential equation, namely the 1D KPZ equation~\eref{eq-free-energy-KPZ}, which has been remarkably tricky to tackle over the last three decades~\cite{corwin_2011_arXiv:1106.1596,quastel_lecture-notes-arizona_2012}.
Nevertheless, the case of the 1D KPZ equation with an uncorrelated noise
--~\textit{i.e.}~our random potential ${V(t,y)}$ in \eref{eq-free-energy-KPZ} with ${\xi=0}$~--
has been recently elucidated, for different initial conditions and in particular for the sharp-wedge condition \eref{eq-free-energy-KPZ-sharpwedge} that is relevant for us.
We refer the interested reader to the recent short review \cite{halpin-healy_takeuchi_2015_JStatPhys160_794}, which retraces the main contributions in the field and their corresponding references.
%and to our recent paper \cite{agoritsas_garcia-garcia_VL_2016_Arxiv-1605.04405} for technical aspects in the quasistatic regime.
%
Thereafter we selectively recall the $\xi=0$ features related to our scalings considerations,
with respect to both the interface roughness and the DP endpoint free-energy fluctuations at large~$\tf$:

%-------------------------

\begin{itemize}

\item[$\bullet$]
%------
%% Steady-state distribution of the free energy (at infinite $\tf$)

At ${\xi=0}$, the steady state distribution of ${F_V(t,y)}$ is purely Gaussian,
as initially stated in \cite{huse_henley_fisher_1985_PhysRevLett55_2924}
and rederived for instance in \cite{halpin-healy_1989_PhysRevLett62_442,agoritsas_2012_FHHtri-analytics} from the Fokker-Planck equation at infinite `time':
\begin{eqnarray}
\label{eq-steady-state-distribution-F-Gaussian-1}
 \bar{\mathcal{P}}_{\rm{st}}\argc{\bar{F}}
 \propto \exp \arga{- \frac{T}{2cD} \int_{\mathbb{R}} dy \, \argc{\partial_y \bar{F}(y)}^2}
\\
\label{eq-steady-state-distribution-F-Gaussian-2} 
\Rightarrow \,
 \overline{\argp{\bar{F}(y_a)-\bar{F}(y_b)}^2}=\frac{cD}{T} \valabs{y_a-y_b}
\end{eqnarray}
In other words, the steady-state free energy is a (`double-sided'-)Brownian process of the coordinate~$y$ and thus scales in distribution as 
%${\bar{F}(a \bar y) \stackrel{(d)}{=} \argp{\frac{cD}{T} a}^{1/2}}{\bar{F}_1( \bar y)}$, where ${\bar{F}_1( \bar y)}$ is a Brownian motion of unit variance.
%
%We will denote below this property in short by: 
${\bar{F}(y) \stackrel{(d)}{\sim} \argp{\frac{cD}{T} y}^{1/2}}$
(see~\sref{sec:powercounting-Flory-free-energy} for a detailed formulation in terms of rescaling properties).

\item[$\bullet$]
%------
%% Approaching the steady-state distribution of the free energy (at asymptotically large $\tf$)

Approaching the steady state, at asymptotically large $\tf$, the distribution ${\bar{\mathcal{P}}\argc{F_V(\tf,y)}}$ is not Gaussian anymore, and it is described by an `Airy' process.
In particular, its two-point correlator displays two regimes in $y$, at small $y$ the same Brownian scaling as in the steady state:
%
% Insert here a figure of a Brownie, with the shape of a scale.
%
\begin{eqnarray}
\label{eq-large-times-two-point-correlator-Cbar}
\fl
\bar{C}(\tf,y)
=\overline{\argp{\bar{F}_V(\tf,y)-\bar{F}_V(\tf,0)}^2}
\mathop{\sim}_{\tf \to \infty}
\frac{cD}{T} \valabs{y},
\quad \textnormal{for} \quad \valabs{y} < \sqrt{B(\tf)}\sim \tf^{2/3} %\sim \argp{\frac{D}{cT}}^{1/3} t^{2/3}
\end{eqnarray}
%with ${B(t) \sim \argp{\frac{D}{cT}}^{1/3} \tf^{2/3}}$,
and at large $y$ it saturates to a plateau ${\frac{cD}{T} \sqrt{B(\tf)}}$, at ${\valabs{y} > \sqrt{B(\tf)}}$.
Moreover, being non-Gaussian, the distribution ${\bar{\mathcal{P}}\argc{F_V(\tf,y)}}$ displays non-trivial higher $n$-point correlators, but for our scaling considerations we can focus solely on ${\bar{C}(\tf,y)}$.
At sufficiently large $\tf$ the Brownian scaling is still the dominant one for the disorder free energy:
${\bar{F}_V(\tf,y) \stackrel{(d)}{\sim} \argp{\frac{cD}{T} y}^{1/2}}$, and this for an increasingly wider range of $y$ as ${\tf \to \infty}$.

\item[$\bullet$]
%------
%% Asymptotic roughness (random-manifold vs thermal)

In \eref{eq-large-times-two-point-correlator-Cbar}, ${B(\tf)}$ is nothing but the DP endpoint variance, and both its asymptotic scalings are known:
\begin{eqnarray}
\label{eq-exact-scalings-asymptotic-roughness-large-tf}
B(\tf)
\mathop{\sim}_{\tf \to \infty}
\argp{\frac{D}{cT}}^{2/3} \tf^{2 \zeta_{\rm{KPZ}}}
& \quad \textnormal{with} \quad \zeta_{\rm{KPZ}}=2/3
\\
\label{eq-exact-scalings-asymptotic-roughness-small-tf}
B(\tf)
\mathop{=}_{\tf \to 0}
B_{\rm{th}}(t)
\stackrel{\eref{eq-thermal-free-energy}}{=}
 \frac{T}{c} \tf^{2 \zeta_{\rm{th}}}
& \quad \textnormal{with} \quad \zeta_{\rm{th}}=1/2
\end{eqnarray}
with a single crossover scale at ${{\tf}_* = T^5/(cD^2)}$.
We can immediately notice that, according to these scalings, the ${T \to 0}$ limit of the ${(\xi=0)}$-roughness is ill-defined, as the amplitude of ${B(\tf)}$ would violently diverge at fixed but large $\tf$.

\end{itemize}

%-------------------------

These different features have been proven to be exact in the case of a completely uncorrelated disorder ($\xi=0$),
yet their mathematical derivation crucially depends on the specific limit ${\xi\to 0}$ (see~\cite{quastel_lecture-notes-arizona_2012} for a pedagogical exposition).
An important issue is thus to assess the robustness of these different scaling features when having simultaneously ${\xi>0}$ and a finite temperature ${T>0}$.
In a series of successive works
\cite{agoritsas_2010_PhysRevB_82_184207,
agoritsas-2012-FHHpenta,
agoritsas_2012_FHHtri-analytics,
agoritsas_2012_FHHtri-numerics,
phdthesis_Agoritsas2013}, we have in fact investigated the consequences of a finite disorder correlation length ${\xi>0}$ as defined in \eref{eq-disorder-variance}-\eref{eq-disorder-correlator-scaling},
using in particular scaling arguments and GVM computation schemes for the roughness,
two complementary types of analytical approaches that we will revisit here.
Regarding the physical picture that has emerged from these works, it can be summarised into three points, that will actually guide the course of our presentation throughout the next sections:
\begin{itemize}
%------
\item[\textit{(i)}]

the previous asymptotic scalings at large $\tf$ seem to be robust to the addition of a finite ${\xi}$, on the one hand the Brownian scaling for the disorder free energy ${\bar{F}_V(\tf,y) \stackrel{(d)}\sim y^{1/2}}$, and on the other hand the KPZ roughness exponent for ${B(\tf) \sim \tf^{4/3}}$;
%the Brownian scaling being in fact confirmed by a non-perturbative renormalization group study of KPZ with $\xi>0$};

%------
\item[\textit{(ii)}]

however, the \emph{prefactors} of these scalings %and their characteristic crossover scales
must display a non-trivial dependence on both $T$ and $\xi$:
\begin{eqnarray}
\label{eq-generic-scaling-KPZ-Brownian}
\bar{F}_V(t,y) \sim (\widetilde{D} y)^{1/2}
\quad \textnormal{and} \quad
B(\tf)
\mathop{\sim}_{\tf \to \infty}
(\widetilde{D}/c^2)^{2/3} \tf^{4/3}
\end{eqnarray}
with the amplitude ${\widetilde{D}}$ of the disorder free energy experiencing a crossover at the characteristic temperature ${T_{\cc}(\xi)=(\xi c D)^{1/3}}$, with the following two asymptotic behaviours:
\begin{eqnarray}
\label{eq-fudging-Dtilde}
 \widetilde{D}=\frac{cD}{T} f(T,\xi)
 \quad \textnormal{with} \quad
 \cases{
 \textnormal{at } T \ll T_{\cc},
 \quad f(T,\xi) \sim T/T_{\cc}
 \\
 \textnormal{at } T \gg T_{\cc},
 \quad f(T,\xi) \to 1 }
\end{eqnarray}
hence the two limits of ${T \to 0}$ and ${\xi \to 0}$ can obviously not be exchanged with impunity;

%------
\item[\textit{(iii)}]

similarly, the typical lengthscale marking the beginning of the asymptotic regime at large $\tf$, 
the so-called `Larkin length'${L_{\cc} (T,\xi)}$, displays the following temperature crossover:
\begin{eqnarray}
\fl
\label{eq-fudging-Larkin-length}
 L_{\cc}(T,\xi) \sim \frac{(T/f(T,\xi))^5}{cD^2},
 \qquad
 L_{\cc} (0,\xi) \sim \frac{T_{\cc}^5}{cD^2},
 \qquad L_{\cc} (T,0) \sim \frac{T^5}{cD^2}
\end{eqnarray}
where its high-temperature behaviour ${L_{\cc} (T,0)}$ coincides as expected with the crossover lengthscale ${{\tf}_*}$ exactly known at ${\xi=0}$ (up to numerical factors that we skip throughout our scaling considerations).

\end{itemize}
%------
%
The only available analytical predictions for the complete crossover in temperature, parametrized by the `fudging' parameter ${f(T,\xi)}$, have been obtained by GVM computation, relating directly this parameter $f$ to the full-replica-symmetry-breaking cutoff \cite{agoritsas_2010_PhysRevB_82_184207,phdthesis_Agoritsas2013}, as we will recall in \sref{sec:GVM}.
We moreover mention that the scaling features in \textit{(i)} can be shown,
within a non-perturbative functional renormalization group study of the 1D KPZ equation at ${\xi>0}$~\cite{NPRG-in-preparation} to be a universal feature of the 1D KPZ equation.

%-------------------------

%% We are considering the continuum model, and beware when dealing with models on lattices!!! Ref. to the appendix of FHH-tri.

At last, we emphasise that these scalings are given for the continuum model of the interface, so that the case of an interface or a DP on a lattice requires a careful translation, as presented for instance in Appendix~E of \cite{agoritsas-2012-FHHpenta}.

%% Appendix~E in \cite{agoritsas-2012-FHHpenta}
%%	"From the discrete to the continuous DP"
%%
%% Appendix~G in \cite{agoritsas_2012_FHHtri-analytics}:
%%	"Scaling laws for a temperature-independent elastic weight"

%_____________________________________________________________________________________________________
%_____________________________________________________________________________________________________
%\newpage
\section{Power countings and Flory arguments}
\label{sec:powercounting-Flory}

%% IN THE INTRODUCTION
%%There is thus a great temptation to perform straightforward power countings on a Hamiltonian or an equation of motion, as in the so-called `Imry-Ma'~\cite{imry-ma_1975_PhysRevLett35_1399,nattermann_1987_EPL4_1241} or `Flory' constructions~\cite{gennes_scaling_1979,halpin-healy_1989_PhysRevLett62_442}, and then to assume that the corresponding scaling behaviours are physically meaningful.

%% IN THE OUTLINE
%% Then in \sref{sec:powercounting-Flory} we systematically list the different options of rescalings based on straighforward power countings, either on the 1D interface Hamiltonian or on the DP endpoint free-energy,  without or with replicas.

%-------------------------

%% List of the different path integrals
%%	with the original Hamiltonian
%%		\eref{eq-pathintegral-geom-fluct-v1}
%%	with the replicated Hamiltonian
%%		\eref{eq-pathintegral-geom-fluct-v2}
%%	with the original free energy
%%		\eref{eq-pathintegral-geom-fluct-DP-v1}
%%	with the replicated free energy
%%	\eref{eq-pathintegral-geom-fluct-DP-v2}

%%We consider the rescalings/power counting of
%%the Hamiltonian of the interface \eref{eq-Hamiltonian-original},
%%and its replicated version \eref{eq-Hamiltonian-replicated-definition},
%%the free energy of the DP endpoint \eref{eq-free-energy-original},
%%and its replicated version \eref{eq-free-energy-replicated-explicit}.

%-------------------------

We now examine systematically the different options of rescalings %of the path-integral representations 
of the statistical averages,
defined either with respect to the interface Hamiltonian in \eref{eq-pathintegral-geom-fluct-v1}-\eref{eq-pathintegral-geom-fluct-v2} (path integrals ${\int \mathcal{D}y(t)}$),
or with respect to the DP endpoint free energy in \eref{eq-pathintegral-geom-fluct-DP-v1}-\eref{eq-pathintegral-geom-fluct-DP-v2} (simple integrals ${\int dy}$).
We focus specifically on the roughness function:
\begin{equation}
\label{eq-def-roughness-both-mappings}
B(t;c,D,T,\xi,\tf)
=\cases{
\overline{\moy{y(t)^2}} \vert_{\arga{c,D,T,\xi,\tf}}
%	${\moydes{\moy{\mathcal{O}\argc{y(\tf)}}}}$
& \textnormal{(static 1D interface)}
\\
\overline{\moy{y^2}_{\tf}} \vert_{\arga{c,D,T,\xi}}
%	${\moydes{\moy{\mathcal{O}(y)}}_{\tf}}$
& \textnormal{(1+1 DP endpoint)}
}
\end{equation}
and we emphasise that we have kept the same set of parameters $\arga{c,D,T,\xi,\tf}$ in the two sides of the mapping in order to avoid any unnecessary confusion.
%although we could have considered the diffusion $\nu$ and ${\lambda}$ for KPZ for instance.

In a nutshell, when we rescale the spatial coordinates
${\arga{t=b \hat{t}, y=a \hat{y}}}$,
we want to review different options that we might have in order to reabsorb the dependence on ${\arga{a,b}}$ into the parameters of the statistical averages \eref{eq-def-roughness-both-mappings}:
\begin{eqnarray}
\label{eq-def-roughness-scaling-function-Bbar}
B(t;c,D,T,\xi,\tf)
= a^2 \, \bar{B}(t/b;c',D',T',\xi',\tf/b)
\end{eqnarray}
As a compromise, from now on we will make a slight abuse of notation in order to discuss these scalings, playing with the set of parameters that are made explicit after $t$ in the roughness function ${B(t;\dots)}$.
The convention will be that we indicate by primes the rescaled parameters within the scaling function ${\bar{B}(\hat{t};\dots)}$.
For instance when rescaling the DP free energy according to the Brownian scaling \eref{eq-generic-scaling-KPZ-Brownian}, we will rather consider:
\begin{eqnarray}
\label{eq-def-roughness-scaling-function-Bbar-Dtilde}
B(\tf;c,\widetilde{D},T,\xi)
= a^2 \, \bar{B}_{\rm{DP}}(\tf/b;c',\widetilde{D}',T',\xi')
\end{eqnarray}
In fact the rewritings %reformulations 
\eref{eq-def-roughness-scaling-function-Bbar}-\eref{eq-def-roughness-scaling-function-Bbar-Dtilde} dictate how the spatial coordinates
${\arga{t=b \hat{t}, y=a \hat{y}}}$
must be conjointly rescaled, in order to rescale with a single overall prefactor the different contributions of the full Hamiltonian ${\mathcal{H} \argc{y(t),V;\tf}}$ or the full free energy ${F_V(\tf,y)}$, along with their associated Boltzmann weights
${\propto \exp \arga{-\mathcal{H} \argc{y(t),V;\tf}/T}}$
and ${\propto \exp \arga{-F_V(\tf,y)/T}}$
in the statistical averages \eref{eq-pathintegral-geom-fluct-v1} and \eref{eq-pathintegral-geom-fluct-DP-v1}.
%with replicas \eref{eq-pathintegral-geom-fluct-v2}-\eref{eq-pathintegral-geom-fluct-DP-v2}

Such power counting corresponds to the so-called `Imry-Ma'~\cite{imry-ma_1975_PhysRevLett35_1399,nattermann_1987_EPL4_1241} or `Flory' constructions~\cite{gennes_scaling_1979,halpin-healy_1989_PhysRevLett62_442}.
Although any rescaling ${a \sim b^{\zeta_{\rm{F}}}}$ compatible with these Flory `rules' is of course allowed,
the physical interpretation of $\zeta_{\rm{F}}$ as being the `true' roughness exponent of the problem is not guaranteed at all.
Nevertheless, we want to emphasise that the scope of such rescalings is broader than the determination of the sole roughness exponent, since they affect the asymptotic behaviour of the scaling functions ${\bar{B}(\hat{t};\cdots)}$ or ${\bar{B}_{\rm{DP}}(\hat{t};\cdots)}$ and might accordingly provide a possibly simpler physical picture, when determining those scaling functions.

We have already partly addressed this issue in \cite{agoritsas_2012_FHHtri-analytics} (in section IV) and in \cite{phdthesis_Agoritsas2013} (in chapter 4 and section 5.5).
Here we recall for reference the power countings on the Hamiltonian and on the free energy (under the Brownian scaling assumption \eref{eq-generic-scaling-KPZ-Brownian}),
and we present in addition their counterparts with replicas.
These different Flory scalings are summarised in \sref{sec:powercounting-Flory-recapitulation}.
In the next \sref{sec:saddlepts-optimaltrajectories} we will focus on three specific cases where a saddle-point analysis of the path integrals allows to confirm or to disqualify the Flory exponent;
we will provide the missing key ingredients that allow to firmly assess what were yet in \cite{agoritsas_2012_FHHtri-analytics,phdthesis_Agoritsas2013} indirect statements, on the existence and properties of saddle points (or from a more physical point of view, of optimal trajectories).

%-------

% \cite{phdthesis_Agoritsas2013}
% Elisabeth Agoritsas, University of Geneva (Switzerland), 2013.
% "Temperature-dependence of a 1D Interface Fluctuations: Role of a Finite Disorder Correlation Length"
%	Chapter 4:		Scaling analysis
% 	Section 5.5: 	Scaling analysis of the DP toymodel
%	Chapter 6:		Gaussian Variational Method (GVM) with replicas
%		And in particular Section 6.6 Concluding remarks, for other GVMs

%-------------------------
\subsection{Hamiltonian ${\mathcal{H} \argc{y(t),V;\tf}}$}
\label{sec:powercounting-Flory-Hamiltonian}

We first recall the expression of the Hamiltonian:
\begin{eqnarray}
\label{eq-Hamiltonian-original-copy1}
 \mathcal{H} \argc{y(t),V;\tf}
 	\stackrel{\eref{eq-Hamiltonian-original}}{=}
 		\mathcal{H}_{\rm{el}} \argc{y(t);\tf} + \mathcal{H}_{\rm{dis}} \argc{y(t),V;\tf}
 \\
\label{eq-Hamiltonian-original-copy2}
 \mathcal{H}_{\rm{el}} \argc{y(t);\tf} \bigg\vert_{c'}
 	= \int_0^{\tf} \!\!\! dt \, \frac{c'}{2} \argp{\partial_t y(t)}^2,
 \\
\label{eq-Hamiltonian-original-copy3}
 \mathcal{H}_{\rm{dis}} \argc{y(t),V;\tf} \bigg\vert_{D',\xi'}
 	= \int_0^{\tf} \!\!\! dt \, V(t,y(t)) \bigg\vert_{D',\xi'}
 \\
 \overline{V(t,y)V(t',y')} \bigg\vert_{D',\xi'} = D' \delta(t-t') R_{\xi'}(y-y')
\end{eqnarray}
where the dependence on the different parameters has been made explicit.
We then rescale the spatial coordinates and the energy
(we set the Boltzmann constant ${k_{\rm{B}}=1}$ so that the temperature has the units of an energy)
according to:
\begin{equation}
 t=b \hat{t}, \quad y = a \hat{y}, \quad T = \widetilde{E} T'
\end{equation}
so that the different parts of the Hamiltonian are rescaled as:
\begin{eqnarray}
\label{eq-rescaling-Ham-part1}
\fl \quad
 \mathcal{H}_{\rm{el}} \argc{y(t);\tf} \bigg\vert_{c'=c}
 	= \frac{ca^2}{b} \, \mathcal{H}_{\rm{el}} \argc{\hat{y}(\hat{t});\hat{\tf}} \bigg\vert_{c'=1}
 \\
\label{eq-rescaling-Ham-part2}
\fl \quad
 \mathcal{H}_{\rm{dis}} \argc{y(t),V;\tf} \bigg\vert_{D'=D,\xi'=\xi}
 %{\begin{array}{l}\scriptstyle{D'=D,}\\ \scriptstyle{\xi'=\xi} \end{array}}
 	\stackrel{(d)}{=} \argp{\frac{bD_0}{a}}^{1/2} \, \mathcal{H}_{\rm{dis}} \argc{\hat{y}(\hat{t}),V;\hat{\tf}} \bigg\vert_{D'=D/D_0,\xi'=\xi/a}
\end{eqnarray}
The rescaling of the disorder Hamiltonian is only valid `in distribution', as indicated by the `$d$';
the random potential of ${V(t,y)}$ is a stochastic variable and as such cannot be rescaled straightforwardly as the deterministic $\mathcal{H}_{\rm{el}}$.
Nevertheless, the scaling of its Gaussian distribution can be deduced from its two-point correlator:
\begin{eqnarray}
\fl \quad
\nonumber
V(t,y)^2 \bigg\vert_{D'=D,\xi'=\xi}
 & \stackrel{(d)}{\sim} &
 \overline{V(t,y) V(t',y')} \bigg\vert_{D'=D,\xi'=\xi}
=
 D \, b^{-1} \delta(\hat{t}-\hat{t}') \, a^{-1} R_{\xi/a}(\hat{y}-\hat{y}')
 \\
\fl \quad
 &=& \frac{D}{ba} \overline{V(\hat{t},\hat{y}) V(\hat{t}',\hat{y}')} \bigg\vert_{D'=1,\xi'=\xi/a}
 \stackrel{(d)}{\sim}
 \frac{D}{ab} V(\hat{t},\hat{y})^2 \bigg\vert_{D'=1,\xi'=\xi/a}
\end{eqnarray}
This scaling in distribution yields a scaling relation on non-fluctuating observables when dealing with statistical averages, after averaging over disorder.

In the statistical averages ${\moydes{\moy{\mathcal{O}\argc{y(\tf)}}}}$,
the Boltzmann weight ${\propto e^{-\frac{1}{T} \mathcal{H} \argc{y(t),V;\tf}}}$ is not modified provided that the elastic and disorder parts of the Hamiltonian scales identically,
\textit{i.e.}~${ca^2/b=(bD/a)^{1/2}}$,
and that the temperature is redefined accordingly:
\begin{eqnarray}
\label{eq-rescaling-Ham-part3}
\fl \quad
 T^{-1} \mathcal{H} \argc{y(t),V;\tf}
 	\bigg\vert_{\!\!\!\! \begin{array}{l} \scriptstyle{c'=c, D'=D,} \\ \scriptstyle{T'=T, \xi'=\xi} \end{array}}
 %\bigg\vert_{c'=c,D'=D,\xi'=\xi}
 	= T'^{-1} \underbrace{\widetilde{E}^{-1} \frac{ca^2}{b}}_{=1} \mathcal{H} \argc{\hat{y}(\hat{t}),V;\hat{\tf}}
 	\bigg\vert_{\!\!\!\! \begin{array}{l} \scriptstyle{c'=1, D'=D/D_0,} \\ \scriptstyle{T'=T/\widetilde{E}, \xi'=\xi/a} \end{array}} 	
 %\bigg\vert_{c'=1,D'=D/D_0,\xi'=\xi/a}
\end{eqnarray}
This `Flory recipe' guarantees that the roughness can thus be rescaled \emph{exactly},
while fixing the relations between the scalings factors ${\lbrace a,b,\widetilde{E} \rbrace}$:
\begin{eqnarray}
\label{eq-rescaling-Ham-Flory-part1a}
\fl \quad
B(t;c,D,T,\xi,\tf)
= a^2 \, \bar{B} \argp{\hat{t}=\frac{t}{b};c'=1,D'=\frac{D}{D_0},T'=\frac{T}{\widetilde{E}},\xi'=\frac{\xi}{a},\hat{\tf}=\frac{\tf}{b}}
\\
\fl \quad
\label{eq-rescaling-Ham-Flory-part1b}
 \qquad  \textnormal{with} \quad \cases{
 a = \argp{D_0^{1/3} c^{-2/3} b}^{3/5}
 \Longleftrightarrow
 b= \argp{D_0^{-1/5} c^{2/5} a}^{5/3},
 \quad \zeta_{\rm{F}}^{(1)}=3/5
 \\
 \widetilde{E} = ca^2/b = \argp{c D_0^2 b}^{1/5} = (c D_0 a)^{1/3}
 }
\end{eqnarray}
%\eref{eq-def-roughness-scaling-function-Bbar}
with $\bar{B}(\hat{t};\dots)$ a scaling function with adimensional parameters, on which the scaling assumptions are actually made.
\emph{So for the Hamiltonian \eref{eq-Hamiltonian-original-copy1}-\eref{eq-Hamiltonian-original-copy2}-\eref{eq-Hamiltonian-original-copy3} the Flory exponent is ${\zeta_{\rm{F}}^{(1)}=3/5}$.}
Nevertheless, we still have two free parameters to fix, and in \eref{eq-rescaling-Ham-Flory-part1a} this would correspond to choose two typical scales and to examine the behaviour of $\bar{B}(\hat{t};\dots)$.

Physically, depending on the regimes we are interested in (low temperature, high temperature, large $\tf$, \dots), we expect that there should be one typical scale and associated rescaling, for which the behaviour of the scaling function $\bar{B}(\hat{t};\dots)$ simplifies radically.
In \eref{eq-rescaling-Ham-Flory-part1a}, we have in fact several natural choices, that we list in the tables~\ref{table-Hamiltonian-part1} and~\ref{table-Hamiltonian-part2}.
Note that in all cases, we choose to rescale the elastic constant to ${c'=1}$.

\begin{table}[h]
\caption{
\label{table-Hamiltonian-part1}
List of the rescaling choices for the Hamiltonian
${ \mathcal{H} \argc{y(t),V;\tf}}$
%of \eref{eq-Hamiltonian-original},
%for ${D_0=D}$.
when imposing ${c'=1}$ and ${D'=1}$ (\textit{i.e.}~${D_0=D}$).
}
%\begin{indented}
%\item[]
\begin{small}
\begin{tabular}{@{} l|cccc|c}
\br
\; \specialcell{\textit{Constraint:} \\~~~~~$D'=1$}	&$b$	&$\widetilde{E}$	&$D_0$	&${a}$	&\specialcell{\textit{Possibly} \\ \textit{relevant for}}
\\
\br
(1a)~~$ T'=1$	&$\frac{T^5}{cD^2}$	&$T$	&$D$	&$\frac{T^3}{cD}$	&${\xi \to 0}$	%`high temperature' ${T \gg T_c}$
\\
\mr
(1b)~~$\xi'=1$	&$\frac{\xi^{1/3}}{D^{5/3} c^{2/3}}=\frac{T_c^5}{cD^2}$	&$(\xi c D)^{1/3}\equiv T_c$	&$D$	&$\xi$	&${T \to 0}$	%`low temperature' ${T \ll T_c}$
\\
\mr
(1c)~~$\hat{\tf}=1$	&$\tf$	&$(c D^2 \tf)^{1/5}$	&$D$	&$\argp{\frac{D^{1/3} \tf}{c^{2/3}}}^{3/5}$	&${\tf \to \infty}$	%`large lenthscale'
\\
\mr
(1d)~~$ T'=T/f$	&$\frac{(T/f)^5}{cD^2}$	&$f$	&$D$	&$\frac{(T/f)^3}{cD}$	&\specialcell{Temperature \\ crossover}
\\
\br
\end{tabular}
\end{small}
%\end{indented}
\end{table}

In \tref{table-Hamiltonian-part1}, we have imposed ${c'=D'=1}$ allowing for a rescaling of the thermal fluctuations (via ${T'=T/\widetilde{E}}$) and of the disorder correlation length (via ${\xi'=\xi/a}$), and as such they are particularly suited for studying the temperature crossover of the roughness \cite{agoritsas_2012_FHHtri-analytics,phdthesis_Agoritsas2013}.
For (1a), (1b) and (1c), $b$ can be identified as the Larkin length ${L_{\cc}(T,\xi)}$ in its different temperature regimes \eref{eq-fudging-Larkin-length}, with in particular ${f=f(T,\xi)}$ the `fudging' parameter describing the complete temperature crossover.
The rescalings (1b) %$\xi'=1$, low temperature
and (1c) %$\hat{\tf} \gg 1$, large lengthscale
will be examined from the point of view of (non-)existent saddle-point of path integrals 
%in \sref{sec:saddlepts-optimaltrajectories-low-temperature}
%and \sref{sec:saddlepts-optimaltrajectories-Flory-failure},
%respectively, and summed up
in~\sref{sec:interpr-saddle-point}.

%------------

\begin{table}[h]
\caption{
\label{table-Hamiltonian-part2}
List of the additional rescaling choices for the Hamiltonian
${ \mathcal{H} \argc{y(t),V;\tf}}$
%of \eref{eq-Hamiltonian-original},
%for ${D_0=D}$.
when imposing ${c'=1}$ and either ${\xi'=1}$ (\textit{i.e.}~${a=\xi}$)
or ${T'=1}$ (\textit{i.e.}~${\widetilde{E}=T}$).
}
%\begin{indented}
%\item[]
\begin{small}
\begin{tabular}{@{} l|cccc|l}
\br
\; \textit{Constraint:}	&$b$	&$\widetilde{E}$	&$D_0$	&${a}$
&\textit{Comment}
\\
\br
(2a)~~${\xi'=1}$, $T'=1$	&$\frac{\xi^2 c}{T}$	&$T$	&$\frac{T^3}{c \xi}$	&$\xi$
&\specialcell{Regime at ${\tf \gg b}$ ill-defined for \\ both limits ${T \to 0}$ and ${\xi \to 0}$.}
\\
\mr
(2b)~~${\xi'=1}$, $\hat{\tf}=1$	&$\tf$	&$\frac{c\xi^2}{T}$	&$\frac{c^2\xi^5}{\tf^3}$	&$\xi$
&\specialcell{Vanishing disorder strength at \\ ${\tf \to \infty}$ $\forall (T,\xi)$ (unphysical!)}
\\
\mr
(2c)~~${T'=1}$, $\hat{\tf}=1$	&$\tf$	&$T$	&$\argp{\frac{T^5}{c \tf}}^{1/2}$	&$\argp{\frac{T \tf}{c}}^{1/2}$
&\specialcell{Vanishing disorder strength and \\ diverging correlation length \\ at ${\tf \to \infty}$.}
\\
\br
\end{tabular}
\end{small}
%\end{indented}
\end{table}

%------------

In \tref{table-Hamiltonian-part2}, we mention a few alternatives that were not presented in \cite{agoritsas_2012_FHHtri-analytics,phdthesis_Agoritsas2013},
which are also valid rescaling with the Flory construction \eref{eq-rescaling-Ham-Flory-part1a}-\eref{eq-rescaling-Ham-Flory-part1b}.
This illustrate the variety of possible typical scales with ${\zeta_{\rm{F}}^{(1)}=3/5}$, although they do not seem relevant for examining the asymptotic roughness to the specific limits we are interested in, namely ${T \to 0}$, ${\xi \to 0}$ or ${\tf \to \infty}$, as briefly commented in the table.

%-------------------------
\subsection{Replicated Hamiltonian ${\widetilde{\mathcal{H}} \argc{y_1(t), \dots ,y_n(t);\tf}}$}
\label{sec:powercounting-Flory-Hamiltonian-replicae}

We now examine the replicated Hamiltonian:
\begin{eqnarray}
\label{eq-Hamiltonian-replicated-copy1}
\fl \quad 
 && \widetilde{\mathcal{H}} \argc{y_1(t), \dots ,y_n(t);\tf}
 	\stackrel{\eref{eq-Hamiltonian-replicated-explicit}}{=}
 		\widetilde{\mathcal{H}}_{\rm{el}} \argc{y_1(t), \dots ,y_n(t);\tf} + \widetilde{\mathcal{H}}_{\rm{dis}} \argc{y_1(t), \dots ,y_n(t);\tf}
 \\
\label{eq-Hamiltonian-replicated-copy2}
\fl
 && \widetilde{\mathcal{H}}_{\rm{el}} \argc{y_1(t), \dots ,y_n(t);\tf} \bigg\vert_{c'}
 	= \int_0^{\tf} \!\!\! dt \, \frac{c'}{2} \sum_{j=1}^{n} \argp{\partial_t y_j(t)}^2
 \\
\label{eq-Hamiltonian-replicated-copy3}
\fl
 && \widetilde{\mathcal{H}}_{\rm{dis}} \argc{y_1(t), \dots ,y_n(t);\tf} \bigg\vert_{D',T',\xi'}
 	=- \frac{D'}{T'} \int_0^{\tf} \!\!\! dt   \sum_{j,k=1}^{n} R_{\xi'}(y_j(t)-y_k(t)) \bigg\vert_{D',T',\xi'}
\end{eqnarray}
where the dependence on the different parameter has been made explicit.
Note that the disorder part has an explicit temperature dependence, contrary to the original Hamiltonian.
We then rescale the spatial coordinates and the energy
according to:
\begin{equation}
 t=b \hat{t}, \quad y = a \hat{y}, \quad T = \widetilde{E} T'
\end{equation}
so that the different parts of the Hamiltonian are rescaled as:
\begin{eqnarray}
\label{eq-rescaling-Ham-replicated-part1}
\fl \quad
 && \widetilde{\mathcal{H}}_{\rm{el}} \argc{y_1(t), \dots ,y_n(t);\tf} \bigg\vert_{c'=c}
 	= \frac{ca^2}{b} \, \widetilde{\mathcal{H}}_{\rm{el}} \argc{\hat{y}_1(\hat{t}),\dots,\hat{y}_n(\hat{t});\hat{\tf}} \bigg\vert_{c'=1}
 \\
\label{eq-rescaling-Ham-replicated-part2}
\fl
 && \widetilde{\mathcal{H}}_{\rm{dis}} \argc{y_1(t), \dots ,y_n(t);\tf} \bigg\vert_{\!\!\!\! \begin{array}{l}\scriptstyle{D'=D,} \\ \scriptstyle{T'=T,} \\ \scriptstyle{\xi'=\xi} \end{array}}
 	= \frac{D_0 \, b}{\widetilde{E} \, a} \, \widetilde{\mathcal{H}}_{\rm{dis}} \argc{\hat{y}_1(\hat{t}),\dots,\hat{y}_n(\hat{t});\hat{\tf}} \bigg\vert_{\!\!\!\! \begin{array}{l}\scriptstyle{D'=D/D_0,} \\ \scriptstyle{T'=T/\widetilde{E},} \\ \scriptstyle{\xi'=\xi/a} \end{array}}
\end{eqnarray}
The scaling of the disorder replicated Hamiltonian is exact, and not only in distribution as in \eref{eq-rescaling-Ham-part2}, but in both those cases the power counting is based on the two-point correlator scaling ${R_{\xi}(y) = a^{-1} R_{\xi/a}(y/a)}$.

Because of the explicit temperature dependence of ${\widetilde{\mathcal{H}}_{\rm{dis}}}$ and its specific dependence on $a$ and $b$, the Flory construction based on the replicated Hamiltonian will have a different Flory exponent. % ${\zeta_{\rm{F}}=2/3}$}.
Indeed, on the one hand its elastic and disorder parts scale identically provided that ${ca^2/b = (D_0 b)/(\widetilde{E} a)}$.
On the other hand, the Boltzmann weight ${\propto e^{-\frac{1}{T} \widetilde{\mathcal{H}} \argc{y_1(t),\dots,y_n(t);\tf}}}$ is not modified if ${\widetilde{E}=ca^2/b}$, for the same reason as in \eref{eq-rescaling-Ham-part3}, although here $a$ and $b$ depend on $\widetilde{E}$ as well.
Thus the `Flory recipe' yields the following rewriting of the roughness and the relations between scaling factors ${\lbrace a,b,\widetilde{E} \rbrace}$:
\begin{eqnarray}
\label{eq-rescaling-Ham-replicated-Flory-part1a}
\fl \quad
B(t;c,D,T,\xi,\tf)
= a^2 \, \bar{B} \argp{\hat{t}=\frac{t}{b};c'=1,D'=\frac{D}{D_0},T'=\frac{T}{\widetilde{E}},\xi'=\frac{\xi}{a},\hat{\tf}=\frac{\tf}{b}}
\\
\fl \quad
\label{eq-rescaling-Ham-replicated-Flory-part1b}
 \qquad  \textnormal{with} \quad \cases{
 a = \argp{\frac{D_0}{c \widetilde{E}}}^{1/3} b^{2/3}
 \Longleftrightarrow
 b= \argp{\frac{c \widetilde{E}}{D_0}}^{1/2} a^{3/2},
 \quad \zeta_{\rm{F}}^{(2)}=2/3
 \\
 \widetilde{E} = ca^2/b = \argp{c D_0^2 b}^{1/5} = (c D_0 a)^{1/3}
 }
\end{eqnarray}
%\eref{eq-def-roughness-scaling-function-Bbar}
with $\bar{B}(\hat{t};\dots)$ a scaling function with adimensional parameters.
\emph{So for the replicated Hamiltonian \eref{eq-Hamiltonian-replicated-copy1}-\eref{eq-Hamiltonian-replicated-copy2}-\eref{eq-Hamiltonian-replicated-copy3}  the apparent Flory exponent is ${\zeta_{\rm{F}}^{(2)}=2/3}$,
however ${\widetilde{E}(a)}$ and ${\widetilde{E}(b)}$ are exactly the same as in \eref{eq-rescaling-Ham-replicated-Flory-part1b}.}
In fact, if we combine the two relations in \eref{eq-rescaling-Ham-replicated-Flory-part1b}, we simply recover their counterparts \eqref{eq-rescaling-Ham-Flory-part1b} for the original Hamiltonian, and in particular that `${a \sim b^{3/5}}$' (\textit{i.e.}~${\zeta_{\rm{F}}^{(1)}=3/5}$).
%
%This last property was \textit{a priori} not evident at all.

In the relations \eref{eq-rescaling-Ham-replicated-Flory-part1a}-\eref{eq-rescaling-Ham-replicated-Flory-part1b}, we still have two free parameters to fix, and in \eref{eq-rescaling-Ham-replicated-Flory-part1a} this would, again, correspond to choose two typical scales and to examine the behaviour of $\bar{B}(\hat{t};\dots)$.
Since the rescaling of the original Hamiltonian and of its replicated counterpart are both based on the same scaling for the disorder, we expect physically that we should find the same values for $\lbrace a,b,\widetilde E \rbrace$ when we fix the two remaining free parameters.
And indeed we find for instance the same values for ${c'=D'=1}$ as those listed in \tref{table-Hamiltonian-part1}, confirming that the crossover lengthscales, such as the Larkin length, should be the same with or without replicas.

Nevertheless, the Flory construction of the replicated Hamiltonian suggests more transparently an additional rescaling, which turns out to correspond exactly to the `physical' scalings of the 1D interface at large lengthscale $\tf$.
%
%This is what we detail thereafter, and it will be used for the starting point of the GVM computation scheme presented from \sref{sec:GVM-upgraded-finite-length} and on.
%
Imposing again ${c'=1}$, we choose to control conjointly $D'$ and $T'$ via their ratio ${D'/T'=1/f}$, $f$ being thus a parameter which controls the amplitude of the disorder replicated Hamiltonian. This choice implies:
\begin{eqnarray}
\label{eq-Ham-replicated-special-Flory-part1}
 \frac{D'}{T'}=\frac{D/D_0}{T/\widetilde{E}} = \frac{1}{f}
 	\Leftrightarrow \frac{D_0}{\widetilde{E}} = \frac{D}{T/f}
\\
\label{eq-Ham-replicated-special-Flory-part2}
 a	= \argp{\frac{D_0}{c \widetilde{E}}}^{1/3} b^{2/3}
 	= \argp{\frac{D}{c T/f}}^{1/3} b^{2/3}
 %\qquad \textnormal{(\textit{i.e.}~KPZ roughness scaling)}
 \\
\label{eq-Ham-replicated-special-Flory-part3}
 \widetilde{E}	= \argc{\argp{\frac{D}{T/f}}^{2} c b}^{1/3}
 	= \argp{\frac{cD}{T/f} a}^{1/2}
 %\qquad \textnormal{(\textit{i.e.}~Brownian free-energy scaling)}
\end{eqnarray}
Imposing moreover that ${\hat{\tf}=1}$, in other words considering the problem at fixed lengthscale ${b=\tf}$, we eventually obtain the KPZ scaling for the roughness and the Brownian scaling for the disorder free energy, with the correct exponents and temperature-dependent prefactors (as recalled in \sref{sec:def-models-previous-results}):
\begin{equation}
\label{eq-Flory-rescaling-winning}
\fl \qquad
 b = \tf,
 \quad
 a	%= \argp{\frac{D_0}{c \widetilde{E}}}^{1/3} b^{2/3}
 	= \argp{\frac{D}{c T/f}}^{1/3} \tf^{2/3},
 \quad
 \widetilde{E}	= \argc{\argp{\frac{D}{T/f}}^{2} c \tf}^{1/3}
\end{equation}
This implies:
\begin{eqnarray}
\fl \qquad
 && T' = \frac{T}{\widetilde{E}}
 	= f \argc{\frac{D^2 c}{(T/f)^5} \tf}^{1/3}
 	\stackrel{\eref{eq-fudging-Larkin-length}}{=} f \argp{\frac{\tf}{L_{\cc}(T,\xi)}}^{1/3}
 	\equiv f \, \hat{\beta}_f^{-1}(\tf)
 \\
\fl
 && D' = \frac{D}{D_0} = \frac{T'}{f} = \hat{\beta}_f^{-1}(\tf)
 \\
\fl
 && \xi' = \frac{\xi}{a} = \frac{\xi}{\argp{\frac{D}{c T/f}}^{1/3} \tf^{2/3}}
    \equiv \mathring{\xi}_f(\tf)
\end{eqnarray}
and coming back at last to the expression for the roughness \eref{eq-rescaling-Ham-replicated-Flory-part1a}:
\begin{eqnarray}
\label{eq-rescaling-Ham-replicated-Flory-winning}
\fl \qquad
\eqalign
B(\tf;c,D,T,\xi,\tf)
	= \argp{\frac{D}{c T/f}}^{2/3} \tf^{4/3} \, \underbrace{\bar{B} \argp{1;1,\hat{\beta}_f^{-1}(\tf),f \, \hat{\beta}_f^{-1}(\tf),\mathring{\xi}_f(\tf),1}}%_{\to \textnormal{cte at } \tf \to \infty \,?}
\end{eqnarray}
This means that, for the asymptotic roughness to be given by
${\argp{\frac{D}{c T/f}}^{2/3} \tf^{4/3}}$,
the scaling function indicated by the underbrace should tend to a numerical constant in the limit ${\tf \to \infty}$. In fact, this condition self-consistently defines the value of ${f=f(T,\xi)}$.
In fact, the specific rescaling \eref{eq-Flory-rescaling-winning} will be used for the starting point of the GVM computation scheme presented from \sref{sec:GVM-upgraded-finite-length} and on, imposing ${f=1}$ and thus better suited for capturing the `high-temperature' regime.

%-------------------------
\subsection{DP free energy ${F_V(t,y)}$}
\label{sec:powercounting-Flory-free-energy}

When we were examining the Hamiltonian, without or with replicas, the scaling of the Gaussian disorder was given by the two-point disorder correlator, which was an input of the model so there was no additional assumption to be made, with respect to this scaling.
On the contrary, when considering the free energy at fixed $\tf$, the scaling of the disorder free energy ${\bar{F}_V(\tf,y)}$ is not known \textit{a priori}, and effectively depends on $\tf$, as mentioned in \sref{sec:def-models-previous-results}.
If we assume nevertheless a dominant Brownian scaling in distribution \eqref{eq-generic-scaling-KPZ-Brownian}
${\bar{F}_V(t,y) \sim (\widetilde{D} y)^{1/2}}$, we can perform the same programme as for the (non-)replicated Hamiltonians.

We first recall the expression for the free energy: %focusing exclusively on its $y$-dependent contributions:
\begin{eqnarray}
\label{eq-free-energy-original-copy}
\fl \quad
F_V(\tf,y) \stackrel{\eref{eq-free-energy-original}}{=} F_{V=0}(\tf,y) + \bar{F}_V(\tf,y)
\, , \qquad
F_{V=0}(\tf,y)
	%= -T \ln \arga{\frac{\exp \argc{- \frac{y^2}{2 B_{\rm{th}}(t)}}}{\sqrt{2 \pi B_{\rm{th}}(t)}}}
	= \frac{c y^2}{2\tf} + \frac{T}{2}\ln \frac{2 \pi T \tf}{c}
\end{eqnarray}
Rescaling once again the spatial coordinates and the energy according to 
${\lbrace t=b \hat{t}, y = a \hat{y}, T = \widetilde{E} T' \rbrace}$,
we focus exclusively on the $y$-dependent contributions to the free energy, which are the sole relevant contributions at fixed $\tf$ with respect to statistical averages \eref{eq-pathintegral-geom-fluct-DP-v1}).
They are rescaled as:
\begin{eqnarray}
\label{eq-rescaling-free-energy-part1}
\fl \quad
 \frac{c y^2}{2\tf} = \frac{c a^2}{b} \frac{\hat{y}^2}{2\hat{\tf}},
 \quad
 \bar{F}_V(\tf,y) \bigg\vert_{\widetilde{D}'=\widetilde{D},\xi'=\xi}
 %{\begin{array}{l}\scriptstyle{D'=D,}\\ \scriptstyle{\xi'=\xi} \end{array}}
 	\stackrel{(d)}{=} \argp{\widetilde{D}_0 a}^{1/2} \, \bar{F}_V(\hat{\tf},\hat{y}) \bigg\vert_{\widetilde{D}'=\widetilde{D}/\widetilde{D}_0,\xi'=\xi/a}
\end{eqnarray}

In the statistical averages ${\moydes{\moy{\mathcal{O}(y)}}_{\tf}}$ in \eref{eq-pathintegral-geom-fluct-DP-v1},
the Boltzmann weight ${\propto e^{-\frac{1}{T} F_V(\tf,y)}}$ is not modified provided that the two previous contributions to the free energy scale
 identically, \textit{i.e.}~${c a^2/b = (\widetilde{D}_0 a)^{1/2}}$,
and that the temperature is redefined accordingly with ${\widetilde{E}=ca^2/b}$ as in \eref{eq-rescaling-Ham-part3}.
This Flory construction on the free energy yields the following rescaled roughness, along with the corresponding relations between the scaling factors ${\lbrace a,b,\widetilde{E} \rbrace}$:
\begin{eqnarray}
\label{eq-rescaling-free-energy-Flory-part1a}
\fl \quad
B(\tf;c,D,T,\xi)
= a^2 \, \bar{B}_{\rm{DP}} \argp{\hat{\tf}=\frac{\tf}{b};c'=1,\widetilde{D}'=\frac{\widetilde{D}}{\widetilde{D}_0},T'=\frac{T}{\widetilde{E}},\xi'=\frac{\xi}{a}}
\\
\fl \quad
\label{eq-rescaling-free-energy-Flory-part1b}
 \qquad  \textnormal{with} \quad \cases{
 a = \argp{\widetilde{D}_0/c^2}^{1/3} b^{2/3}
 \Longleftrightarrow
 b= \argp{c^2 /\widetilde{D}_0}^{1/2} a^{3/2},
 \quad \zeta_{\rm{F}}^{(3)}=2/3
 \\
 \widetilde{E} = ca^2/b = \argp{\widetilde{D}_0^2 b/c}^{1/3} = (\widetilde{D}_0 a)^{1/2}
 }
\end{eqnarray}
%\eref{eq-def-roughness-scaling-function-Bbar}
with $\bar{B}_{\rm{DP}}(\hat{\tf};\dots)$ a scaling function with adimensional parameters, based on the assumption of a dominant Brownian scaling of the DP disorder free energy.

Among the different options of typical parameters, that could be listed as in tables~\ref{table-Hamiltonian-part1} and~\ref{table-Hamiltonian-part2}, we can highlight one option, which consists in identifying the Flory constructions of the replicated Hamiltonian and of the free energy,
respectively
\eref{eq-rescaling-Ham-replicated-Flory-part1b}
and \eref{eq-rescaling-free-energy-Flory-part1b},
as they have in common that ${\zeta_{\rm{F}}^{(2)}=\zeta_{\rm{F}}^{(3)}=2/3}$.
\begin{eqnarray}
\fl \quad
	\frac{\widetilde{D}_0}{c^2} \equiv \frac{D_0}{c \widetilde{E}}
	\Longleftrightarrow \widetilde{D}_0 = \frac{c D_0}{\widetilde{E}}
	\quad \Rightarrow \,
	\cases{
	 a = \argp{\frac{D_0}{c \widetilde{E}}}^{1/3} b^{2/3}
 		\Longleftrightarrow
		 b= \argp{\frac{c \widetilde{E}}{D_0}}^{1/2} \!\!\! a^{3/2}
	 \\
	 \widetilde{E} =\argp{c D_0^2 b}^{1/5} = (c D_0 a)^{1/3}
	}	
\end{eqnarray}
This choice consistently yields the same scalings as for the replicated Hamiltonian \eqref{eq-rescaling-Ham-replicated-Flory-part1b}, as expected.

%-------------------------
\subsection{Replicated DP free energy ${\widetilde{F}(t,y_1, \dots, y_n)}$}
\label{sec:powercounting-Flory-free-energy-replicae}

At last, we examine the Flory construction starting from the replicated free energy,
keeping only its $y$-dependent contributions, which are the sole relevant contributions at fixed $\tf$ with respect to statistical averages \eref{eq-pathintegral-geom-fluct-DP-v2},
and its two-point cumulant, under the Brownian scaling assumption:
\begin{eqnarray}
\label{eq-free-energy-replicated-explicit-copy}
\fl \qquad
\widetilde{F} (\tf, y_1, \dots, y_n)
	\stackrel{\eref{eq-free-energy-replicated-explicit}}{=}
	\sum_{j=1}^{n} \frac{c y^2}{2 t}
		+ \frac{1}{4T} \sum_{j,k=1}^{n} \bar{C}(\tf,y_j-y_k)\bigg\vert_{\widetilde{D},\xi}
	+\dots
\end{eqnarray}
with ${\bar{C}}$ defined just after \eref{eq-free-energy-replicated-explicit}, and with a behaviour at large $\tf$
sketch for instance in~\eref{eq-large-times-two-point-correlator-Cbar}.
Rescaling once again the spatial coordinates and the energy according to 
${\lbrace t=b \hat{t}, y = a \hat{y}, T = \widetilde{E} T' \rbrace}$,
the first two parts of this free energy are rescaled as:
\begin{eqnarray}
\label{eq-rescaling-free-energy-replicated-part1}
\fl \quad
 && \sum_{j=1}^{n} \frac{c' y^2}{2 \tf} \bigg\vert_{c'=c}
 =  \sum_{j=1}^{n} \frac{\hat{y}^2}{2 \hat{\tf}} \bigg\vert_{c'=c}
 \\
\fl
 && \frac{1}{4T'} \sum_{j,k=1}^{n} \bar{C}(\tf,y_j-y_k)
 \bigg\vert_{\begin{array}{l}\scriptstyle{\widetilde{D}'=\widetilde{D},} \\ \scriptstyle{T'=T,}\\ \scriptstyle{\xi'=\xi} \end{array}}
 %\bigg\vert_{\widetilde{D}', T'=T,\xi'=\xi}
 = \frac{\widetilde{D}_0 a}{\widetilde{E}} \frac{1}{4T/\widetilde{E}} \sum_{j,k=1}^{n} \bar{C}(\hat{\tf},\hat{y}_j-\hat{y}_k)
 \bigg\vert_{\begin{array}{l}\scriptstyle{\widetilde{D}'=\widetilde{D}/\widetilde{D}_0,} \\ \scriptstyle{ T'=T/\widetilde{E},} \\ \scriptstyle{\xi'=\xi/a} \end{array}}
 %\bigg\vert_{\widetilde{D}'=\widetilde{D}/\widetilde{D}_0, T'=T/\widetilde{E},\xi'=\xi/a}
\end{eqnarray}

Similarly to the non-replicated free-energy case discussed in \sref{sec:powercounting-Flory-free-energy},
we impose on the one hand that these two parts scale identically,
\textit{i.e.}~${ca^2/b =\widetilde{D}_0 a/\widetilde{E}}$,
and on the other hand we consistently redefine the overall temperature in the Boltzmann weight with~${\widetilde{E}=ca^2/b}$.
This `Flory recipe' thus yields:
%for the replicated free energy under the Brownian assumption:
%the following rescaled roughness, along with the corresponding relations between the scaling factors ${\lbrace a,b,\widetilde{E} \rbrace}$:
%
\begin{eqnarray}
\label{eq-rescaling-free-energy-replicated-Flory-part1a}
\fl \quad
B(\tf;c,D,T,\xi)
= a^2 \, \bar{B}_{\rm{DP}} \argp{\hat{\tf}=\frac{\tf}{b};c'=1,\widetilde{D}'=\frac{\widetilde{D}}{\widetilde{D}_0},T'=\frac{T}{\widetilde{E}},\xi'=\frac{\xi}{a}}
\\
\fl \quad
\label{eq-rescaling-free-energy-replicated-Flory-part1b}
 \qquad  \textnormal{with} \quad \cases{
 a = \frac{\widetilde{D}_0}{c \widetilde{E}} b
 \Longleftrightarrow
 b= \frac{c \widetilde{E}}{\widetilde{D}_0} a\:,
 \quad \zeta_{\rm{F}}^{(4)}=1
 \\
 \widetilde{E} = ca^2/b = \argp{\widetilde{D}_0^2 b/c}^{1/3} = (\widetilde{D}_0 a)^{1/2}
 }
\end{eqnarray}
%\eref{eq-def-roughness-scaling-function-Bbar}
with $\bar{B}_{\rm{DP}}(\hat{\tf};\dots)$ a scaling function with adimensional parameters, based on the assumption of a dominant Brownian scaling of the DP disorder free energy.

So this last Flory construction, based on the replicated DP free energy assuming a Brownian scaling, yields a new Flory exponent ${\zeta_{\rm{F}}^{(4)}=1}$, however ${\widetilde{E}(a)}$ and ${\widetilde{E}(b)}$ are exactly the same as in \eref{eq-rescaling-free-energy-Flory-part1b}.
In fact, if we combine the two relations in \eref{eq-rescaling-free-energy-replicated-Flory-part1b}, we simply recover their counterparts \eqref{eq-rescaling-free-energy-Flory-part1b} for the original DP free energy, and in particular that `${a \sim b^{2/3}}$' (\textit{i.e.}~${\zeta_{\rm{F}}^{(3)}=2/3}$).

%-------------------------
\subsection{Summary of the different power countings}
\label{sec:powercounting-Flory-recapitulation}

Throughout this section, we have explored the different power countings and Flory rescalings,
either on the 1D interface Hamiltonian or on the DP free energy,
without or with replicas,
based on a rescaling of the spatial coordinates with $(a,b)$ and of the energy with $\widetilde{E}$.
These different power countings are listed in \tref{table-carte-floryste}.

%% La carte du Floryste ;-)
%------------

\begin{table}[h]
\caption{
\label{table-carte-floryste}
List of the Flory power countings presented in \sref{sec:powercounting-Flory}.
}
%\begin{indented}
%\item[]
\begin{small}
\begin{tabular}{@{} cc|cc|c}
\br
\textit{Starting point}
	& \textit{Ref.}
		& \textit{Power counting}
			& $\zeta_{\rm{F}}$ %\textit{\specialcell{Flory \\ exponent}}
				& ${\widetilde{E} = ca^2/b}$
\\
\br
${\mathcal{H} \argc{y(t),V;\tf}}$
	& \eref{eq-rescaling-Ham-Flory-part1a}-\eref{eq-rescaling-Ham-Flory-part1b}
		& ${a = \argp{D_0^{1/3} c^{-2/3} b}^{3/5}}$,
				& $3/5$ %${\zeta_{\rm{F}}^{(1)}=3/5}$
					& ${\argp{c D_0^2 b}^{1/5} = (c D_0 a)^{1/3}}$
\\
\mr
${\widetilde{\mathcal{H}} \argc{y_1(t), \dots ,y_n(t);\tf}}$
	& \eref{eq-rescaling-Ham-replicated-Flory-part1a}-\eref{eq-rescaling-Ham-replicated-Flory-part1b}
		& ${a = \argp{\frac{D_0}{c \widetilde{E}}}^{1/3} b^{2/3}}$,
			& $2/3$ %${\zeta_{\rm{F}}^{(2)}=2/3}$
				& ${\argp{c D_0^2 b}^{1/5} = (c D_0 a)^{1/3}}$
\\
\mr
${F_V(\tf,y)}$
	& \eref{eq-rescaling-free-energy-Flory-part1a}-\eref{eq-rescaling-free-energy-Flory-part1b}
		& ${a = \argp{\widetilde{D}_0/c^2}^{1/3} b^{2/3}}$,
			& $2/3$ %${\zeta_{\rm{F}}^{(3)}=2/3}$
				& ${\argp{\widetilde{D}_0^2 b/c}^{1/3} = (\widetilde{D}_0 a)^{1/2}}$
\\
\mr
${\widetilde{F} (\tf, y_1, \dots, y_n)}$
	& \eref{eq-rescaling-free-energy-replicated-Flory-part1a}-\eref{eq-rescaling-free-energy-replicated-Flory-part1b}
		& ${a = \frac{\widetilde{D}_0}{c \widetilde{E}} b}$,
			& $1$ %${\zeta_{\rm{F}}^{(4)}=1}$
				& ${\argp{\widetilde{D}_0^2 b/c}^{1/3} = (\widetilde{D}_0 a)^{1/2}}$
\\
\br
\end{tabular}
\end{small}
%\end{indented}
\end{table}

We have seen that these quantities have different values for the Flory exponent,
defined by the identical scaling of the two parts of each quantity:
${\zeta_{\rm{F}}^{(1)}=\frac35, \zeta_{\rm{F}}^{(2)}=\frac23, \zeta_{\rm{F}}^{(3)}=\frac23, \zeta_{\rm{F}}^{(4)}=1}$.
When we impose moreover that the Boltzmann weight in statistical averages should not be modified by a Flory rescaling, we recover the same expressions for the Hamiltonians, and similarly for the DP free energies.

Since all these power countings are based on quantities which are different incarnations of the same model, these different rescalings turn out to be equivalent, and it is possible to recover one from each other.
Nevertheless, they are nothing more than power countings at this stage, and an additional physical input is required in order to assess if a given Flory exponent corresponds (or not) to a physical exponent, and which features of the Flory construction are to be trusted (or not).
For that purpose, in the next section we will discuss how path-integral saddle points can precisely provide such an input.

%_____________________________________________________________________________________________________
%_____________________________________________________________________________________________________
%\newpage
\section{Saddle points and optimal trajectories in path integrals}
\label{sec:saddlepts-optimaltrajectories}

In this section we examine, using a saddle-point asymptotic analysis, the relation between the roughness exponent at large ${\tf}$
and the Flory exponents arising from the well-chosen scalings allowing for
%and exponents arising from well-chosen scalings – the Flory ones, which allow
a common rescaling of elastic and disordered contributions to the free energy or to the Hamiltonian,
as we have just discussed in \sref{sec:powercounting-Flory}.
We first justify why the Flory rescaling of the free energy gives the correct roughness exponent (in \sref{sec:saddlepts-optimaltrajectories-large-tf}),
while the Flory rescaling of the Hamiltonian does not (in \sref{sec:saddlepts-optimaltrajectories-Flory-failure})
–~noticing %identifying in passing
the crucial role of the disorder correlation length $\xi$ in the KPZ problem.
The key ingredient %tool
is the Lax-Oleinik principle~\cite{lax_hyperbolic_1957,oleinik,e_invariant_2000,bec_burgers_2007_PhysRep447_1}, which gives a condition for the existence of optimal point-to-line trajectories in the zero-temperature limit.
We then describe how the use of this principle makes it possible to identify the explicit dependency in $\xi$ of the asymptotic roughness, in the zero-temperature limit,
from a saddle-point analysis on the Hamiltonian at ${T \to 0}$ (in \sref{sec:saddlepts-optimaltrajectories-low-temperature}).
Last (in \sref{sec:interpr-saddle-point}), we discuss these three cases from the perspective of the scaling function ${\bar{B}(\hat{t};\dots)}$, as introduced in \sref{sec:powercounting-Flory} for the corresponding Flory power countings.

%"Saddle point on the free energy at large lengthscale $\tf$" % \meevid{(valid!)}
%\sref{sec:saddlepts-optimaltrajectories-large-tf}
%\\
%"Saddle point on the Hamiltonian with the Flory scaling" %\meevid{(wrong!)}
%\sref{sec:saddlepts-optimaltrajectories-Flory-failure}
%\\
%"Saddle point on the Hamiltonian at low temperature $T$" %\meevid{(valid!)}
%\sref{sec:saddlepts-optimaltrajectories-low-temperature}
%\\
%"Interpretation of the saddle-points asymptotics for the scalings"
%\sref{sec:interpr-saddle-point}

%-------------------------
% "Saddle points and optimal trajectories in path integrals"
\subsection{Saddle point on the free energy at large lengthscale $\tf$} % \meevid{(valid!)}
\label{sec:saddlepts-optimaltrajectories-large-tf}

As we have recalled in \sref{sec:def-models-previous-results},
%As we pointed out in section~\ref{sec:powercounting-Flory},
in the large-$\tf$ regime and at ${\xi=0}$,
the disorder free energy $\bar F(\tf,y)$ rescales as a Brownian process in the coordinate~$y$, cf. \eref{eq-steady-state-distribution-F-Gaussian-2}.
Moreover, we pointed out in \sref{sec:powercounting-Flory-free-energy} that the associated power counting assuming such a Brownian scaling gives a Flory exponent $2/3$. %\meevid{(${a \sim b^{2/3}}$)}.
In fact, such a Flory-type argument was precisely invoked by Huse, Henley and Fisher in Ref.~\cite{huse_henley_fisher_1985_PhysRevLett55_2924} to identify this exponent as the asymptotic roughness exponent for the disordered interface problem.
%and on the other hand the roughness scales with an exponent
%%and this implies that the associated power-counting argument gives the roughness exponent
%$\zeta=\frac 23$.
%
%In fact, this was precisely the Flory-type argument provided by Huse, Henley and Fisher in Ref.~\cite{huse_henley_fisher_1985_PhysRevLett55_2924} to justify the value of this exponent for the disordered interface problem.
%
In~\cite{agoritsas_2012_FHHtri-analytics}, we have proposed a procedure explaining how %why
this power counting generalises at ${\xi>0}$
and why such a Flory-type arguments holds for predicting the correct roughness exponent,
by using a large-$\tf$ saddle-point analysis that we recall here for reference.
Indeed, it offers a good starting point to understand which physical reasons underpin the matching or not between the Flory exponent and the physical %actual
roughness exponent, seen in the light of a saddle-point asymptotic analysis.

Using the Brownian scaling of the disordered free energy at large $\tf$ rederived in \sref{sec:powercounting-Flory-free-energy}, one performs the following rescaling
\begin{equation}
\fl
\qquad\qquad
  t=\tf \hat t ,
  \qquad 
  y= (\widetilde{D}/c^2)^{\frac13}\, \tf^{\frac 23}\,\hat y ,
  \qquad
  \bar F_V(t,y) \stackrel{(d)}{=} (\widetilde{D}^2 \tf/c)^{1/3} \hat F (\hat t, \hat y)
  \label{eq-rescalingsB-for-lowT-toymodel_hat}
\end{equation} 
where $\hat F $ is a Brownian motion of the coordinate $ \hat y$ with unit variance.
It implies from~\eref{eq-pathintegral-geom-fluct-DP-v1} for $\mathcal O(y)=y^2$ that, in the explicit expression of the roughness function, the elastic and disorder contribution share a common prefactor as follows:
%$B(t;c,\widetilde D,T,\tilde{\xi})$ in \eref{eq-def-moments-pdf-replicas-1}:
\begin{eqnarray}
\fl
\qquad\qquad
 B(\tf) 
\ \mathop{\sim}_{\tf\to\infty}\ 
\Big[\frac{\widetilde D }{c^2}\Big]^{\frac 23}\tf^{\frac 43} 
\
%\bar B(\tf)
b_1(\tf)
  \label{eq-rescaling_B_zeta23_0}
\\[2mm]
\fl
\qquad\qquad
%\bar B(\tf)
b_1(\tf)
\ =\
 \overline{
  \
  \frac{\displaystyle\int_{\mathbb R} d\hat y\:\hat y^2 \exp\Big\{\!- \tfrac 1T \big(\tfrac {\widetilde{D}^2 }{c}\tf\big)^{\frac 13}\Big[ \tfrac{\hat y^2}{2} + 
  \hat F(\hat t,\hat y)\Big]\Big\}}
  {\displaystyle \int_{\mathbb R} d\hat y\:  \exp\Big\{\!- \tfrac 1T \big(\tfrac {\widetilde{D}^2 }{c}\tf\big)^{\frac 13}\Big[ \tfrac{\hat y^2}{2} + 
  \hat F(\hat t,\hat y)\Big]\Big\}}
  \
  }
 \label{eq-rescaling_B_zeta23}
\end{eqnarray}
where the overline denotes the average over $\hat{F}$.
%(\textit{i.e.}~the disorder average over $\bar{F}_V$).
%
At fixed $\hat{F}$ (\textit{i.e.}~at fixed disorder configuration), one can evaluate the integrals over $\hat y$ using a saddle-point asymptotic analysis in the large-$\tf$ limit. 
The numerator and the denominator of~\eref{eq-rescaling_B_zeta23} are dominated by the same value $\hat y^\star[\hat F]$ of $\hat y$ which minimises the rescaled energy
${ \frac{\hat y^2}{2} + \hat F(\hat t,\hat y) }$.
We emphasise %One notes
that this implies that $\hat y^\star[\hat F]$ \textit{is independent of $\tf$},
and that
${b_1(\tf)=\overline{(\hat y^\star[\hat F])^2}\sim \tf^0}$.
We thus deduce from \eref{eq-rescaling_B_zeta23} that
\begin{equation}
 B(\tf) 
\ 
\mathop{\sim}_{\tf\to\infty}
\ 
\overline{(\hat y^\star[\hat F])^2} 
\
(\widetilde D/c^2 )^{\frac 23}\,\tf^{\frac43}
 \label{eq-Basympt-scaling-prediction}
\end{equation}
which yields as announced the roughness exponent ${\zeta=\frac 23}$.
So, on the one hand the power counting based on the Brownian scaling puts as a prefactor $\tf^{4/3}$ prescribed by the Flory scaling,
and on the other hand the independence on~$\tf$ of the saddle point $\hat y^\star[\hat F]$ selects the Flory scaling as the `true' physical one.

This procedure provides an example where the naive Flory power counting gives a correct prediction.
A key point in the reasoning is that the minimiser $\hat y^\star[\hat F]$ of ${ \frac{\hat y^2}{2} + \hat F(\hat t,\hat y) }$ does exist and has a finite variance, which can be justified mathematically.
In fact, in the uncorrelated disorder case ($\xi=0$), the Brownian scaling of $\bar F_V$ can be extended to include the large- but finite-$\tf$ regime where the fluctuations in the coordinate $\hat y$ are described by the Airy process~\cite{praehofer-spohn_2002_JStatPhys108_1071}.
Then, a similar rescaling procedure follows and leads to the same expression of the asymptotic roughness as~(\ref{eq-rescaling_B_zeta23_0}-\ref{eq-rescaling_B_zeta23}) with now $\hat F$ being the opposite of the Airy$_2$ process $\mathcal A_2(\hat y)$.
The same saddle-point analysis can be performed where $\hat y^\star$ is now the minimiser of ${ \frac{\hat y^2}{2} -  \mathcal A_2(\hat y) }$, which does exist and whose distribution has been characterised in Refs.~\cite{schehr_2012_JStatPhys149_385,flores_quastel_remenik_2012_CommunMathPhys317_363,baik_2012_JMathPhys53_083303}.
In particular, this allows one to evaluate the numerical constant $\overline{(\hat y^\star[\hat F])^2} $ in the prefactor of the asymptotic roughness~\eref{eq-Basympt-scaling-prediction}.

As for the correlated disorder case ${\xi>0}$, as long as the Brownian scaling is the dominant one in the evaluation of the path-integral saddle point, this argument remains valid and yields the KPZ roughness exponent.
However, it cannot yield more information than the value of this exponent, and in particular it does not give access to the temperature dependence of the amplitude ${\widetilde{D}}$, which controls both the disorder free-energy and the roughness amplitudes, according to \eref{eq-generic-scaling-KPZ-Brownian}-\eref{eq-fudging-Dtilde}.
%
%The specific limit of zero temperature will be discussed in \sref{sec:saddlepts-optimaltrajectories-low-temperature}.
%
Nevertheless,
a non-perturbative functional renormalisation study of the 1D KPZ at ${\xi>0}$ supports the assumption of the dominant Brownian scaling of the free-energy \cite{NPRG-in-preparation}, in agreement with previous numerical studies \cite{agoritsas-2012-FHHpenta,agoritsas_2012_FHHtri-numerics}, assessing furthermore the validity of the present argument for the roughness exponent ${\zeta=\frac23}$.

We now try to implement the same construction for the Hamiltonian description of the roughness and explain why it fails.

%-------------------------
% "Saddle points and optimal trajectories in path integrals"
\subsection{Saddle point on the Hamiltonian with the Flory scaling} %\meevid{(wrong!)}
\label{sec:saddlepts-optimaltrajectories-Flory-failure}

As we pointed out in \sref{sec:powercounting-Flory-Hamiltonian}, from the Flory rescaling of the Hamiltonian
\begin{equation}
  t = \tf\,\hat t ,
\qquad 
  y =  \tf^{\zeta_{\FF}} \big(\tfrac{D}{c^2}\big)^{\frac 15} \hat y ,
\qquad 
  \zeta_{\FF} = \frac 35
\label{eq:explicit-rescaling-F}
\end{equation}
one gets from the path integral~\eref{eq-pathintegral-geom-fluct-v1}
for ${\mathcal{O}\argc{y(t)}=y(t)^2}$
an expression of the roughness in which the prefactors of the elastic and disorder contributions, in the Hamiltonian, are rescaled with a common prefactor as follows:
%\begin{small}
  \begin{eqnarray}
    \fl
\qquad
    B(\tf)
\ =\
    \Big[\frac{D}{c^2}\Big]^{\frac 25}\:
    \tf^{2\zeta_{\FF}} \:
    b_2(\tf)
    \label{eq:BtfFlory-rescaled_0}
\\[2mm]
\fl
\qquad
 b_2(\tf)
\  =\
%    \EE_{\hat V}
\overline{
\
    \frac
    {\displaystyle\int_{\hat y(0)=0}\hspace*{-8mm}\mathcal D\hat y(\hat t)
      \:\hat y(1)^2
      \exp\Big\{\!-\tfrac {(cD^2)^{\frac 15}}T \tf^{\frac 15}\!\displaystyle\int_{0}^{1} \!\!d\hat t\:\Big[ \tfrac 12 (\partial_{\hat t}\hat y)^2+\hat V_{\hat\xi(\tf)}(\hat t,\hat y(\hat t))\Big]\Big\}}
    {\displaystyle\int_{\hat y(0)=0}\hspace*{-8mm}\mathcal D\hat y(\hat t)
      \exp\Big\{\!-\tfrac {(cD^2)^{\frac 15}}T \tf^{\frac 15}\!\displaystyle\int_{0}^{1} \!\!d\hat t\:\Big[ \tfrac 12 (\partial_{\hat t}\hat y)^2+\hat V_{\hat\xi(\tf)}(\hat t,\hat y(\hat t))\Big]\Big\}}
\
}
    \label{eq:BtfFlory-rescaled}
  \end{eqnarray}
%\end{small}
%
Here, the (\emph{à la} Flory-)rescaled disorder correlation length reads
\begin{equation}
  {\hat\xi_{\FF}(\tf)}= \frac{\xi}{\tf^{\zeta_{\FF}} \big(\tfrac{D}{c^2}\big)^{\frac 15}}
  \label{eq:defxiFF}
\end{equation}
and the random potential $\hat  V_{\hat \xi}$ has a correlation length $\hat \xi$ and a disorder strength equal to $1$:
\begin{equation}
  \overline{\hat V_{\hat \xi}(\hat t,\hat y)\hat V_{\hat \xi}(\hat t',\hat y')}=\delta(\hat t'-\hat t)\hat R_{\hat \xi}(\hat y'-\hat y)
\end{equation}

We consider at first the case of a strictly uncorrelated disorder ($\xi=0$) where in~\eref{eq:BtfFlory-rescaled} the arguments of exponentials take the form $\tf^{1/5}$ times a $\tf$-independent expression.
The path integral \eref{eq:BtfFlory-rescaled} thus takes precisely a form which (in appearance) is amenable to a saddle-point analysis at $\tf\to\infty$,
similarly to the expression \eref{eq-rescaling_B_zeta23_0} that we have recalled in the previous subsection. %~\ref{sec:saddlepts-optimaltrajectories-large-tf}.
By analogy, let us precisely assume that there exists an optimal trajectory $\hat y^\star_{\hat V}(\hat t) $ that minimises the integral of the rescaled energy
$\int_{0}^{1} \!d\hat t\:\big[ \tfrac 12 (\partial_{\hat t}\hat y)^2+\hat V_{\hat{\xi}=0}(\hat t,\hat y(\hat t))\big]$,
with the initial condition $\hat y(0)=0$.
Then, because in~\eref{eq:BtfFlory-rescaled} this minimiser would be the same in the numerator and in the denominator, one would obtain 
\begin{equation}
 B(\tf) 
\quad \mathop{\sim}_{\tf\to\infty}^{\rm{wrong!}\vphantom{|_|}}\quad 
\overline{(\hat y_{\hat V}^\star(1))^2} \;\,
    \big(\tfrac{D}{c^2}\big)^{\frac 25}\:
    \tf^{\frac 65} \:
 \label{eq-Basympt-scaling-prediction_wrongFloryHam}
\end{equation}
where the overline denotes the average over the rescaled random potential $\hat V$.
This reasoning, that would lead to a roughness exponent $\zeta=\frac 35$, is in fact wrong: the optimal trajectory $\hat y^\star_{\hat V}(\hat t) $ does not exist and the saddle-point analysis that we have sketched is invalid, because %at $\xi=0$ the disorder $\hat V$ is too irregular.
the uncorrelated disorder $\hat V_{\hat{\xi}=0}$ is too irregular,
for the assumption of the existence of an optimal trajectory to be valid.

Indeed, for a minimiser of the Hamiltonian~\eref{eq-Hamiltonian-original} to exist, according to the Lax-Oleinik principle, the disorder has to be smooth enough: in our context, this requires to have a non-zero correlation length, see for instance Ref.~\cite{bec_burgers_2007_PhysRep447_1} for a discussion of this optimisation principle in the context of the noisy Burgers equation (the original variational principle was designed for the noiseless Burgers equation~\cite{lax_hyperbolic_1957,oleinik,kruzkov_generalized_1975} and was later generalised to the noisy one~\cite{e_invariant_2000,bec_burgers_2007_PhysRep447_1}).
Hence, let us consider as a second step the case of correlated disorder ($\xi>0$) and try to implement a saddle-point analysis.
After the Flory rescaling~\eref{eq:explicit-rescaling-F} leading to the reformulation \eref{eq:BtfFlory-rescaled}, the distribution of the rescaled disorder $\hat V_{\hat\xi(\tf)}(\hat t,\hat y)$ \emph{depends on $\tf$ via its correlation length}, according to~\eref{eq:defxiFF}.
In other words, the rescaled Hamiltonian
\begin{equation}
\HHh_{\hat V}\big[\hat y(\hat t),\hat\xi_{\FF}(\tf)\big]=
\displaystyle\int_{0}^{1} d\hat t\:\Big[ \tfrac 12 (\partial_{\hat t}\hat y)^2+\hat V_{\hat\xi_{\FF}(\tf)}(\hat t,\hat y(\hat t))\Big]  
\label{eq:rescaled-ham-Flory-for-saddle} 
% NB: en passant, il est 12h51, et les trois lettres "ham" me donnent faim.
%
\end{equation}
depends on $\tf$ and even though, thanks to the Lax-Oleinik principle, a minimising trajectory $\hat y^\star_{\hat V}(\hat t;\tf)$ does exist, it actually depends on $\tf$. In the end, this means that one cannot use the corresponding saddle-point asymptotics
\begin{equation}
 B(\tf) 
\quad \mathop{\sim}_{\tf\to\infty}^{\rm{true!}}\quad 
\overline{(\hat y_{\hat V}^\star(1;\tf))^2} \;\,
    \big(\tfrac{D}{c^2}\big)^{\frac 25}\:
    \tf^{\frac 65} \:
 \label{eq-Basympt-scaling-prediction_with-tf}
\end{equation}
to infer directly the value of the roughness exponent.
In this expression, in fact, one necessarily has $\overline{(\hat y_{\hat V}^\star(1;\tf))^2}\sim \tf^{2/15}$ in order to recover the correct KPZ exponent,
as known from \eref{eq-Basympt-scaling-prediction} for instance,
and thus to be self-consistently compatible with the Brownian scaling of $\bar{F}_V$.
The physical interpretation of this result is the following: the variance of the endpoint fluctuation of the optimal trajectory ${b_2(\tf)=\hat y^\star_{\hat V}(\hat t;\tf)}$ of the rescaled Hamiltonian~\eref{eq:rescaled-ham-Flory-for-saddle} depends on $\tf$, and this occurs only through the rescaled disorder correlation length~${\hat\xi_{\FF}(\tf)}\sim \tf^{-3/5}$ given in~\eref{eq:defxiFF}. 
As $\tf\to\infty$, this variance diverges as $\tf^{2/15}$: this is a manifestation that the  $\xi\to 0$ limit of the optimal trajectory of the Hamiltonian is ill-defined.

This very fact is, as we have discussed, at the core of the invalidity of the Flory roughness exponent $\zeta_{\FF}=3/5$ for the KPZ fluctuations~; as we have seen, this mismatch is due to the singular scaling properties of the $\xi=0$ uncorrelated disorder. In pictorial words, the naive (Flory) power counting performed on the Hamiltonian yields a `bare' (or `dimensional' ---~and incorrect) roughness exponent $\zeta_{\FF}=3/5$ and amounts to neglecting the existence of a microscopic length $\xi$. This very length, in turn, if correctly taken into consideration, modifies the bare dimensional exponent and `dresses' $\zeta_{\FF}$ to give the valid $\zeta= 2/3$ KPZ exponent.
In \sref{sec:saddlepts-optimaltrajectories-large-tf}, the assumption of the dominant Brownian scaling of the disorder free energy ${\tf \to \infty}$ was the key that allowed us to take a successful shortcut and circumvent this pitfall: otherwise we are simply not able to guess specifically that ${b_2(\tf) \sim \tf^{2/15}}$.

%-------------------------
% "Saddle points and optimal trajectories in path integrals"
\subsection{Saddle point on the Hamiltonian at low temperature $T$} %\meevid{(valid!)}
%\subsection{Saddle point on the Hamiltonian}
\label{sec:saddlepts-optimaltrajectories-low-temperature}

We just used the Lax-Oleinik principle to explain why the Flory power counting on the Hamiltonian fails to yield the correct roughness exponent.
Here we invoke this principle again, but this time in order to determine the low-temperature asymptotics $T\to 0$ of our problem.
%combining a scaling compatible with this Flory power counting with the assumption of ${\zeta=\frac23}$ for the asymptotic roughness --~assumption that is supported by the saddle-point argument presented in \sref{sec:saddlepts-optimaltrajectories-large-tf}.
%
%There is still a rescaling which makes it possible to use the Lax-Oleinik principle and to obtain an interesting property of the low-temperature asymptotics $T\to 0$ of our problem (see section IV.B.1 of Ref.~\cite{agoritsas_2012_FHHtri-analytics}).
%
In fact, the Lax-Oleinik principle is the missing ingredient that completes and thus confirms the saddle-point argument given in section IV.B.1 of Ref.~\cite{agoritsas_2012_FHHtri-analytics}.

We consider the following Flory rescaling, as given in \sref{sec:powercounting-Flory-Hamiltonian}:
\begin{equation}
\fl \qquad
  t = \frac{T_{\cc}^5}{ cD^2}\,\hat t ,
\qquad 
  \tf = \frac{T_{\cc}^5}{ cD^2}\,\hat \tf ,
\qquad 
  y =  \frac{T_{\cc}^3}{cD}\, \hat y,
\qquad
  T_{\cc} = (\xi c D)^{1/3}
\label{eq:explicit-rescaling-lowT-ham}
\end{equation}
with the characteristic temperature that has been previously defined between \eref{eq-generic-scaling-KPZ-Brownian} and \eref{eq-fudging-Dtilde}.
It allows to factor out the dependency of the elastic and disorder contribution to the Hamiltonian into a common prefactor, while fixing the correlation length of the disorder to $1$ as follows:
\begin{eqnarray}
  \fl\qquad
  B(\tf)
  \ =\
  \xi^2
    b_3(\hat \tf)
    \label{eq:BtfT-rescaled_0}
\\[2mm]
  \fl\qquad
  b_3(\hat \tf)
  \  =\
\overline{
\
    \frac
    {\displaystyle\int_{\hat y(0)=0}\hspace*{-8mm}\mathcal D\hat y(\hat t)
      \:\hat y(\hat\tf)^2
      \exp\Big\{\!-\tfrac {T_{\cc}}{T}\!\displaystyle\int_{0}^{\hat \tf} \!\!d\hat t\:\Big[ \tfrac 12 (\partial_{\hat t}\hat y(\hat t))^2+\hat V_{1}(\hat t,\hat y(\hat t))\Big]\Big\}}
    {\displaystyle\int_{\hat y(0)=0}\hspace*{-8mm}\mathcal D\hat y(\hat t)
      \exp\Big\{\!-\tfrac {T_{\cc}}{T}\!\displaystyle\int_{0}^{\hat \tf} \!\!d\hat t\:\Big[ \tfrac 12 (\partial_{\hat t}\hat y (\hat t))^2+\hat V_{1}(\hat t,\hat y(\hat t))\Big]\Big\}}
\
}
    \label{eq:BtfT-rescaled}
  \end{eqnarray}
Here  the disorder $ \hat V_{1}$ has a strength $D$ equal to $1$ and is correlated with a correlation length $\hat \xi$ equal to 1 :
\begin{equation}
  \overline{\hat V_1(\hat t,\hat y)\hat V_1(\hat t',\hat y')}=\delta(\hat t'-\hat t)\hat R_{\hat\xi=1}(\hat y'-\hat y)
\end{equation}
and the rescaled Hamiltonian
\begin{equation}
\int_{0}^{\hat \tf} \!\!d\hat t\:\Big[ \tfrac 12 (\partial_{\hat t}\hat y (\hat t))^2+\hat V_{1}(\hat t,\hat y(\hat t))\Big]  
\label{eq:rescHamlowT}
\end{equation}
is independent of the temperature  $T$, which means that~\eref{eq:BtfT-rescaled} is amenable to a well-posed low-temperature asymptotic analysis thanks the large $T_{\cc}/T$ prefactor in the argument of the exponentials.

The numerator and the denominator of \eref{eq:BtfT-rescaled} are both dominated by the same minimising trajectory $\hat{y}^\star_{\hat V}(\hat t;\hat \tf)$ whose existence, this time, is guaranteed by the Lax-Oleinik principle, because the rescaled correlation length of the disorder is finite ($\hat \xi=1$).
Since~\eref{eq:rescHamlowT} is explicitly independent of the physical parameters, $\hat{y}^\star_{\hat V}(\hat t;\hat \tf)$ depends only on $\hat t$ and $\hat \tf$ and on no other parameters.
In particular, at ${\hat \tf =1}$ %or equivalently ${\tf=\frac{T_{\cc}^5}{cD^2}}$,
the path integral hidden in ${b_3(1)}$ yields a plain numerical constant, which is finite thanks to the Lax-Oleinik principle, and the corresponding roughness is:
\begin{equation}
 B(\tf = L_{\cc}(0,\xi)) =  \xi^2,
 \qquad
 L_{\cc} (0,\xi) = \frac{T_{\cc}^5}{cD^2}
 \label{eq:BtfT-rescaled_0b}
\end{equation}
emphasising the special role played by the zero-temperature Larkin length ${L_{\cc} (0,\xi)}$, presented in \eref{eq-fudging-Larkin-length} as the typical lengthscale marking the beginning of the asymptotic regime at large $\tf$.
Coming back to~\eref{eq:BtfT-rescaled}, if we assume that above the Larkin length, at large ${\hat \tf}$, we have a scale invariance in the form of a power law, this means that
\begin{equation}
  b_3(\hat \tf) 
  \ \mathop{\sim}_{\hat \tf\to\infty}\ 
  \hat A\:
  \hat \tf^{\,2\zeta}
\end{equation}
where $\hat A=\overline{(\hat y_{\hat V}^\star(1;\hat\tf))^2}$ is a numerical prefactor, independent of the physical parameters.
The reasoning exposed above does not allow to extract the value of the exponent $\zeta$, but
%because the problem is still in the KPZ universality class
the assumption of ${\zeta=\frac23}$ is supported for instance by the saddle-point argument presented in \sref{sec:saddlepts-optimaltrajectories-large-tf}.
%
%(as confirmed by numerical analysis, see Ref.~\cite{agoritsas_2012_FHHtri-numerics}), one has $\zeta=2/3$.
%combining a scaling compatible with this Flory power counting with the assumption of ${\zeta=\frac23}$ for the asymptotic roughness --~assumption that is supported by the saddle-point argument presented in \sref{sec:saddlepts-optimaltrajectories-large-tf}
%
From~\eref{eq:BtfT-rescaled}-\eref{eq:BtfT-rescaled_0b} it follows that
\begin{eqnarray}
  B(\tf)
 \ \mathop{\sim}_{\tf\to\infty}\ 
\hat A \
\xi^2\
\Big(\frac{T_{\cc}^5}{ cD^2}\Big)^{-\frac 43}
\
\tf^{\frac 43}
\;,\quad
\textnormal{in the $T\to 0$ asymptotics}
\end{eqnarray}
By direct identification with the result~\eref{eq-Basympt-scaling-prediction}, this allows us to identify the low-temperature asymptotic behaviour of the amplitude $\widetilde D$ of the disorder free-energy fluctuations as
\begin{equation}
\widetilde D
  \ \mathop{\sim}_{T\to 0}\ 
\frac{c D}{T_{\cc}}
= c^{2/3} D^{2/3} \xi^{-1/3}
\label{eq:result-Dtilde-lowT}
\end{equation}
We can immediately see, by comparing it to the exact result $\lim_{\xi\to 0}\widetilde D = cD/T$
recalled in \eref{eq-steady-state-distribution-F-Gaussian-2},
that the two limits ${T\to 0}$ and ${\xi\to 0}$ do not commute.

The relation \eref{eq:result-Dtilde-lowT} is one of the few known asymptotic results valid for a $\xi>0$ correlated disorder~; it as been supported by a variety of other analytical approaches~\cite{agoritsas_2012_FHHtri-analytics,phdthesis_Agoritsas2013} as well as checked numerically~\cite{agoritsas_2012_FHHtri-numerics}. 
It is one of the essential ways of characterising the `low-temperature' phase of the KPZ equation, which illustrates the importance of keeping track of the finite disorder correlation length, especially as the thermal fluctuations vanish in this specific limit.
%for this problem the non-commutation of two limits $T\to 0$ and $\xi\to 0$ (this is seen \emph{e.g.}~by comparing~\eref{eq:result-Dtilde-lowT} to the exact result $\lim_{\xi\to 0}\widetilde D = cD/T$).

%-------------------------
\subsection{Interpretation of the saddle-points asymptotics for the scalings}
\label{sec:interpr-saddle-point}

We can revisit the three rescalings that we have presented in the previous subsections by rewriting them with the explicit dependencies of the rescaled roughness functions $\bar B$ and $\bar{B}_{\rm{DP}}$ in the physical parameters,
as defined by \eref{eq-def-roughness-both-mappings}-\eref{eq-def-roughness-scaling-function-Bbar-Dtilde} in \sref{sec:powercounting-Flory}.
Note that we consider ${t=\tf}$, so we have removed thereafter the explicit last parameter giving the length of the interface. %${B(\tf;c,\widetilde D,T,\xi,\tf)}$.

%%%%%%%%%%%%%%%%
%\meevid{
%\begin{eqnarray}
%\fl
%\quad
%  B(\tf;c,\widetilde D,T,\xi)
%\ 
%\stackrel{\eref{eq-rescaling_B_zeta23_0}}=
%\
%\Big[\frac{\widetilde D }{c^2}\Big]^{\frac 23}\tf^{4/3} 
%%
% \bar{B} \Big(\tf=1;c=1,\widetilde D=1,\frac{T}{(\widetilde D^2\tf/c)^{\frac 13}},\frac{\xi}{(\widetilde D\tf^2/c^2)^{\frac 13}}\Big)
%\label{eq:resc1}
%\\[2mm]
%%
%\fl
%\quad
%  B(\tf;c,D,T,\xi)
%\ 
%\stackrel{\eref{eq:BtfFlory-rescaled_0}}=
%\
%  \Big[\frac{D}{c^2}\Big]^{\frac 25}\:
%  \tf^{6/5} \:
%  \bar{B} \Big(\tf=1;c=1,D=1,\frac{T}{(cD^2\tf)^{\frac 15}},\frac{\xi}{(D\tf^3/c^2)^{\frac 15}}\Big)
%%
%\label{eq:resc2}
%\\[2mm]
%%
%\fl
%\quad
%  B(\tf;c,D,T,\xi)
%\ 
%\stackrel{\eref{eq:BtfT-rescaled_0}}=
%\
%  \xi^2
%\
%  \bar{B}\Big(\frac{\tf}{T_{\cc}^5/(cD^2)};c=1,D=1,\frac{T}{T_{\cc}},\xi=1\Big)
%\label{eq:resc3}
%\end{eqnarray}
%}
%
%%%%%%%%%%%%%%%%

\begin{itemize}

%----
\item[$\bullet$]
%"Saddle point on the free energy at large lengthscale $\tf$" % \meevid{(valid!)}
%\sref{sec:saddlepts-optimaltrajectories-large-tf}

The rescaling of \sref{sec:saddlepts-optimaltrajectories-large-tf} is based on a Flory power counting of the free energy, first assuming a dominant Brownian scaling of the disorder free energy, and secondly setting ${\hat \tf =1}$:
\begin{equation}
\fl
  B(\tf;c,\widetilde D,T,\xi)
\ 
\stackrel{\eref{eq-rescaling_B_zeta23_0}}=
\
\Big(\frac{\widetilde D }{c^2}\Big)^{\frac 23}\tf^{4/3} 
 \bar{B}_{\rm{DP}} \Big(1;1,1,\frac{T}{(\widetilde D^2\tf/c)^{\frac 13}},\frac{\xi}{(\widetilde D\tf^2/c^2)^{\frac 13}}\Big)
\label{eq:resc1}
\end{equation}
The existence of a large-$\tf$ saddle point can then be reformulated as:
\begin{equation}
\fl
  b_1(\tf)
  \stackrel{\eref{eq-rescaling_B_zeta23}}{\equiv}
  \bar{B}_{\rm{DP}} \Big(1;1,1,\frac{T}{(\widetilde D^2\tf/c)^{\frac 13}},\frac{\xi}{(\widetilde D\tf^2/c^2)^{\frac 13}}\Big)
  \mathop{=}_{\tf \to \infty}
  \bar{B}_{\rm{DP}} \Big(1;1,1,0,0 \Big)
  \sim \tf^0
  \label{eq:resc1-reformulation}
\end{equation}
In other words, the limit ${\bar{B}_{\rm{DP}} (1;1,1,0,0)}$ is a well-defined \emph{finite} numerical constant, so the roughness scaling can be read straightforwardly from the prefactor in \eref{eq:resc1} at large $\tf$: ${B(\tf;c,\widetilde D,T,\xi) \sim (\widetilde D /c^2 )^{\frac 23}\tf^{4/3}}$. 

%----
\item[$\bullet$]

%"Saddle point on the Hamiltonian with the Flory scaling" %\meevid{(wrong!)}
%\sref{sec:saddlepts-optimaltrajectories-Flory-failure}

Similarly, the rescaling of \sref{sec:saddlepts-optimaltrajectories-Flory-failure} is based on a Flory power counting of the Hamiltonian, setting again ${\hat \tf=1}$:
\begin{equation}
\fl
  B(\tf;c,D,T,\xi)
\ 
\stackrel{\eref{eq:BtfFlory-rescaled_0}}=
\
  \Big[\frac{D}{c^2}\Big]^{\frac 25}\:
  \tf^{6/5} \:
  \bar{B} \Big(1;1,1,\frac{T}{(cD^2\tf)^{\frac 15}},\frac{\xi}{(D\tf^3/c^2)^{\frac 15}}\Big)
\label{eq:resc2}
\end{equation}
The large-$\tf$ limit can be reformulated as:
\begin{equation}
\fl
  b_2(\tf)
  \stackrel{\eref{eq:BtfFlory-rescaled}}{\equiv}
  \bar{B} \Big(1;1,1,\frac{T}{(cD^2\tf)^{\frac 15}},\frac{\xi}{(D\tf^3/c^2)^{\frac 15}}\Big)
  \mathop{\sim}_{\tf \to \infty}
  \tf^{\frac{2}{15}}
  \neq \lim_{\epsilon\to 0} \bar{B} \Big(1;1,1,\epsilon^{\frac 15},\epsilon^{\frac 35} \Big)
\label{eq:resc2-reformulation}
\end{equation}
In fact, this limit is \emph{not} defined, as ${b_2(\tf)}$ diverges at ${\tf \to \infty}$. It can only be characterised indirectly by comparison with \eref{eq:resc1-reformulation}.

%----
\item[$\bullet$]
%"Saddle point on the Hamiltonian at low temperature $T$" %\meevid{(valid!)}
%\sref{sec:saddlepts-optimaltrajectories-low-temperature}

The last rescaling, of \sref{sec:saddlepts-optimaltrajectories-low-temperature},
is based on the same Flory power counting of the Hamiltonian:
\begin{equation}
\fl
  B(\tf;c,D,T,\xi)
\ 
\stackrel{\eref{eq:BtfT-rescaled_0}}=
\
  \xi^2
\
  \bar{B}\Big(\frac{\tf}{L_{\cc}(0,\xi)};1,1,\frac{T}{T_{\cc}},1\Big)
\label{eq:resc3}
\end{equation}
Setting this time ${\hat \tf=\tf/L_c(0,\xi)}$, the zero-temperature limit can consequently be reformulated as:
\begin{equation}
\fl
  b_3(\hat \tf)
  \stackrel{\eref{eq:BtfT-rescaled}}{\equiv}
  \bar{B}\Big(\frac{\tf}{L_{\cc}(0,\xi)};1,1,\frac{T}{T_{\cc}},1 \Big)
  \mathop{=}_{T \to 0}
  \bar{B} \Big(\frac{\tf}{L_{\cc}(0,\xi)};1,1,0,1 \Big)
  \mathop{\sim}_{\tf \to \infty}
  \argp{\frac{\tf}{L_{\cc}(0,\xi)}}^{\frac43}
\label{eq:resc3-reformulation}
\end{equation}
So the limit ${\bar{B} (\hat \tf;1,1,0,1)}$ is well-defined and behaves asymptotically as a power law.
Physically, this means that there exists an optimal trajectory in the random potential, thanks to the Lax-Oleinik principle, whose fluctuations are finite at fixed $\hat \tf$ and scale-invariant at large ${\tf \gg L_{\cc}(0,\xi)}$.

%----
\end{itemize}

The three rescalings \eref{eq:resc1}, \eref{eq:resc2} and \eref{eq:resc3} take the form of a prefactor multiplying the roughness function rescaled in an explicit manner.
In the $\tf\to\infty$ asymptotic regime, the two first relations correspond to a joint $\xi\to 0 $ and $T\to 0$ limit, which, as we have discussed, has to be considered with care.
The saddle-point analysis allowed us to show that the first relation gives the correct KPZ exponent (the rescaled roughness goes to a constant).
In the second relation \eref{eq:resc2}, it would be tempting to first send $T$ to 0 and then to send $\tf$ to $+\infty$, assuming that the rescaled roughness would go to a constant as the rescaled correlation length~$\hat\xi(\tf)={\xi}/{(D\tf^3/c^2)^{\frac 15}}$ goes to zero. However, as we have seen, this is in fact wrong: in~\eref{eq:resc2}, the rescaled roughness keeps full memory of~$\hat\xi(\tf)$ even as $\hat\xi(\tf)\to 0$.
In fact, the rescaling \eref{eq:resc2} was mentioned in \cite{agoritsas_2010_PhysRevB_82_184207}, stating that obviously the simultaneous limit of ${T \to 0}$ and ${\xi \to 0}$ begin ill-defined, as these limits are not exchangeable. Here, thanks to the Lax-Oleinik principle, we can give a physical meaning to this mathematical statement: the existence (or not) of an optimal trajectory with a finite variance, which is guaranteed only in a smooth enough random potential.
As for the last relation, \eref{eq:resc3}, it is more amenable to a well-defined zero-temperature limit, since its correlation length is kept finite, equal to $1$. Assuming the value of the exponent $\zeta$ gives the complete prefactor of the asymptotic roughness in the zero-temperature limit.

We mention finally that an extension of such rescalings and saddle-point analysis gives access to the non-linear response of the interface to an external force driving it out of equilibrium~\cite{agoritsas_garcia-garcia_VL_2016_Arxiv-1605.04405}, characterised by the so-called `creep' law relating the interface velocity to the force.

%_____________________________________________________________________________________________________
%_____________________________________________________________________________________________________
%\newpage
\section{Gaussian Variational Method (GVM)}
\label{sec:GVM}

As we have discussed in the previous section, combining scaling and saddle-point arguments can yield information about the asymptotic behaviour of the roughness function. However, in order to have access to its full lengthscale dependence, alternative analytical tools are required, and the GVM approach precisely provides a framework for computing an approximate expression for $B(\tf)$.
As a starting point, thermal and disorder averages of observables are expressed within the replica approach (see equation~\eref{eq-pathintegral-geom-fluct-v2}) through a replicated Hamiltonian~\eref{eq-Hamiltonian-replicated-explicit} that enters in the definition of its associated Boltzmann weight~\eref{eq-Hamiltonian-replicated-definition}.
In general, the computation of such averages is difficult because the replicated Hamiltonian is non-quadratic.
The GVM computation scheme~\cite{mezard_parisi_1991_replica_JournPhysI1_809,bouchaud_1991_JPhysA24_L1025} consists in finding the `best' quadratic Hamiltonian representing the replicated one, according to a well-defined extremalisation criterion.
It is equivalent to performing a Hartree-Fock approximation on the field theory associated to the replicated Hamiltonian~\cite{mezard_parisi_1991_replica_JournPhysI1_809,shakhnovich_frozen_1989_JPhysA22_1647}.
It has been applied to a variety of systems belonging to the class of elastic manifolds in random media~\cite{mezard_parisi_1991_replica_JournPhysI1_809,mezard_parisi_1992_JPhysI02_2231,goldschmidt-blum_1993_PhysRevE48_161,PhysRevLett72_1530,giamarchi_ledoussal_1995_PhysRevB52_1242}.

Here we first recall in \sref{sec:GVM-previous} the GVM results obtained in previous works~\cite{agoritsas_2010_PhysRevB_82_184207,agoritsas-2012-FHHpenta,phdthesis_Agoritsas2013} on the interface with a short-range elasticity and in a random-bond correlated disorder (as defined in \sref{sec:def-models-Hamiltonian-1Dinterface}).
For the Hamiltonian description of the infinite interface, such a GVM approach yields the Flory exponent $\zeta_{\FF}=3/5$ for the roughness at large $\tf$ instead of the correct KPZ exponent $\zeta=2/3$~\cite{agoritsas_2010_PhysRevB_82_184207,phdthesis_Agoritsas2013}.
We then show in sections~\ref{sec:GVM-upgraded-finite-length}, \ref{sec:GVM-upgraded-finite-length_ii} and~\ref{sec:GVM-upgraded-finite-length_iii} that, in fact, by considering a \emph{finite} interface instead of an infinite one, the GVM computation is strongly modified and yields the correct KPZ exponent for the asymptotic roughness. We finally discuss in \sref{sec:GVM-discussion} the physical interpretation of those results.
In addition, the details regarding the analytical and numerical computations have been gathered in the appendix~\ref{sec:details-gvm-comp} and appendix~\ref{sec:app-iter-num_GVM}, respectively.
%Note that we have gathered in appendix~\ref{sec:details-gvm-comp} and appendix~\ref{sec:app-iter-num_GVM} the details of a GVM computation and of a numerical procedure used to analyse GVM variational equations.

%-------------------------
\subsection{Previous GVM approximation schemes}
\label{sec:GVM-previous}

%\cite{agoritsas-2012-FHHpenta}

% \cite{phdthesis_Agoritsas2013}
% Elisabeth Agoritsas, University of Geneva (Switzerland), 2013.
% "Temperature-dependence of a 1D Interface Fluctuations: Role of a Finite Disorder Correlation Length"
%	Chapter 6:		Gaussian Variational Method (GVM) with replicas
%		And in particular Section 6.6 Concluding remarks, for other GVMs, with refs ad hoc.

In a first computation scheme,
for an infinite interface ($\tf=+\infty$) in the Hamiltonian description, the replicas ${\mathbf y}=(y_a(q))_{1\leq a\leq n}$ are described by continuous Fourier modes with $q\in\mathbb R$, for which a quadratic trial Hamiltonian is defined as follows: 
\begin{equation}
\HHt_0\big[{\mathbf y}\big]
=
\frac 12 \int_{\mathbb R} \dbar q \sum_{a,b=1}^{n}  y_a(-q) {G}^{-1}_{ab}(q) y_b(q)
\qquad ({\rm{with }}\ \dbar q\equiv\tfrac{dq}{2\pi})
\label{eq:Ht0_GVMH}
\end{equation}
with ${G_{ab}^{-1}(q)}$ a ${n \times n}$ a `hierarchical matrix', whose lines and columns are obtained by permutations of the first line \cite{book_beyond-MezardParisi,mezard_parisi_1991_replica_JournPhysI1_809,castellani-cavagna_2005_JStatMechP05012,agoritsas_2010_PhysRevB_82_184207}.
Note that the trial Hamiltonian ${\HHt_0\big[{\mathbf y}\big]}$ does not couple the different Fourier modes.
The Gibbs-Bogoliubov variational principle which characterises~\cite{mezard_parisi_1991_replica_JournPhysI1_809,agoritsas_2010_PhysRevB_82_184207} the best trial Hamiltonian takes the form $\delta \mathcal F_{\var}/\delta G_{ab}(q)=0$ ($\forall a,b,q$), for the variational free energy $\mathcal F_{\var}$ defined as
\begin{equation}
  \mathcal F_{\var}
=
  \mathcal F_0 + \big\langle \HHt - \HHt_0 \big\rangle_0
\label{eq:Fvar_GVMH}
\end{equation}
Here $\langle \cdot \rangle_0$ denotes the average with respect to the variational Boltzmann weight ${\propto e^{-\frac 1T \HHt_0[{\mathbf y}]}}$.
The Hamiltonians $\HHt$ and $\HHt_0$ are respectively given by~\eref{eq-Hamiltonian-replicated-explicit} and~\eref{eq:Ht0_GVMH}.
 Besides 
\begin{equation}
\mathcal F_0 
=
-\frac 1T \log Z_0  
\label{eq:F0_GVMH}
\end{equation}
is the free energy corresponding to the partition function $Z_0$ which normalises the Boltzmann weight $e^{-\frac 1T \HHt_0[{\mathbf y}]}$.
Once the optimal `hierarchical matrices' $G_{ab}^{-1}(q)$ and $G_{ab}(q)$ are found, the roughness is reconstituted from the structure factor computed in the GVM approximation as
\begin{eqnarray}
  S(q)=\int dq' \big\langle y_1(-q') y_1(q)\big\rangle_0
  \stackrel{\rm{(GVM)}}{=} T \lim_{n\to 0} G_{11}(q)
  \label{eq:Sq_GVMH}
  \\
  B(t) =\int_{\mathbb{R}} \dbar q \, 2 \argc{1-\cos (q t)} \, S(q) %\lim_{n\to 0} G_{11}(q)
  \label{eq:Sq_GVMH-Bt-corresp}
\end{eqnarray}
The details of the computation are given in~\cite{agoritsas_2010_PhysRevB_82_184207}. We summarise here the structure of the results for comparison to the ones mentioned or derived in the next subsections.
The quadratic form defined by the replica matrix $G_{ab}$, whose first line is chosen for the definition of the other lines by permutation, is parametrized as
\begin{equation}
  {G}^{-1}_{ab}(q)
=
  c q^2 \delta_{ab}-\sigma_{ab}
\end{equation}
In absence of disorder, ${\sigma_{ab}=0}$ so ${S(q)=\frac{T}{c q^2}}$, and we recover the pure thermal roughness ${B_{\rm{th}}(t)=Tt/c}$ (recalled for instance in \eref{eq-exact-scalings-asymptotic-roughness-small-tf}).
With disorder, the first line of the matrix $\sigma_{ab}$ is represented by an increasing function $\sigma(u)$ of a continuous parameter $u\in[0,1]$, or, equivalently, by the self-energy function
\begin{equation}
  \left[  \sigma \right] (u)
 = u\:  \sigma (u) - \int_0^u dv \,  \sigma (v) 
\label{eq:sigmac_GVMH} 
\end{equation}
Writing explicitly the variational equation and solving it yields \cite{agoritsas_2010_PhysRevB_82_184207} that:
\begin{equation}
  \label{eq:sigmau_GVMH}
  [\sigma](u) = \cases{
    A\, u^{10} 
    &   if $\ u<u_{\cc}$
\\
    [\sigma](u_{\cc})
    &   if $\ u>u_{\cc}$
\\}
\end{equation}
where $A$ is a constant that depends on the physical parameters $c, D$ and $T$.
The cut-off~$u_{\cc}$ depends on $\xi$ and on the other parameters only through the ratio $T/T_{\cc}$.
The power-law behaviour for $u<u_{\cc}$ corresponds to a full replica symmetry breaking (full-RSB) regime, which is often encountered in glassy systems where it indicates the occurrence of an hierarchical organisation of an infinite number of metastable states into valleys and sub-valleys~\cite{book_beyond-MezardParisi}.
%which is often encountered in systems presenting many metastable states, typically in glassy systems.
%
It appears that the exponent $10$ in~\eref{eq:sigmau_GVMH}, when reconstituting the roughness function from the structure factor~\eref{eq:Sq_GVMH}, dictates that the roughness $B(t)$ obtained in the GVM approximation behaves as $B(t)\sim t^{2\zeta_{\FF}}$ at large $t$, with ${\zeta_{\FF}=3/5}$ being the Flory exponent.
This point will be thoroughly discussed in \sref{sec:GVM-discussion} in the comparison between different GVM approaches.

In a second computation scheme, the GVM approach can be also implemented at the free-energy level, as described in~\cite{agoritsas_2010_PhysRevB_82_184207,phdthesis_Agoritsas2013}.
In this construction, for a large but finite interface ($\tf<+\infty$), the replicas ${\mathbf y}=(y_a)_{1\leq a\leq n}$ describe the position of the end-point of the replicated polymer.
Its replicated free energy is given by~\eref{eq-free-energy-replicated-explicit}.
At fixed~$\tf$, the quadratic trial replicated free energy is defined as follows: 
\begin{equation}
\widetilde{F}_0 (\tf,{\mathbf y})
=
\frac 12 \sum_{a,b=1}^{n} y_a {G}^{-1}_{ab}(\tf) y_b
\label{eq:Ht0_GVMtm}
\end{equation}
The Bogoliubov variational principle characterising the best $G_{ab}$ takes the form $\partial \mathcal F_{\var}/\partial G_{ab}(\tf)=0$ ($\forall a,b$), for the variational free energy $\mathcal F_{\var}$ defined as in~(\ref{eq:Fvar_GVMH}-\ref{eq:F0_GVMH}) but now for the corresponding Boltzmann weight $e^{-\frac 1T \widetilde{F}_0 (\tf,{\mathbf y})}$.
Once the optimal hierarchical matrix $G_{ab}^{-1}(\tf)$ is found, the GVM approximation of the roughness function is reconstituted directly from
\begin{equation}
  B(\tf) \stackrel{\rm{(GVM)}}{\simeq} \big\langle (y_1)^2\big\rangle_0 =  T \lim_{n\to 0} G_{11}(\tf)
  \label{eq:Sq_GVMtm}
\end{equation}
The structure of the solution of the variational equation is rather different from the infinite-$\tf$ Hamiltonian one: in contrast to~\eref{eq:sigmau_GVMH} one now has two plateaus
\begin{equation}
  \label{eq:sigmau_GVMtm}
  [\sigma](u) = \cases{
    0
    &   if $\ u<u_{*}(\tf)$
\\
    A\, u^{2} 
    &   if $\ u_*(\tf)<u<u_{\cc}$
\\
    [\sigma](u_{\cc})
    &   if $\ u>u_{\cc}$
\\}
\end{equation}
The form of the solution yields a roughness ${B(\tf)\sim\tf^{4/3}}$ (at $\tf\to\infty$) with the KPZ roughness exponent. This GVM approximation %description %yielding $\zeta=2/3$ 
however relies on the assumption that the disorder free energy $\bar F_V(\tf,y)$
%is close to that of a Brownian motion
fluctuates as a Brownian process in the coordinate $y$. This assumption is only approximate at $\xi>0$ and $\tf<\infty$, as discussed in \sref{sec:def-models-previous-results}.
Nevertheless, we have actually checked in a third GVM computation scheme that including the saturation of the two-point correlator ${\bar{C}(\tf,y)}$ for large $y$ \eref{eq-large-times-two-point-correlator-Cbar}, and thus breaking the Brownian scaling, still self-consistently predicts the same asymptotic roughness behaviour, see \cite{agoritsas-2012-FHHpenta} and appendix~D of \cite{phdthesis_Agoritsas2013}.

In the next subsections, we present a new GVM approach, based on the Hamiltonian description at finite $\tf$ (hence not based on such a Brownian approximation) while still allowing us to recover the correct KPZ exponent.

%-------------------------
\subsection{Beyond the Flory scaling: a  finite-$\tf$ GVM scheme in the Hamiltonian description}
\label{sec:GVM-upgraded-finite-length}

Our aim is to adapt the Hamiltonian GVM of the infinite interface to the description of a finite interface of length $\tf$.
To proceed, we first identify the rescaling which makes it possible to extract as a prefactor of $B(\tf)$ the dominant behaviour at large~$\tf$.
Instead of using the Flory rescaling~\eref{eq:explicit-rescaling-F} of the (non-replicated) Hamiltonian~\eref{eq-Hamiltonian-original},
we use the following rescaling:
%
% If only my anguish could be weighed
%     and all my misery be placed on the scales!
% It would surely outweigh the sand of the seas—
%     no wonder my words have been impetuous.
% [Job, 6:1-3]
%
\begin{equation}
  t = \tf\,\hat t ,
\qquad 
  y(t) =  \tf^\zeta \big(\tfrac{D}{c T}\big)^{\frac 13}\, \tilde y(\hat t) ,
\qquad 
  \zeta = \frac 23
\label{eq:explicit-rescaling-F-repH}
\end{equation}
which, in the \emph{replicated} Hamiltonian~\eref{eq-Hamiltonian-replicated-explicit}, absorbs the coefficients of the elastic and disorder contribution into a common prefactor:
%
%
%
%\begin{small}
  \begin{eqnarray}
\fl\qquad
    B(\tf) 
    &=
    \bigg(\frac{D}{cT}\bigg)^{\frac 23}
    \tf^{2\zeta}\
%
%    \frac
      \displaystyle
      \lim_{n\to 0}
      \int_{\tilde {\mathbf y}(0)=0}\hspace*{-7mm}
      \mathcal D\hat{\mathbf y}(\hat t)
      \:{\tilde y}_1(1)^2\,
      \exp
      \left\{
        \!-\!\left[\frac{\tf}{\tfrac{T^5}{cD^2}}\right]^{\frac 13}
\HHt\big[\tilde {\mathbf y}(\hat t),\mathring\xi(\tf)\big]
      \right\}
\label{eq:exprBtfKPZreplicae}
  \end{eqnarray}%
%\end{small}%
with
\begin{equation}
\fl \qquad
\HHt\big[\tilde {\mathbf y}(\hat t),\mathring\xi(\tf)\big]
=
\int_0^1 \!\! d\hat t\:
         \bigg[
           \sum_a \tfrac 12 (\partial_{\hat t}\tilde y_a)^2
          -  
           \sum_{a<b}
           \hat R_{\mathring\xi(\tf)}\Big(\tilde y_b(\hat t)-\tilde y_a(\hat t)\,\Big)
         \bigg]
\label{eq:replicatedHamKPZresc}
\end{equation}
Now instead of the Flory-rescaled correlation length~\eref{eq:defxiFF}, the correlator is taken at a KPZ-rescaled correlation length
\begin{equation}
\fl \qquad
  {\mathring\xi(\tf)}= \frac{\xi}{\tf^{\zeta} \big(\tfrac{D}{cT}\big)^{\frac 13}}
  = \frac{\xi}{\sqrt{B_{\rm{asympt}}(\tf)}},
  \qquad B_{\rm{asympt}}(\tf) = \argp{\frac{D}{cT}}^{2/3} \tf^{\frac43}
  \label{eq:defxiKPZ}
\end{equation}
We will denote the rescaled inverse temperature appearing in~\eref{eq:exprBtfKPZreplicae} as
\begin{equation}
\fl \qquad
\hat\beta (\tf)
	= \left[\frac{\tf}{\tfrac{T^5}{cD^2}}\right]^{\frac 13}
	= \left[\frac{\tf}{L_c (T,0)}\right]^{\frac 13}
\label{eq:defbetahat}
\end{equation}
%
%Both  ${{\mathring\xi(\tf)}}$ and ${\hat\beta (\tf)}$ have an explicit dependence on $\tf$, but from now we will skip it to simplify the notations.
%
We emphasise that in the large-$\tf$ regime, we have ${{\mathring\xi(\tf)} \sim \tf^{-2/3} \ll 1}$ and ${\hat\beta (\tf)^{-1}\sim \tf^{- 1/3} \ll 1}$, so that studying the asymptotic roughness at $\tf\to\infty$ amounts here to taking a zero-temperature-like limit, since the reduced temperature $T/T_{\cc}$ reads $1/(\hat\beta\mathring\xi^{1/3})\sim\tf^{-2/9}\ll 1$.

Although, in~\eref{eq:exprBtfKPZreplicae}, the correct large-$\tf$ behaviour in $\tf^{4/3}\!$ appears as a prefactor of the roughness, it is not obvious how to extract from the expression~\eref{eq:exprBtfKPZreplicae} the large-$\tf$ asymptotics of $B(\tf)$. Indeed, performing a saddle-point asymptotic analysis in the $\tf\to\infty$ regime is not compatible with the $n\to 0$ limit, and a full computation of the replicated roughness would involve the understanding of a complete $\tf$-dependent Bethe Ansatz solution~\cite{dotsenko_2010_EPL90_20003,dotsenko_2010_JStatMech_P03022,calabrese_2010_EPL90_20002,ledoussal_calabrese_2012_JStatMech2012_P06001} ---~which is out of reach of the presently available technology and would also only be possible at $\xi=0$.
Here, we will resort to a study of the roughness~\eref{eq:exprBtfKPZreplicae} based on the GVM approximation.
The choice \eref{eq:explicit-rescaling-F-repH} is the Flory scaling obtained in \sref{sec:powercounting-Flory-Hamiltonian-replicae}, and as already mentioned,  this power counting coincides with the high-temperature (or equivalently $\xi=0$) asymptotic roughness scaling. As such, it includes explicitly the high-temperature Larkin length ${L_{\cc}(T,0) \sim T^5/(cD^2)}$ of \eref{eq-fudging-Larkin-length}.
This means in particular that, for our GVM computation to predict the correct asymptotic roughness, we must have at ${T \gg T_c(\xi)}$:
\begin{equation}
\fl\qquad
      \displaystyle
      \lim_{n\to 0}
      \int_{\tilde {\mathbf y}(0)=0}\hspace*{-7mm}
      \mathcal D\hat{\mathbf y}(\hat t)
      \:{\tilde y}_1(1)^2\,
      \exp
      \left\{
        \!-\!\left[\frac{\tf}{\tfrac{T^5}{cD^2}}\right]^{\frac 13}
\HHt\big[\tilde {\mathbf y}(\hat t),\mathring\xi(\tf)\big]
      \right\}
\mathop{\to}_{\tf \to \infty} {\rm{cte}} < \infty
\label{eq-GVM-expectations}
\end{equation}

The complete GVM study of this problem is itself rather cumbersome: one has to separate the fluctuations of the endpoint (at $t=\tf$) from the bulk ones ($0<t<\tf$) in order to use a Fourier representation of $\tilde y(\hat t)$.
This requires to use a `dual GVM' scheme, with distinct Gaussian variational Ansätze for the endpoint and bulk fields~\cite{GVM-in-preparation}.
We restrict our study here to a simplified case yielding physically equivalent results: from now on, we constrain the interface to start from 0 in 0 and to end in 0 in $\tf$.
The roughness will be measured at any intermediate point $0<t<\tf$ (for instance in $t=\tf/2$).
In the large-$\tf$ limit, the corresponding roughness exponent is not affected by such a boundary condition. The analysis of the dual GVM will be exposed in a future study~\cite{GVM-in-preparation}.

Because the interface has a  finite length ${\hat{t} \in [0,1]}$, the Fourier modes are discrete, and we indexed them $\omega\in 2\pi\mathbb Z$:
\begin{equation}
  \tilde y(\omega) = \int_0^1d\hat t\: e^{-i\omega\hat t}\: \tilde y(\hat t)
\qquad
  \tilde y(\hat t) = \sum_{\omega} e^{i\omega\hat t}\: \tilde y(\omega)
\label{eq:def_TFdisc}
\end{equation}
The rescaled replicated Hamiltonian~\eref{eq:replicatedHamKPZresc} rewrites
\begin{eqnarray}
\label{eq-Hreplic-part1}
\HHt\big[\tilde {\mathbf y}(\omega),\mathring\xi\big]
&\ =\
\HHt_\text{el}\big[\tilde {\mathbf y}(\omega) \big]
+
\HHt_\text{dis}\big[\tilde {\mathbf y}(\omega) ,\mathring\xi\big]
\\[3mm]
\label{eq-Hreplic-part2}
\HHt_\text{el}\big[\tilde {\mathbf y}(\omega) \big]
&\ =\
\sum_a \sum_\omega \tfrac 12  \omega^2 \tilde y_a(-\omega) \tilde y_a(\omega)
\\
\label{eq-Hreplic-part3}
\HHt_\text{dis}\big[\tilde {\mathbf y}(\omega) ,\mathring\xi\big]
&\ =\
-
\int_{\mathbb{R}} \dbar \lambda
R_{\mathring \xi}(\lambda) 
\int_0^1 \!\! d\hat t\:
           \sum_{a<b}\:
           e^{i\lambda\left[\tilde y_b(\hat t)-\tilde y_a(\hat t)\right]}
\end{eqnarray}
Since we work at fixed $\tf$, from now on we will skip the explicit dependence on $\tf$ of both ${{\mathring\xi(\tf)}}$ and ${\hat\beta (\tf)}$, in order to simplify the notations.
%The Fourier transformations are defined as in~\eref{eq:def_TFdisc}. \\

Similarly to the GVM presented in \sref{sec:GVM-previous}, the hierarchical replica matrix $G_{ab}^{-1}(\omega)$,
with ${a,b \in \arga{1, \dots, n}}$ and ${\omega \in 2 \pi \mathbb{Z}}$,
is parametrized as
\begin{equation}
  {G}^{-1}_{ab}(\omega)
=
  \omega^2 \delta_{ab}-\sigma_{ab}
\end{equation}
For its inverse, one has on the one hand the diagonal coefficient
\begin{equation}
G_{aa}(\omega)= \tilde G(\omega)
\end{equation}
which encodes the structure factor and hence the roughness according to \eref{eq:Sq_GVMH}-\eref{eq:Sq_GVMH-Bt-corresp}.
On the other hand, for $a\neq b$, as in \sref{sec:GVM-previous}, the $G_{ab}(\omega)$ are described by a function  $G(\omega,u)$ of a continuous parameter $u\in[0,1]$.
As derived in \sref{sec:contparamvareq_app} of appendix~\ref{sec:details-gvm-comp}, the variational equations take the following form
\begin{eqnarray}
\fl\qquad
 \sigma (u)
=
\frac{2}{\sqrt\pi}
\hat\beta^{\frac 32}
\bigg\{
\hat\beta
\mathring\xi^2+
  \sum\limits_\omega
  \big[\tilde G(\omega) - G(\omega,u)\big]
\bigg\}^{-\frac 32}
\label{eq:eqvar_iGVM_txt}
% \end{eqnarray}
% %
% where from the hierarchical inversion relations
\\
\fl\qquad
    \sum_\omega  \big[\tilde G(\omega) - G(\omega,u)\big]
%    &
= \frac{1}{u} \:
      \frac{\coth \argp{\tfrac 12\sqrt{[ \sigma](u)}}}{2\sqrt{[ \sigma](u)}}
      - \int^1_u \frac{dv}{v^2} \:
      \frac{\coth \argp{\tfrac 12\sqrt{[ \sigma](v)}}}{2\sqrt{[ \sigma](v)}}
\label{eq:sum_omega_inv_iGVM_txt}
\end{eqnarray}
The variational equations of the infinite-$\tf$ Hamiltonian GVM~\cite{agoritsas_2010_PhysRevB_82_184207} present a very similar structure: the only difference is that the expressions $\coth \argp{\tfrac 12\sqrt{[ \sigma](\cdot)}}$ in~\eref{eq:sum_omega_inv_iGVM_txt} are then replaced by $1$.
As we will explain, this difference crucially implies that the two GVMs possess distinct scalings.
From a technical point of view, we will see that the singular behaviour of $\coth x$ as
$x\to 0$ affects the full scaling of the GVM solution.
This corresponds to the fact that the small-$[\sigma](u)$ regime governs the large physical scales, and this is precisely the regime where $\coth \argp{\tfrac 12\sqrt{[ \sigma](u)}}$ behaves very differently from $1$.

\begin{figure}[!t]
  \centering
  \includegraphics[width=.4\columnwidth]{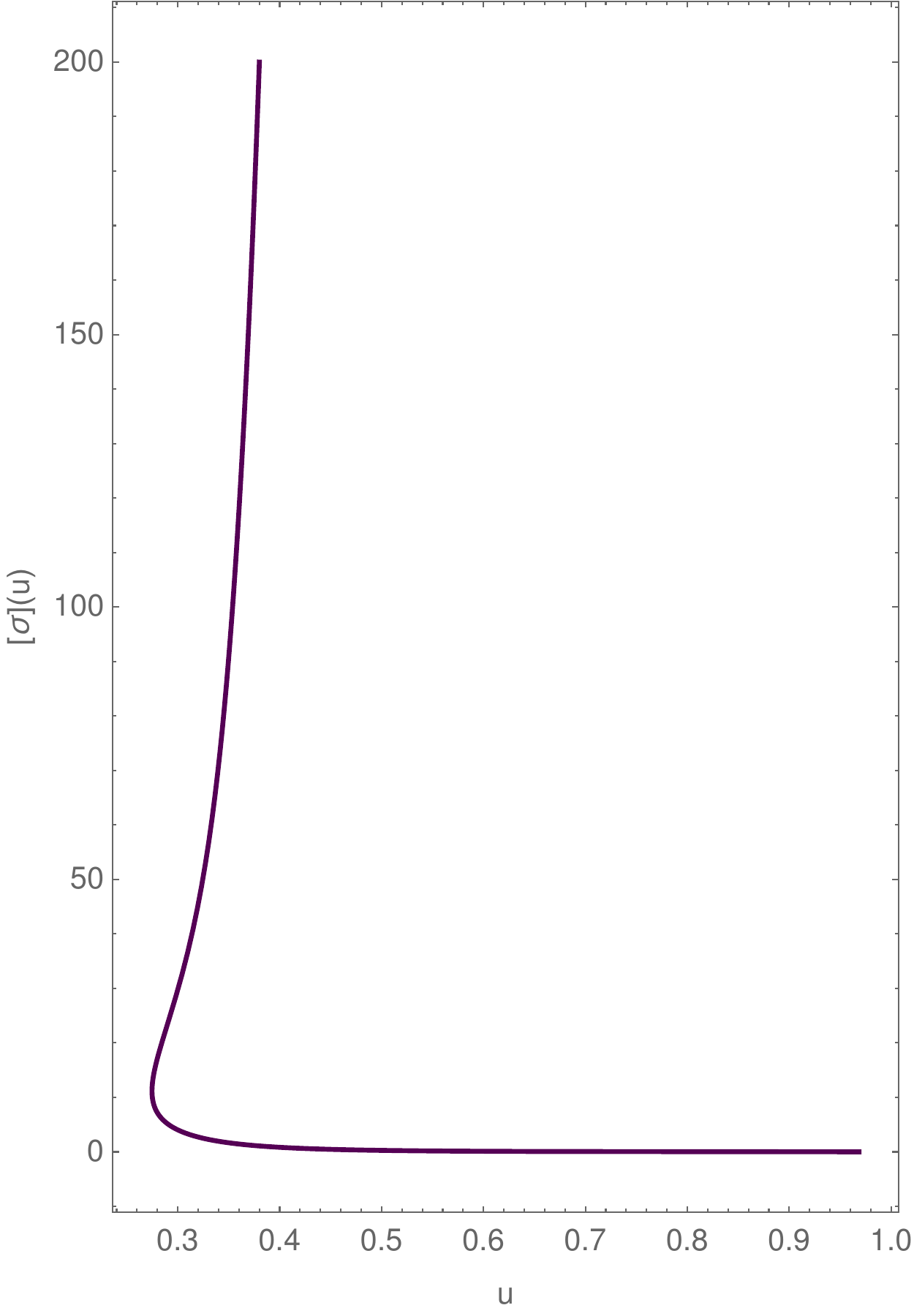}\qquad
  \includegraphics[width=.4\columnwidth]{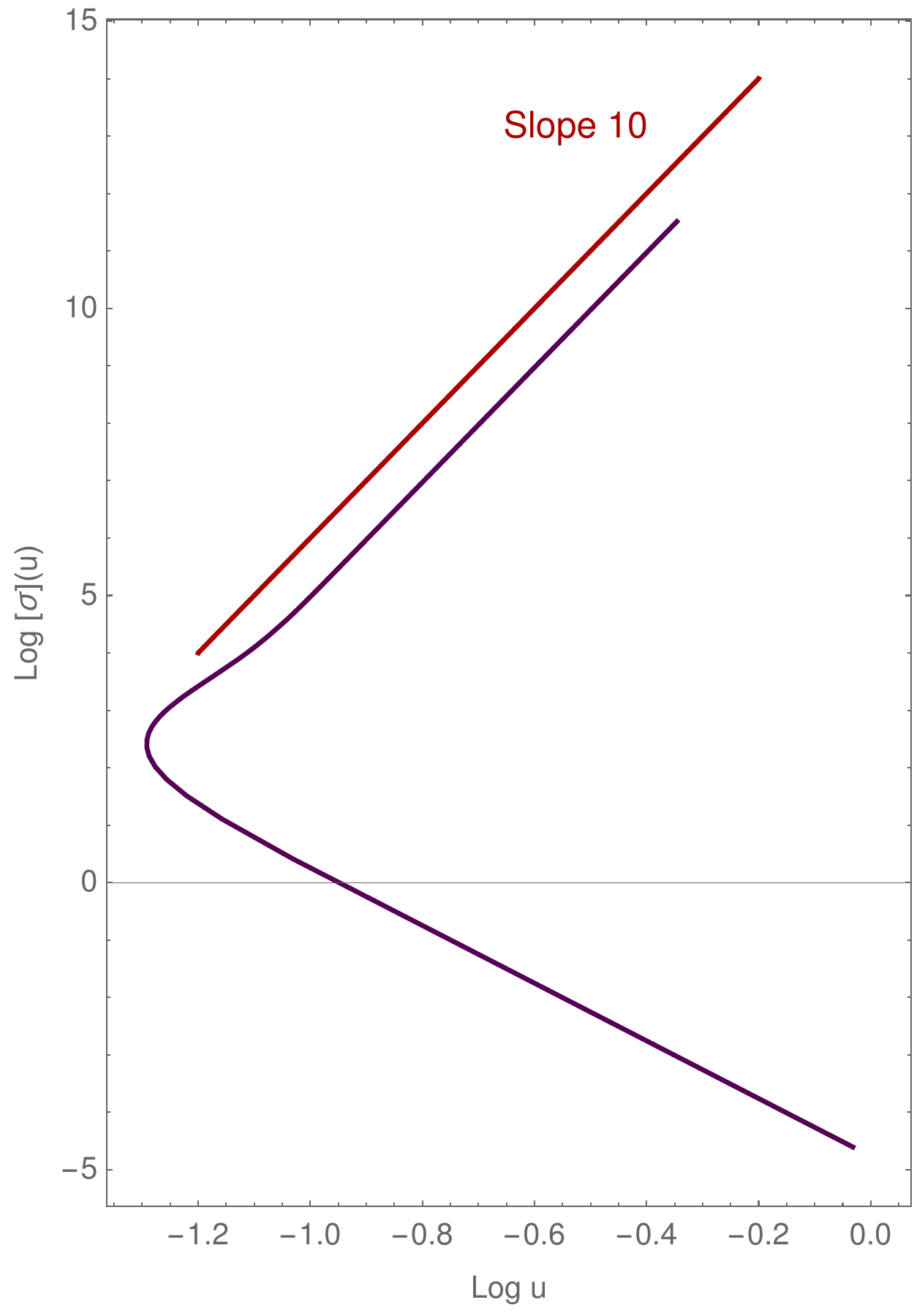}
  \caption{
  Possible value of $[\sigma](u)$ deduced from the parametric equation~\eref{eq:sigmaCimplicit_txt} for the finite-$\tf$ Hamiltonian GVM, when ${\sigma'(u) \neq 0}$.
(\textbf{Left})~Normal scale. (\textbf{Right})~Log-Log scale.
The increasing branch of $[\sigma](u)$ is the physical branch. %since ${\sigma(u)}$ is defined as being a monotonous function.
At large $[\sigma](u)$, one recovers a behaviour of the form $[\sigma](u)\sim A\,u^{10}$, akin to the infinite-$\tf$ GVM result~\eref{eq:sigmau_GVMH}.
%finite-$\tf$ result
However, for small $u$ there is no solution such that ${\sigma'(u) \neq 0}$, which implies that $\sigma(u)$ has a plateau on an interval $[0,u_\star]$ with $u_\star>0$.
This fact marks the main physical difference between the finite-$\tf$ and the infinite-$\tf$ GVM solutions.
The numerical parameter is $\hat\beta=10$.
\label{fig:sigmaCimplicit}% 
}
\end{figure}

%-------------------------
\subsection{Results of  the finite-$\tf$ GVM scheme }
\label{sec:GVM-upgraded-finite-length_ii}

As detailed in \sref{sec:compar_hGVM_iGVM}  of appendix~\ref{sec:details-gvm-comp}, one infers from the variational equations~(\ref{eq:eqvar_iGVM_txt}-\ref{eq:sum_omega_inv_iGVM_txt}) that whenever $\sigma'(u)\neq 0$ (\textit{i.e.}~when the solution is not a plateau), one has:
\begin{eqnarray}
\fl \quad
1
=
\frac 35
\Big(\frac{27\times 2^9}{\pi}\Big)^{\frac 15}
\hat\beta^{\frac 35} \:u\: \{[\sigma](u)\}^{-\frac{1}{10}}
\nonumber
\\
\fl\qquad \quad\;
\times
\underbrace{
\bigg[
  \frac
  {\sinh^2 \argp{\tfrac 12\sqrt{[ \sigma](u)}}}
  {\sinh \argp{\sqrt{[ \sigma](u)}}+\sqrt{[ \sigma](u)}}
  \bigg]^{\frac 35}
% \nonumber
% \\
% &\quad\qquad\qquad\qquad\;\quad\;
%\times
\;
\left\{
1+
\frac 13
  \frac
  {[\sigma](u)\coth \argp{\tfrac 12\sqrt{[ \sigma](u)}}}
  {\sinh \argp{\sqrt{[ \sigma](u)}}+\sqrt{[ \sigma](u)}}
\right\}
}_{ =1~\text{for~the~infinite{\mbox{\footnotesize{-}}}}\tf\text{~GVM}%_{\mathcal H}
}
\label{eq:sigmaCimplicit_txt}
\end{eqnarray}
This equation takes the form $u=\hat\beta^{-\frac 35} \mathcal G\big([\sigma](u)\big)$ from which one infers a parametric form for $[\sigma](u)$.
For the infinite-$\tf$ {GVM}, the solution is simply $[\sigma](u)\propto u^{10}$, as recalled in \sref{sec:GVM-previous}.
We represent on figure~\ref{fig:sigmaCimplicit} the form of $[\sigma](u)$ implied by the parametric equation~\eref{eq:sigmaCimplicit_txt} for the finite-$\tf$ GVM.
Only one branch is physically allowed: the one with $[\sigma](u)$ increasing. In the $[\sigma](u)\gg 1$ regime, the result becomes equivalent to $[\sigma](u)\propto u^{10}$. This is self-consistently checked from~\eref{eq:sigmaCimplicit_txt}, where for $[\sigma](u)\gg 1$ the underbrace goes to 1.

A striking feature of the parametric equation~\eref{eq:sigmaCimplicit_txt} is that there is no solution for $[\sigma](u)$ in the regime $u\to 0$.
It marks a strong difference with the \mbox{infinite-$\tf$} Hamiltonian GVM.
This is seen from~\eref{eq:sigmaCimplicit_txt}, for instance by looking for a solution of the form $[\sigma](u)\sim u^{\nu}\ll 1$ (with $\nu>0$) in the limit $u\to 0$.
One finds $\nu=-5$ which is self-contradictory with the assumption $[\sigma(u)]\ll 1$. %, since ${\sigma(u)}$ is defined as being a monotonous function.
Also, a direct numerical study of the parametric equation~\eref{eq:sigmaCimplicit_txt}, read as $u=\hat\beta^{-\frac 35} \mathcal G\big([\sigma](u)\big)$, shows that the function $\mathcal G(\cdot)$ only takes values which are bounded away from~$0$ --~see figure~\ref{fig:sigmaCimplicit}.
This implies that
\textit{(i)}~there is necessarily a non-empty interval $[0,u_\star[$ on which $[\sigma](u)$ is constant, since \eref{eq:sigmaCimplicit_txt} is valid only for $\sigma'(u)\neq 0$\,;
and that \textit{(ii)}~$[\sigma](u)$ presents a discontinuity (\emph{i.e.}~a step) in $u_\star$, since the minimal value that a non-constant $[\sigma](u)$ can take is strictly positive according to \eref{eq:sigmaCimplicit_txt}.

Note that the discontinuous behaviour of $[\sigma](u)$ that we have described as resulting from the parametric equation~\eref{eq:sigmaCimplicit_txt} can be traced back to the singular behaviour of  $\coth \big({\sqrt{[\sigma](u)}}\,\big)$ as $[\sigma](u)\to 0$ in the variational equation~\eref{eq:sum_omega_inv_iGVM_txt}. In turn, the $\coth$ function arises from the sum over the \emph{discrete} Fourier modes $\omega\in 2\pi\mathbb Z$, \emph{i.e.}~from the very fact that the interface that we consider has a finite length $\tf$.
A sum over continuous Fourier modes yields $1$ instead of the $\coth$ function.

%-------------------------
\subsection{Scaling analysis in the $\xi\to 0$ regime}
\label{sec:GVM-upgraded-finite-length_iii}

\begin{figure}[!t]
  \centering
  \includegraphics[width=.48\columnwidth]{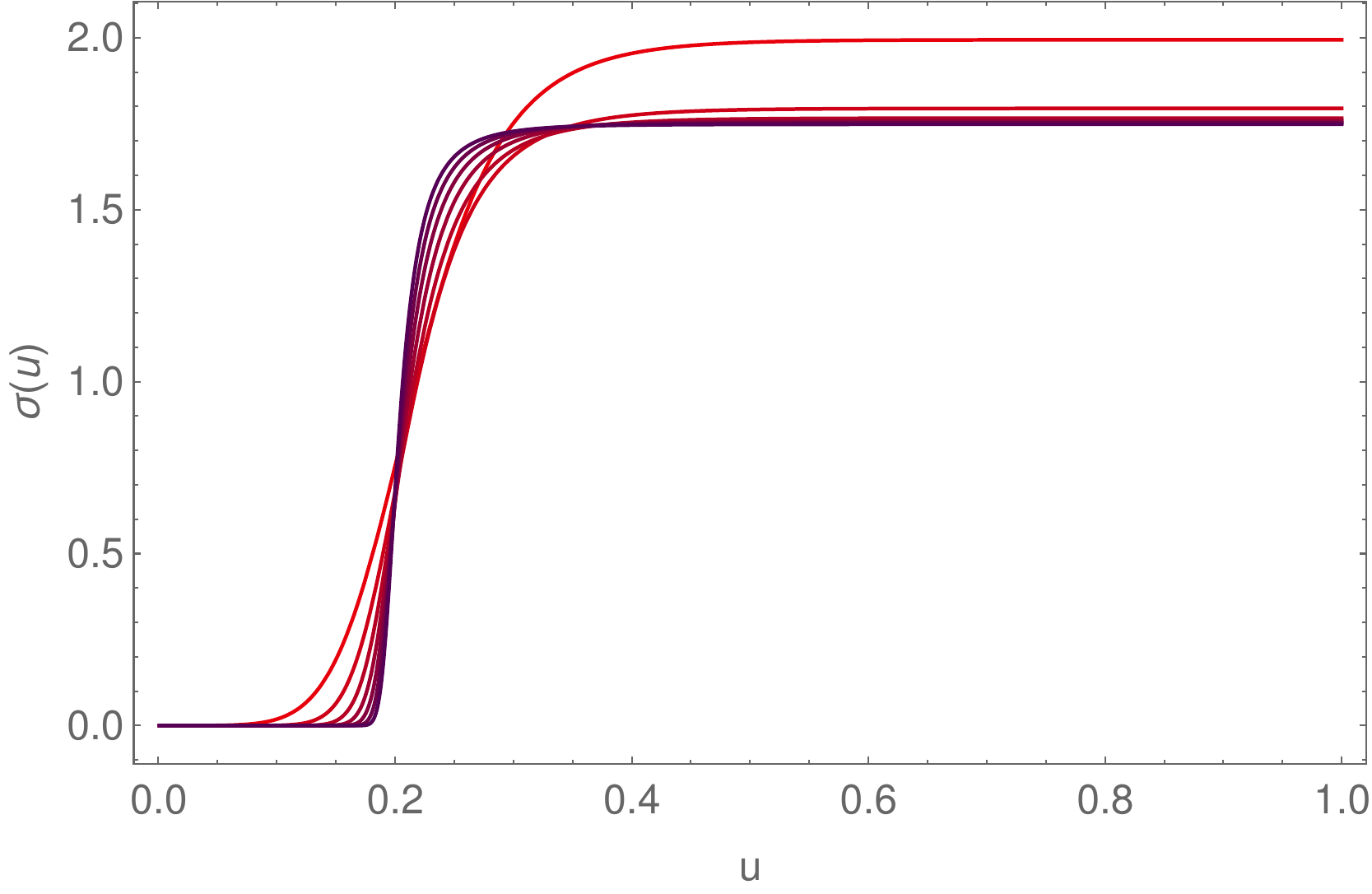}
\hfill
  \includegraphics[width=.50\columnwidth]{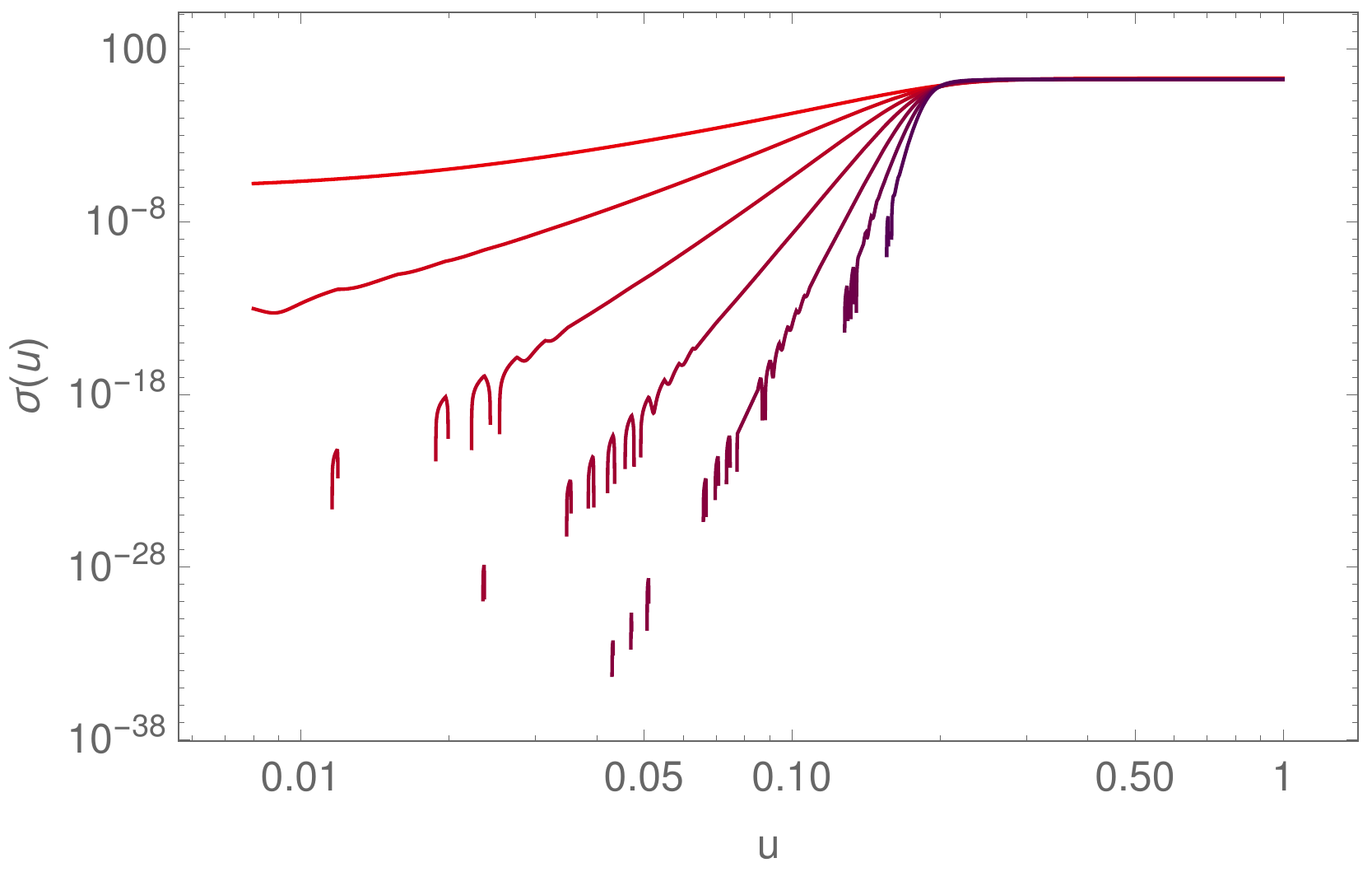}
  \caption{
$7$ iterations of the fixed-point procedure deduced from~(\ref{eq:eqvar_iGVM_txt}-\ref{eq:sum_omega_inv_iGVM_txt}) (and described in appendix~\ref{sec:iternum_iGVM}) for the finite-$\tf$ Hamiltonian GVM. It supports the convergence of $\sigma_k(u)$ to a \mbox{1-step RSB} form, with $\sigma(u<u_{\cc})=0$ and  $\sigma(u>u_{\cc})={\rm{cte.}}$, corresponding to~\eref{eq:1RSB_iGVM_txt}.
The index $k$ of the iteration increases from the red to the dark purple curves.
(\textbf{Left})~$\sigma_k(u)$. 
(\textbf{Right}~$\sigma_k(u)$ in log-log scale.  One observes that in the region before the cut-off $u_{\cc}$, $\sigma_k(u)$ does not converge to any stable power-law regime, contrarily to what is observed for other GVM procedures (see figures~\ref{fig:GVM_ham_iterations} and~\ref{fig:GVM_toy_iterations}).
The numerical parameters are $\hat\beta=10$ and $\mathring\xi=0.8$.
\label{fig:GVM_toy_iterations_i}%
}
\end{figure}
%%
%% from ~/recherche/GVM-finite-time_interface_iteration_intermediate-GVM.nb
%% from ~/recherche/GVM-finite-time_interface_iteration_intermediate-GVM_better.nb
%%

Solving the variational equations (\ref{eq:eqvar_iGVM_txt}-\ref{eq:sum_omega_inv_iGVM_txt}) and finding the stable optimal solution is a more complex task for the finite-$\tf$ Hamiltonian GVM than for the infinite-$\tf$ one, % Hamiltonian GVM.
for which ${[\sigma](u)}$ is a pure power law with a plateau at ${u \in [u_{\cc},1]}$.
To help finding the form of the solution, we have developed a numerical iterative procedure, as exposed in appendix~\ref{sec:app-iter-num_GVM}.
As a benchmark, we have tested this procedure on the infinite-$\tf$ Hamiltonian GVM (see \sref{sec:interGVMH}) and on the free-energy GVM (see \sref{sec:numericGVMtm}), whose analytical solutions are fully known \cite{agoritsas_2010_PhysRevB_82_184207,phdthesis_Agoritsas2013}.
We have obtained that this iterative procedure successfully recovers the analytical results recalled in \sref{sec:GVM-previous}.

\begin{figure}[!t]
  \centering
  \includegraphics[width=.47\columnwidth]{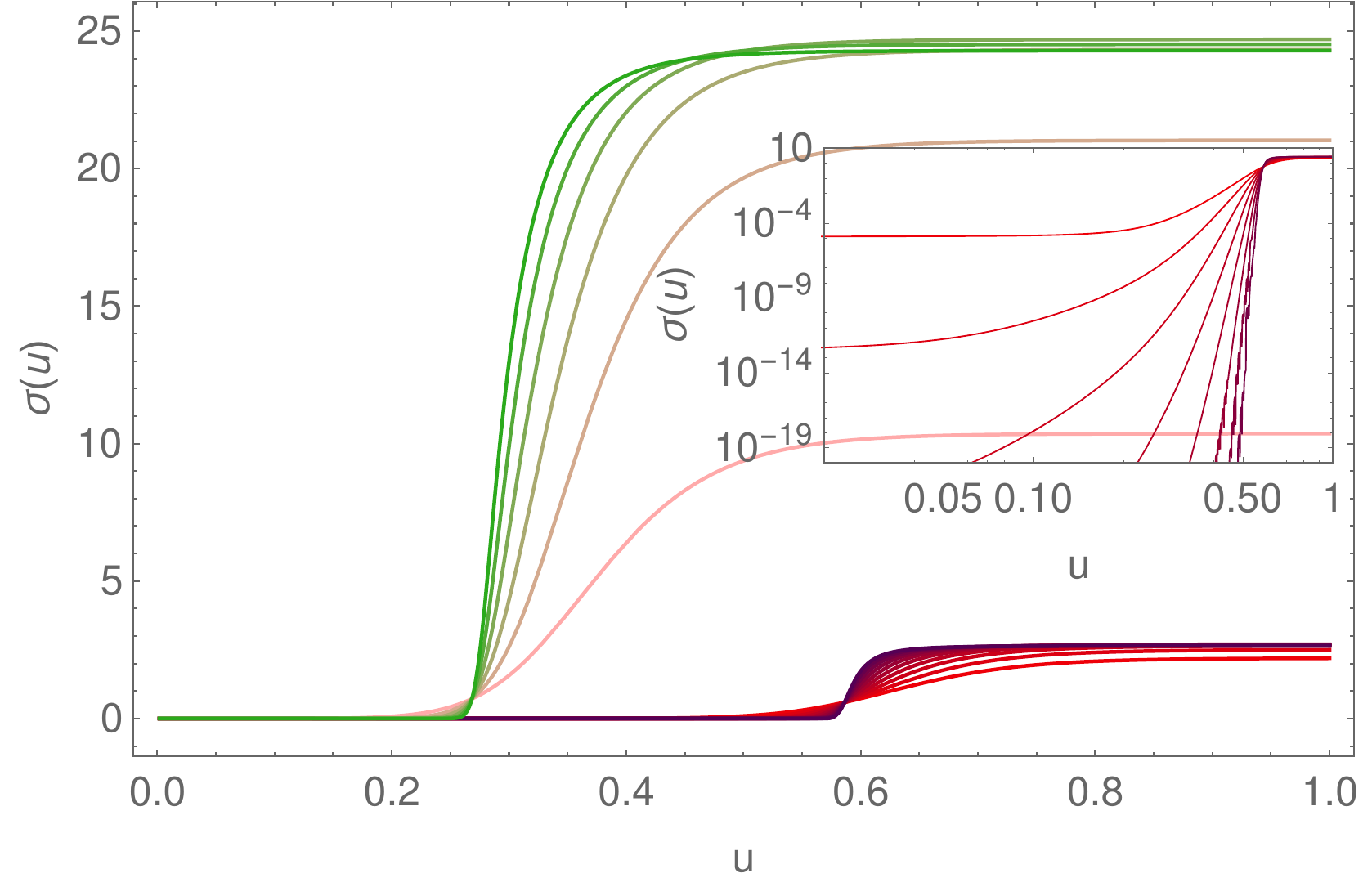}
\quad
  \includegraphics[width=.47\columnwidth]{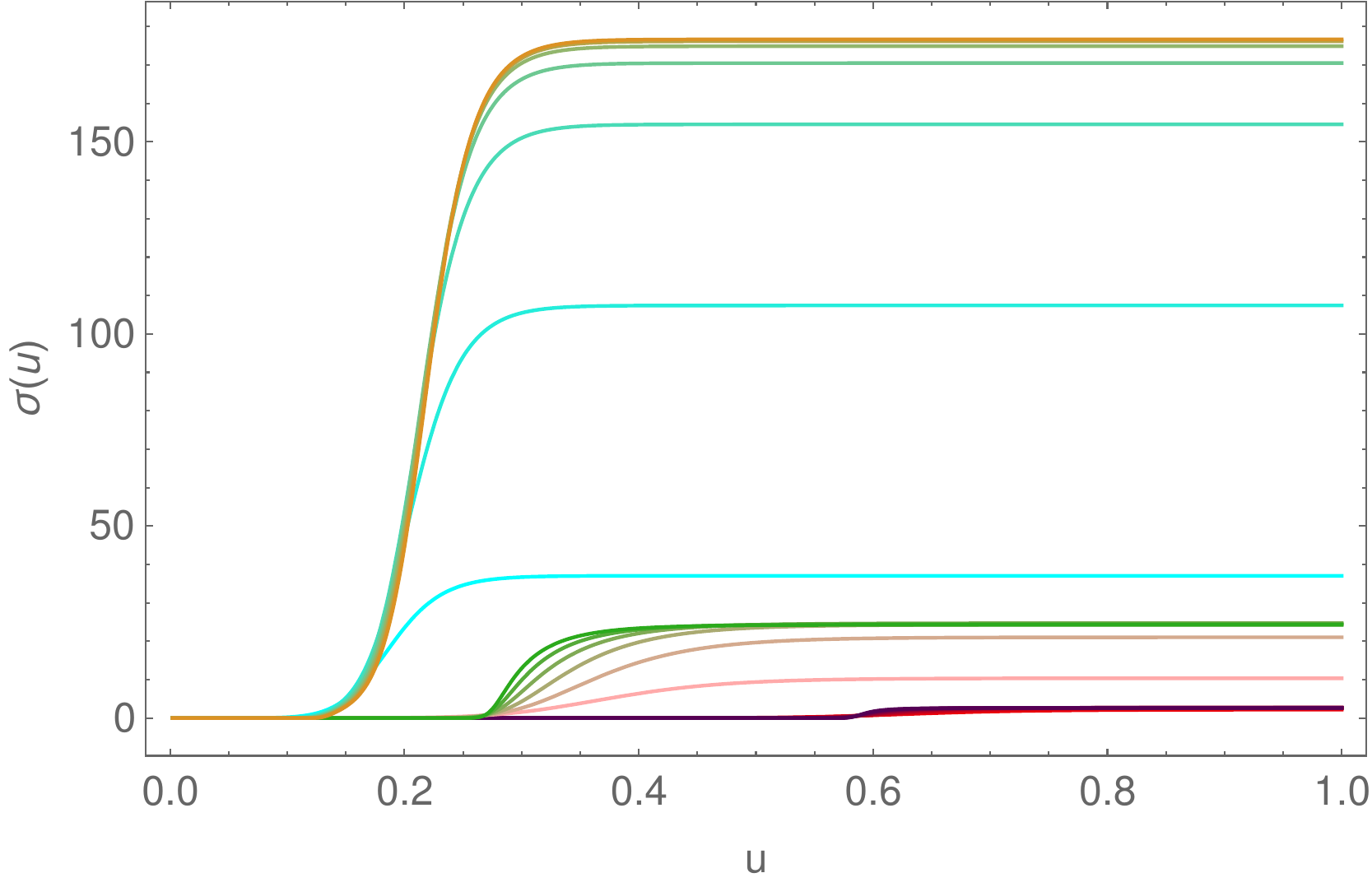}
  \caption{
Iterations $\sigma_k(u)$ of the fixed-point procedure for $\sigma(u)$ (described in \sref{sec:iternum_iGVM}) for the finite-$\tf$ GVM, in the small-$\xi$ regime, where $\sigma_k(u)$ converges to a 1-step RSB form.
The iterations $\sigma_k(u)$ are evaluated for different values of $\tf\in\{\tf^{(1)},\tf^{(2)}=2^3 \tf^{(1)},\tf^{(3)}=2^6 \tf^{(1)}\}$.
(\textbf{Left})~Results for $\tf^{(1)}$ ($k$ increasing from red to violet) and $\tf^{(2)}$ ($k$ increasing from rose to green). The inset shows the log-log representation for $\tf^{(1)}$, illustrating the absence of power-law regime.
(\textbf{Right})~Results for $\tf^{(1)}$, $\tf^{(2)}$ and $\tf^{(3)}$ ($k$ increasing from cyan to orange).
The scaling~\eref{eq:uc_Sigma1_largetf_iGVM_txt} implies that $u_{\cc}$ is divided by $2$ from $\tf^{(i)}$ to $\tf^{(i+1)}$, as observed. Similarly, the height~$\Sigma_1$ of the plateau $\sigma(u>u_{\cc})$ is multiplied by~$8$.
This supports the scaling relation~\eref{eq:uc_Sigma1_largetf_iGVM_txt}: $\hat \beta u_{\cc}\sim \tf^0$ that is essential for the determination of the KPZ exponent $\zeta=\frac 23$, see~\eref{eq:iBt1tf_iGVM_Poisson}.
\label{fig:GVM_toy_iterations_iGVM_scaling}%
}
\end{figure}

We have applied the same numerical procedure to the finite-$\tf$ Hamiltonian GVM, for small values of the disorder correlation length~$\xi$ (which yield the most stable numerical results).
As shown on figure~\ref{fig:GVM_toy_iterations_i},
the iterative procedure for the GVM equations~(\ref{eq:eqvar_iGVM_txt}-\ref{eq:sum_omega_inv_iGVM_txt}), described in appendix~\ref{sec:iternum_iGVM}, supports a \emph{1-step RSB form} for $\sigma(u)$, instead of a full-RSB solution.
It corresponds for $[\sigma](u)$ to a step function, with a plateau $\Sigma_1$ after a cut-off $u_{\cc}$:
\begin{equation}
[\sigma](u)
=
\cases{
0 & $u<u_{\cc}    $
\\
\Sigma_1 & $u>u_{\cc}    $
}
\label{eq:1RSB_iGVM_txt}
\end{equation}
One indeed observes on figure~\ref{fig:GVM_toy_iterations_i} (see also the inset of~\fref{fig:GVM_toy_iterations_iGVM_scaling}~(left) that, in a region $[0,u_{\cc}]$, the successive iterations $\sigma_k(u)$ converge to  zero, while for $u>u_{\cc}$, the $\sigma_k(u)$'s converge to a constant plateau.
This means that the iterations $\sigma_k(u)$ converge to the a step function, without developing any stable power-law intermediate regime, in opposition to the results presented in figures~\ref{fig:GVM_ham_iterations} and~\ref{fig:GVM_toy_iterations} for the other versions of the GVM.

We have thus studied analytically the behaviour of a 1-step RSB form~\eref{eq:1RSB_iGVM_txt} of the solution to the GVM variational equations~(\ref{eq:eqvar_iGVM_txt}-\ref{eq:sum_omega_inv_iGVM_txt}).
The determination of the optimal values (according to the variational principle) of the parameters $\Sigma_1$ and $u_{\cc}$ is rather complex.
It is detailed in sections~\ref{sec:study-interm-gvm} and~\ref{sec:scaling-1-step_iGVM} of appendix~\ref{sec:details-gvm-comp}, in a self-consistent large~$\tf$ asymptotics.
One finds that, as $\tf\to\infty$, and as long as $\xi$ can be neglected, one has:
\begin{equation}
  u_{\cc}
  \sim
  \hat\beta^{-1}
  \sim
  \Big({\frac{T^5}{cD^2}}\Big)^{\frac 13}
  \:\tf^{-\frac 13}
\qquad\quad
\Sigma_1
\sim
  \hat\beta^{2}
\sim
  \Big({\frac{T^5}{cD^2}}\Big)^{-\frac 23}
  \:\tf^{\frac 23}
\label{eq:uc_Sigma1_largetf_iGVM_txt} 
\end{equation}
These two asymptotic behaviour also include a purely ($\xi,c,D,T$-independent) numerical prefactor that we omit here for clarity.
We conjecture that, at finite $\xi$, the optimal solution to the variational equations consists also in two plateaus, but separated by a full-RSB branch belonging to the non-constant $[\sigma](u)$ plotted in figure~\ref{fig:sigmaCimplicit}.
We reserve the study of such a solution to a future study~\cite{GVM-in-preparation}~; it would make it possible to track the role of $\xi$ in the scalings.
Since $\xi$ is neglected before obtaining the solution \eref{eq:uc_Sigma1_largetf_iGVM_txt},
it is not surprising that the combination of the parameters $\arga{c,D,T}$ that rescales the length $\tf$ of the interface is precisely the high-temperature Larkin length ${L_{\cc}(T,0)}=\frac{T^5}{cD^2}$ defined in~\eref{eq-fudging-Larkin-length}.

%%
%% from ~/recherche/GVM-finite-time_interface_iteration_intermediate-GVM_systematic-study.nb
%%

% \subsection{Roughness at intermediate scales}
We now determine the scaling of the roughness implied by the asymptotic behaviour~\eref{eq:uc_Sigma1_largetf_iGVM_txt}.
One has, defining a roughness function $B(t_1;\tf)$ at an intermediate position $t_1$:
\begin{eqnarray}
  B(t_1;\tf) 
  &
  \:\,\equiv\:\,
    \overline{\big\langle \big[y(t_1)-y(0)\big]^2\big\rangle} & \text{with~} 0<t_1<\tf
\label{eq:defB1Btf}
\\
  &
  \stackrel{\eref{eq:explicit-rescaling-F-repH}}= 
    \bigg(\frac{D}{cT}\bigg)^{\frac 23}
    \tf^{2\zeta}\
    \overline{\big\langle \big[\tilde y(\hat t_1)\big]^2\big\rangle} \qquad & \text{with~} \hat t_1 = \frac{t_1}{\tf}
\label{eq:relBt1tf_iGVM}
\end{eqnarray}
In the GVM approximation, one uses the following estimate:
\begin{eqnarray}
  \overline{\big\langle \big[\tilde y(\hat t_1)\big]^2\big\rangle}
  &
  \simeq 
  \big\langle \big[\tilde y_1(\hat t_1)\big]^2\big\rangle_0
\end{eqnarray}
Coming back to the Fourier representation, one has
\begin{eqnarray}
\big\langle \big(\tilde y_1(\hat t_1)\big)^2\big\rangle_0  
&
\stackrel{\phantom{\eref{eq:Gtitildesigma1_iGVM}}}
=
  \hat\beta^{-1}
  \sum_\omega
  \argp{1-\cos \argp{\omega\,\hat t_1}} \:\tilde G(\omega)
\\
&
\stackrel{\eref{eq:Gtitildesigma1_iGVM}}
=
  \hat\beta^{-1}
  \sum_\omega
  \frac{  \argp{1-\cos \argp{\omega\,\hat t_1}}}{\omega^2}
  \:\frac{\omega^2+\tilde\Sigma_1/u_{\cc}}{\omega^2+\tilde\Sigma_1}
\end{eqnarray}
Using the Poisson summation formula for summing over the discrete Fourier modes, one finds
\begin{eqnarray}
\fl\quad
\big\langle \big(\tilde y_1(\hat t_1)\big)^2\big\rangle_0  
%\stackrel{?}
=
&\
(u_{\cc} \hat\beta)^{-1} \frac{\hat t_1}{2}
+
\hat\beta^{-1}
\left(1-u_{\cc}\right)
 \frac{   
  \sinh \Big(\frac{1}{2} \left(1-\hat t_1\right) {\Sigma}_1^{\frac 12}\Big) 
  \sinh \Big(\frac{1}{2} \hat t_1 {{\Sigma}_1^{\frac 12}}\Big)}
  {
  \sqrt{{\Sigma }_1} u_{\cc}\;
  \text{sinh}\Big(\frac 12 {\Sigma}_1^{\frac 12}\Big)
  }
\label{eq:iBt1tf_iGVM_Poisson}
\end{eqnarray}
With the large-$\tf$ asymptotics~\eref{eq:uc_Sigma1_largetf_iGVM_txt}, one obtains
\begin{eqnarray}
\big\langle \big(\tilde y_1(\hat t_1)\big)^2\big\rangle_0  
%\stackrel{?}
=
&\
C_0 \frac{\hat t_1}{2}
+
C_1   \Big({\frac{T^5}{cD^2}}\Big)^{\frac 13} \tf^{-\frac 13}
\end{eqnarray}
where $C_0$ and $C_1$ are numerical constants, independent of $\tf$.
At dominant order, we thus have $\big\langle \big(\tilde y_1(\hat t_1)\big)^2\big\rangle_0\sim \tf^0$,
and coming back to the original roughness function through~\eref{eq:relBt1tf_iGVM}, one finally obtains $\zeta=\frac 23$.
In other words, the Flory rescaling of the replicated Hamiltonian \eref{eq:explicit-rescaling-F-repH} has put as a prefactor of the roughness~\eref{eq:exprBtfKPZreplicae} the high-temperature asymptotic roughness~$(\frac{D}{cT})^{2/3}    \tf^{4/3}$, and the GVM procedure has shown that at leading order in the limit ${\tf \to \infty}$ the correction to this scaling is only a numerical factor, as expected from \eref{eq-GVM-expectations}.

% \subsection{Numerical test of scaling in $\tf$ of the 1-step RSB solution}
% \label{sec:numerical-scaling-1_iGVM}

\smallskip
The numerical iterative procedure for the GVM equations, described in \sref{sec:iternum_iGVM}, allows one to test the previously obtained scalings~\eref{eq:uc_Sigma1_largetf_iGVM_txt} of the cut-off $u_{\cc}\sim \tf^{-1/3}$ and of the amplitude $\sigma_1=\Sigma_1/u_{\cc}\sim \tf$ of $\sigma(u)$ for $u>u_{\cc}$, in the regime where a 1-step RSB form for $\sigma(u)$ is valid (\textit{i.e.}~at sufficiently small $\xi$ for our approximation to be self-consistent).
As shown on figure~\ref{fig:GVM_toy_iterations_iGVM_scaling}, the predicted scalings~\eref{eq:uc_Sigma1_largetf_iGVM_txt} are in agreement with the numerical observations.

\smallskip
Last, one remarks that the 1-step RSB Ansatz~\eref{eq:1RSB_iGVM_txt} also describes the small $\tf$ regime: one checks from~\eref{eq:eqvar_iGVM_1step} and~\eref{eq:varFreplicated_iGVM} that as $\tf\to 0$, one has $u_{\cc}\to 1$ and $\Sigma_1\to 0$.
One recovers a replica symmetric solution, with ${\sigma (u) \to 0 \, \forall u \in [0,1]}$, so that the trial replicated Hamiltonian has only an elastic contribution.
From the expression~\eref{eq:iBt1tf_iGVM_Poisson} of the rescaled roughness, one finds that
$\langle \big(\tilde y_1(\hat t_1)\big)^2\rangle_0 \sim \hat\beta^{-1} \hat t_1$  and inserting this results in~\eref{eq:relBt1tf_iGVM} one recovers the expected thermal roughness $B(\tf)\sim T \tf/c$ as $\tf\to 0$.
In particular, the typical lengthscale separating the thermal and the KPZ regimes of the roughness, defined as the intersection point between those two regimes, is the $(\xi=0)$-Larkin length~\eref{eq-fudging-Larkin-length} $L_{\cc}(T,0)=\frac{T^5}{cD^2}$.
To study the lower temperature regime of the Larkin length, one would have to solve the finite-$\tf$ Hamiltonian GVM equations at $\xi>0$~\cite{GVM-in-preparation}.

%-------------------------
\subsection{Discussion}
\label{sec:GVM-discussion}

We now explain the reason why taking an interface of finite length~$\tf$, compared to the infinite-$\tf$ case, modifies the structure of the Hamiltonian GVM  in a  way which is significant enough to induce a change of the asymptotic roughness exponent $\zeta$ from $\zeta_{\FF}=3/5$ to $\zeta_{\KPZ}=2/3$.
Let us come back to the roughness $B(t_1,\tf)$ at position $t_1$, defined in~\eref{eq:defB1Btf}, and to its expression in terms of the diagonal coefficient $\tilde G(\omega)$ of the hierarchical matrix \cite{mezard_parisi_1991_replica_JournPhysI1_809}:
\begin{eqnarray}
B(t_1,\tf)
&
\;\,\stackrel{\eref{eq:explicit-rescaling-F-repH}}
=
\;
  \bigg(\frac{D}{cT}\bigg)^{\frac 23}
  \tf^{2\zeta}\
  \hat\beta^{-1}
  \sum_\omega
  (1-\cos \omega\,\hat t_1) \:\tilde G(\omega)
\label{eq:Bt1tf_explicit}
\\
  \tilde G(\omega)
& 
\stackrel{\eref{eq:tilde-replica-03}}
=
 \frac{1}{ \omega^2} \left( 1 + \int_0^1 \frac{dv}{v^2} \frac{[  \sigma](v)}{ \omega^2 + [  \sigma](v)} 
%+ \frac{  \sigma (0)}{ \omega^2} 
\right) 
\label{eq:Gtildeomega_rappel}
\end{eqnarray}
Because of the denominator $\omega^2+[  \sigma](v)$, one finds that the large-scale regime ($t_1\to\infty$, $\tf\to\infty$ with finite $\hat t_1=t_1/\tf$ ) is governed by the behaviour of $[\sigma](u)$ at small values of $[\sigma](u)$, and hence at small $u$.
In fact, in Ref.~\cite{agoritsas_2010_PhysRevB_82_184207} (section~IV.E) it was shown by analysing~(\ref{eq:Bt1tf_explicit}-\ref{eq:Gtildeomega_rappel}) that the large-$\tf$ behaviour of the roughness in $\sim\tf^{2\zeta_{\rm{asympt}}}$ is governed as follows whenever $[\sigma](u)$ behaves as a power-law for small $u$:
\begin{equation}
 [\sigma](u)
\ \mathop{\sim}_{u\to 0}\ 
 u^\nu
\quad
\Longrightarrow
\quad
\zeta_{\rm{asympt}} = \frac 12 + \frac 1\nu
\label{eq:relation_nu_zeta-asympt}
\end{equation}
For the infinite-$\tf$ Hamiltonian GVM, the function $[\sigma](u)$ behaves as a power-law as $u\to 0$.
This is found by solving explicitly the variational equations
(\ref{eq:varGVMham}-\ref{eq:ronron-replica-04_ham}),
but this can also be obtained heuristically in a simple way.
Inserting
$ [\sigma](u)
\sim
 u^\nu
$
into~\eref{eq:ronron-replica-04_ham}, one gets
  \begin{eqnarray}
    \int_{\mathbb{R}}\! \dbar q\:
    \big[\tilde G(q) - \ G(q,u)  \big]
    &
      \ \mathop{\sim}_{u\to 0}\ 
      u^{-\frac{2+\nu}{2}}
      \label{eq:ronron-replica-04_ham_devsmallu}
  \end{eqnarray}
and thus from~\eref{eq:varGVMham}
%  \begin{equation}
$
    u^{\nu-1} \sim u^{\frac{3}{2} \frac{2+\nu}{2}}
$
%  \end{equation}
  which imposes $\nu-1 = {\frac{3}{2} \frac{2+\nu}{2}}$ and one finds $\nu=10$.
Hence, using \eref{eq:relation_nu_zeta-asympt}, one finally finds that the infinite-$\tf$ Hamiltonian GVM is bound to have a roughness scaling with the Flory exponent $\frac 12+\frac 1{10}=\frac 35$ at large scales.

For the finite-$\tf$ Hamiltonian GVM, on the other hand, as discussed in \sref{sec:GVM-upgraded-finite-length_ii} and shown on~\fref{fig:sigmaCimplicit}, the function $[\sigma](u)$ has to start by a plateau at small $u$. This implies that the previous reasoning cannot be applied.
Since the GVM variational equations~(\ref{eq:eqvar_iGVM_txt}-\ref{eq:sum_omega_inv_iGVM_txt}) intertwine all values of $u$, one obtains in the end a different GVM structure.
In the $\xi\to 0$ regime discussed in~\sref{sec:GVM-upgraded-finite-length_iii}, one gets a \mbox{$1$-step} RSB solution instead of a full-RSB one.
In other words, formally, the limits $u\to 0 $ and $\tf\to\infty$ do not commute: the $u\to 0 $ regime (which governs the large physical scales) is different in the finite- and in the infinite-$\tf$ Hamiltonian GVMs approximations.
The physical interpretation of this statement is as follows: if we discard from the beginning the existence of a finite interface length~$\tf$, we forbid the GVM to take into account the dependence in~$\tf$ in its self-consistent variational equation, therefore missing the correct roughness exponent at the end of the day.
%and this makes that one misses the correct roughness exponent.
%
In particular, one may wonder if the behaviour $\sigma[u]\sim u^{10}$ for $u$ far enough from $0$ illustrated on figure~\ref{fig:sigmaCimplicit}, in the regime where $\sigma[u]$ is non-constant, might prevent the GVM from yielding the correct roughness exponent $2/3$. However, since $\sigma[u]$ has to start by a plateau for small $u$, the previous reasoning based on~\eref{eq:ronron-replica-04_ham_devsmallu} and~\eref{eq:varGVMham} cannot be applied and the roughness exponent is thus not constrained to be equal to the $3/5$ Flory one in our approach.

%\begin{itemize}
%
% \item Toymodel
%
% The behaviour of $\sigma(u)$ is more complex that in the Hamiltonian GVM.
% If one assumes that
% \begin{equation}
% \sigma(u)\sim
% \cases{
%      \sigma(u_*) & if $0<u<u_*$ \\
%      u^\alpha+\text{constant} & if $u_*<u<u_{\cc}$ \\
%      \sigma(u_{\cc}) & if  $u_{\cc}<u<1$
% }
% \end{equation}
% then, in the regime $u_*<u<u_{\cc}$, one has $[\sigma](u)\sim u^{\alpha+1}$.
% %
% Thus, as read from~\eref{eq:toy-replica-04} (see figure~\ref{fig:GVM_toy_GGu_iterations}),
% \begin{equation}
% \tilde G-G(u)\sim
% \cases{
%      \tilde G-G(u_*) & if               $0<u<u_*$ \\
%      u^{-(2+\alpha)}+\text{constant} & if $u_*<u<u_{\cc}$ \\
%      \tilde G-G(u_{\cc}) & if            $u_{\cc}<u<1$
% }
% \end{equation}
% Finally, \eref{eq:varGVMtoy} implies $u^\alpha\sim u^{\frac12(2+\alpha)}$, whence $\alpha=2$.

% \end{itemize}

As a final remark, we emphasise  that the free-energy GVM computation was successful in catching ${\zeta=\frac23}$ because we had put by hand from the beginning, as a shortcut, the Brownian scaling of the disorder free-energy.
The infinite-$\tf$ Hamiltonian GVM cannot catch this value of the exponent, because it predicts that the structure factor at small Fourier modes displays a Flory scaling, as we have just discussed. %(at ${u \to 0}$).
For the free-energy GVM at fixed $\tf\to\infty$, we need some additional input regarding the scaling of the free-energy,
namely, that it has a Brownian distribution~\cite{agoritsas_2010_PhysRevB_82_184207,phdthesis_Agoritsas2013}.
If one accounts for the finite-$\tf$ correction to its pure Brownian scaling, the GVM approximation can be shown to be self-consistent~\cite{agoritsas-2012-FHHpenta}.
Keeping track of the \emph{finite} $\tf$ within a Hamiltonian GVM approach thus appears to be the correct way in order to avoid the Flory pitfall in this computation scheme.

%_____________________________________________________________________________________________________
%_____________________________________________________________________________________________________
%\newpage
\section{Concluding remarks}
\label{sec:conclusion}

In this work,
we first revisited %the power countings and
the standard Flory arguments
regarding the geometrical fluctuations of the static 1D interface with a short-range elasticity and a random-bond disorder, at finite temperature and with a finite disorder correlation length.
This specific problem can be exactly mapped on the free-energy fluctuations of a growing 1+1 DP, which evolve according to a KPZ equation, and as such is relevant for the whole 1D KPZ universality class.
Comparing different possible power countings, performed either on the 1D interface Hamiltonian or on the 1+1 DP free energy (without or with replicas),
we identified the physically meaningful power countings through a saddle-point analysis of path integrals:
we found that the validity of exponents found by power counting arises as a consequence of the existence of optimal trajectories with finite variance, either at zero temperature or at asymptotically large timescales.
Moreover, we related the failure of the Flory power counting on the Hamiltonian
on the one hand to the absence of such a saddle point, %and optimal trajectories,
and on the other hand to having wrongly neglected the scaling of the disorder correlation length $\xi$. 
%or the finite size $\tf$ of an interface segment. [this is for the GVM - next paragraph]

Secondly, using our new insights on Flory arguments, power countings, and optimal trajectories, we devised a GVM approximation scheme for the Hamiltonian description of the interface, taking into account the finite length $\tf$ of the interface.
In the large-$\tf$ regime, it allowed us to compute the interface roughness with its correct KPZ asymptotic scaling (${B(t) \sim t^{4/3}}$), avoiding the usual Flory pitfall (${B(t) \sim t^{6/5}}$).
We were thus able to address one of the remaining open issues of Ref.~\cite{agoritsas_2010_PhysRevB_82_184207}, rehabilitating the GVM predictions of scaling exponents.
We identified the precise features of the GVM solution which allow for a non-Flory roughness exponent to emerge.
Another advantage, compared to the free-energy GVM procedure, is that we do not rely here on a STS decomposition, which is rather specific of the 1+1 DP, opening perspectives to apply the proposed procedure to other systems.
Our solution, however, is for the moment restricted to the $\xi\to 0$ regime.
The understanding of the GVM dependence in $\xi$ is an open perspective: we conjecture that instead of a 1-step RSB solution, the GVM variational equation presents a full-RSB solution surrounded by two plateaus~\cite{GVM-in-preparation}. 
Such an approach should allow to capture the full temperature dependence of the interface characteristic crossover length- and energy-scales, taking into account the finite disorder correlation length ${\xi}$.
Once this issue is settled, a natural application of our finite-length GVM framework would be to determine the polymer endpoint momenta and distribution, following the ideas presented in~\cite{goldschmidt-blum_1993_PhysRevE48_161,bouchaud_mezard_parisi_1995_PhysRevE52_3656}.
Of particular interest are moment ratios such as the kurtosis or the skewness, which could be computed in principle from this distribution: they are independent of the amplitude of the fluctuations and their value could be compared to known numerical, experimental and analytical results~\cite{halpin-healy_directed_1991,takeuchi_evidence_2012,tracy_level-spacing_1994,tracy_orthogonal_1996}.

More generally, studying disordered systems, we showed that it can be very useful to reformulate standard scaling arguments directly on the underlying path integrals of observable averages, in order to validate (or to invalidate) the corresponding scaling predictions regarding a given observable, averaged over disorder and thermal fluctuations.
In that respect, the Lax-Oleinik principle played a key role in setting a rigorous framework for the existence of optimal trajectories.
Such a procedure could of course be generalised to disordered elastic systems beyond the static 1D case that we have considered.
For instance, it can be used for identifying the validity range of the so-called quasistatic `creep' regime, reformulating the corresponding standard scaling arguments
\cite{ioffe_vinokur_1987_JPhysC20_6149,nattermann_1987_EPL4_1241}
into the analysis of a  saddle point small-force asymptotic behaviour of the system velocity (represented as path integral), %at low temperature and for a large system size,
as we have recently done with other co-authors in Ref.~\cite{agoritsas_garcia-garcia_VL_2016_Arxiv-1605.04405}.
%
%% Connection with the creep regime in the zero-force limit:
% \cite{agoritsas_garcia-garcia_VL_2016_Arxiv-1605.04405}
% Elisabeth Agoritsas, Reinaldo García-García, Vivien Lecomte, Lev Truskinovsky, and Damien Vandembroucq
% "Driven interfaces: from flow to creep through model reduction"
%
%% Standard scalings arguments for the creep
% \cite{ioffe_vinokur_1987_JPhysC20_6149}
% L. B. Ioffe AND V. M. Vinokur
% "Dynamics of interfaces and dislocations in disordered media"
% 
% \cite{nattermann_1987_EPL4_1241}
% T. Nattermann
% "Interface Roughening in Systems with Quenched Random Impurities"
%

Besides, the approach we have presented can be extended to interfaces with more general boundary conditions: instead of pinning the two extremities of the interface to~0, one can free its endpoint and include the study of its fluctuations within the GVM approximation scheme (this requires to device a `dual GVM' description with two GVM Ansätze capturing the fluctuations of both the bulk and the extremities of the interface~\cite{GVM-in-preparation}).
Other perspectives include for instance the study of the DP in dimensions higher than 1+1~\cite{halpin-healy_2012_PhysRevLett109_170602,halpin-healy_2013_PhysRevE88_042118}.
The connection to the `self-consistent expansion' approach~\cite{katzav_roughness_2007,schwartz_ideas_2008} is also worth investigating, as it corresponds at minimal order to an Hartree-Fock approximation and as it as been applied to the study of the dynamics of the KPZ equation~\cite{katzav_structure_2006}.

In conclusion, although scalings arguments can be very powerful shortcuts to potentially long and cumbersome computations, they rely on implicit assumptions which must be independently validated. In that respect, identifying saddle points of path integrals describing observable averages is a possible strategy worth keeping in mind.

%_____________________________________________________________________________________________________
%_____________________________________________________________________________________________________

\section*{Acknowledgements}

We thank Thierry Giamarchi for numerous and fruitful discussions at the early stages of this work.
E.~A.~acknowledges financial support from ERC grant ADG20110209 and by a Fellowship for Prospective Researchers Grant No P2GEP2-15586 from the Swiss National Science Foundation.
E.~A.~and V.~L.~ackowledge support by the National Science Foundation under Grant No. NSF PHY11-25915 during a stay at KITP, UCSB where part of this research was performed.

%_____________________________________________________________________________________________________
%_____________________________________________________________________________________________________

\appendix

%_____________________________________________________________________________________________________
\renewcommand{\thesection}{A}
\section{The finite-$\tf$ Hamiltonian GVM approach}
\label{sec:details-gvm-comp}

In this section, we provide the details of the finite-$\tf$ Hamiltonian GVM computations presented in \sref{sec:GVM-upgraded-finite-length} and~\sref{sec:GVM-upgraded-finite-length_ii}.
The starting point is the rescaled and replicated Hamiltonian \eref{eq:replicatedHamKPZresc}:
\begin{equation}
\fl \qquad
\HHt\big[\tilde {\mathbf y}(\hat t),\mathring\xi(\tf)\big]
%\stackrel{\eref{eq:replicatedHamKPZresc}}{=}
= \int_0^1 \!\! d\hat t\:
         \bigg[
           \sum_a \tfrac 12 (\partial_{\hat t}\tilde y_a)^2
          -  
           \sum_{a<b}
           \hat R_{\mathring\xi(\tf)}\Big(\tilde y_b(\hat t)-\tilde y_a(\hat t)\,\Big)
         \bigg]
\label{eq:replicatedHamKPZresc-bis}
\end{equation}
whose different contributions are defined by \eref{eq-Hreplic-part1}-\eref{eq-Hreplic-part2}-\eref{eq-Hreplic-part3}.
We consider the trial Hamiltonian \eref{eq:Ht0_GVMH} but with \emph{discrete} Fourier modes (${\omega \in 2 \pi \mathbb{Z}}$) instead of continuous Fourier modes (${q \in \mathbb{R}}$):
\begin{equation}
\fl \qquad
\HHt_0\big[{\mathbf y} , \tf \big]
=
\frac 12 \sum_{\omega} \sum_{a,b=1}^{n}  y_a(-\omega) {G}^{-1}_{ab}(\omega) y_b(\omega)
\qquad ({\rm{with }}\ \dbar q\equiv\tfrac{dq}{2\pi})
\label{eq:Ht0_GVMH-bis}
\end{equation}
with ${G^{-1}_{ab}(\omega)}$ a hierarchical matrix.
We emphasise that the successive steps of the derivation are similar to those presented in \cite{agoritsas_2010_PhysRevB_82_184207} until \eref{eq:eqvar_iGVM}.
As we detail, having  \emph{discrete} modes instead of continuous ones modifies the GVM solution and affects drastically the scaling properties at large lengthscales.
%instead of continuous Fourier modes (${q \in \mathbb{R}}$) we deal now with sum over \emph{discrete} Fourier modes (${\omega \in 2 \pi \mathbb{Z}}$) because of the finite interface length $\tf$.

% \subsection{Settings}
% \label{sec:settings-iGVM}

%The modes $\tilde y_a(\omega)$ are complex; the only constraint arising from the $\tilde y_a(\hat t)$ being real is that $\tilde y_a(\omega)^*=\tilde y_a(-\omega)$.

%The roughness exponent can be reconstituted from the intermediate-time roughness $B(t_1)$ (with $0<t_1<\tf$) as in~\eref{eq:rougness-at-intermediate-time}.

%-------------------------
\subsection{Variational equations}
\label{sec:vari-equat_iGVM}

The Bogoliubov variational principle takes the form ${\partial \mathcal F_\text{var}/\partial  G_{ab}(\omega)=0}$ ($\forall a,b,\omega$) for the variational free energy
\begin{equation}
  \mathcal F_\text{var}
=
  \mathcal F_0 + \langle \HHt - \HHt_0 \rangle_0
\label{eq:Fvar_iGVM}
\end{equation}
with
\begin{equation}
\mathcal F_0 
=
-\hat\beta^{-1} \log Z_0  
=
-\frac 12 \sum_\omega \log \det \tilde G(\omega) + \text{const.}
\label{eq:expression_Fvar0_log-det}
\end{equation}
and
\begin{eqnarray}
\fl\qquad
\langle \HHt_\text{el} - \HHt_0 \rangle_0
&=
\frac 12
{\sum\limits_\omega} 
\Big\langle 
\sum_{a=1}^{n} 
 \omega^2 \tilde y_a(-\omega) \tilde y_a(\omega) 
+
 \sum_{a,b=1}^{n} 
  \tilde y_a(\omega) {G}^{-1}_{ab}(\omega)\tilde y_b(\omega)
\Big\rangle_0
\\
&=
\frac 12
\sum_\omega
\sum_{a,b=1}^{n}
\big[
 \omega^2 \delta_{ab}
+
{G}^{-1}_{ab}(\omega)
\big]
\langle
\tilde y_a(-\omega) \tilde y_b(\omega) \rangle_0
\\
&=
\frac{\hat\beta^{-1}}2
\sum_\omega
\sum_{a,b=1}^{n}
\big[
 \omega^2 \delta_{ab}
+
{G}^{-1}_{ab}(\omega)
\big]
 G_{ab}(\omega)
\\
&=
\frac{\hat\beta^{-1}}2
\sum_\omega
\sum_{a=1}^{n}
 \omega^2 G_{aa}(\omega)
\
+
\
\text{const.}
\label{eq:HelH0_0}
\end{eqnarray}
For a Gaussian quenched random potential, with a generic two-point correlator of Fourier transform $R_{\mathring\xi}(\lambda)$, the averaged disorder Hamiltonian is:
%The average disorder Hamiltonian is, for a generic Gaussian correlator of Fourier transform $R_{\mathring\xi}(\lambda)$:
\begin{eqnarray}
\fl\quad
\big\langle
  \HHt_\text{dis}\big[\tilde {\mathbf y}(\omega),\mathring\xi(\tf)\big]
\big\rangle_0
&=
-
\int_{\mathbb{R}} \dbar \lambda
R_{\mathring \xi}(\lambda) 
  \sum_{a<b}\:
  \Big\langle
  e^{i\lambda\left[\tilde y_b(\hat t)-\tilde y_a(\hat t)\right]}
  \Big\rangle_0
\label{eq:Hdis-for-Fvar_iGVM}
\\
&=
-
\int_{\mathbb{R}} \dbar \lambda
R_{\mathring \xi}(\lambda) 
\int_0^1 \!\! d\hat t\:
  \sum_{a<b}\:
  e^{-\frac {\lambda^2}2\left\langle\left[\tilde y_b(\hat t)-\tilde y_a(\hat t)\right]^2\right\rangle_0}
\end{eqnarray}
Then, using that in the GVM
\begin{eqnarray}
  \big\langle [\tilde y_b(\hat t)- \tilde y_a(\hat t)]^2\big\rangle_0
&=
  \hat\beta^{-1}
\sum_\omega
  \big[ G_{aa}(\omega)+ G_{bb}(\omega)-2 G_{ab}(\omega)\big]
\end{eqnarray}
one obtains finally
\begin{eqnarray}
\fl\qquad
\big\langle
  \HHt_\text{dis}\big[\tilde {\mathbf y}(\omega),\mathring\xi(\tf)\big]
\big\rangle_0
&=
-
\int_{\mathbb{R}} \dbar \lambda
R_{\mathring \xi}(\lambda) 
  \sum_{a<b}\:
  e^{-\frac{\lambda^2}{2\hat\beta}
  \sum\limits_\omega
  \left[ G_{aa}(\omega)+ G_{bb}(\omega)-2 G_{ab}(\omega)\right]
}
\label{eq:Hdis_ave0_iGVM}
\end{eqnarray}
The extremalisation of $\mathcal F_{\var}$ with respect to~$ G_{ab}(\omega)$ ($a\neq b$) yields that $ G^{-1}_{a\neq b}(\omega)=-\sigma_{ab}$ is independent of $\omega$:
\begin{eqnarray}
\fl\quad
& 0
=
-\frac{1}{2\hat \beta}  G^{-1}_{ab}(\omega)
-
\int_{\mathbb{R}} \dbar \lambda
R_{\mathring \xi}(\lambda) 
\frac{\lambda^2}{\hat \beta}
  e^{-\frac{\lambda^2}{2\hat\beta}
  \sum\limits_{\omega'}
  \left[ G_{aa}(\omega')+ G_{bb}(\omega')-2 G_{ab}(\omega')\right]
}
\\
\fl\quad
\Longleftrightarrow 
\quad
&\sigma_{ab}
=
2
\int_{\mathbb{R}} \dbar \lambda
R_{\mathring \xi}(\lambda) 
\lambda^2
  e^{-\frac{\lambda^2}{2\hat\beta}
  \sum\limits_{\omega}
  \left[ G_{aa}(\omega)+ G_{bb}(\omega)-2 G_{ab}(\omega)\right]
}
\end{eqnarray}
Introducing
$
   G_{aa}(\omega) = \tilde G (\omega)
$
the variational equation $\partial \mathcal F_\text{var}/\partial  G_{ab}(\omega)=0$ writes
\begin{eqnarray}
\sigma_{ab}
&=
2
\int_{\mathbb{R}} \dbar \lambda
R_{\mathring \xi}(\lambda) 
\lambda^2
  e^{-\frac{\lambda^2}{\hat\beta}
  \sum\limits_\omega
  \big[\tilde G(\omega) - G_{a\neq b}(\omega)\big]
}
\label{eq:coupled-replicae-tilde}
\end{eqnarray}
The variation with respect to~$ G_{aa}(\omega)$ (in which $\langle \HHt_\text{el} - \HHt_0 \rangle_0$, through~\eref{eq:HelH0_0}, gives a non-zero contribution) yields
\begin{eqnarray}
    0
&=
\underbrace{
-\frac{1}{2\hat \beta}  G^{-1}_{aa}
+\frac{1}{2\hat \beta} \omega^2 \delta_{aa}
}_{
=\frac{1}{2\hat \beta}\sigma_{aa}
}
+
\frac{1}{2\hat \beta}\sum_{a'(\neq a)}  \sigma_{aa'}
\end{eqnarray}
where for the last term we recognised $\sigma_{aa'}$ from~\eref{eq:coupled-replicae-tilde}. One thus obtains that
\begin{equation}
  \sigma_{aa} = - \sum_{a' (\neq a)}  \sigma_{aa'}
%\equiv \tilde\sigma
\end{equation}
This implies that the sum of the coefficients on each line or column of the hierarchical matrix, namely its `connected part', stems solely from the elastic part of the Hamiltonian:
\begin{equation}
\fl \qquad
 G_{\cc}^{-1}(\omega)
 \equiv \sum_{a=1}^{n} G_{ab}^{-1}(\omega) = \omega^2
 \quad \stackrel{\argp{G_{\cc}^{-1}  G_{\cc} =1}}{\Longleftrightarrow} \quad
  G_{\cc}(\omega)
 \equiv \sum_{a=1}^{n} G_{ab}(\omega) = 1/\omega^2
 \label{eq-connected-parts-STS}
\end{equation}

%-------------------------
\subsection{Continuous parametrization of the variational equation}
\label{sec:contparamvareq_app}

Then, introducing a continuous parametrization of the space of replicas, we map ${a \in \arga{1, \dots n}}$ to ${u \in [0,1]}$ and \eref{eq:coupled-replicae-tilde} becomes:
\begin{eqnarray}
\sigma (u)
&=
2
\int_{\mathbb{R}} \dbar \lambda
R_{\mathring \xi}(\lambda) 
\lambda^2
  e^{-\frac{\lambda^2}{\hat\beta}
  \sum\limits_\omega
  \big[\tilde G(\omega) - G(\omega,u)\big]
}
\label{eq:coupled-replicae-tilde_u}
\end{eqnarray}
We now use that $G_{ab}$ verifies the replica algebra of hierarchical matrices (see appendix~II of Ref.~\cite{mezard_parisi_1991_replica_JournPhysI1_809}, 
or, for notations similar to the ones used here, appendix~B of Ref.~\cite{agoritsas_2010_PhysRevB_82_184207}):
\begin{eqnarray}
  \partial_u\big[\tilde G(\omega)-  G(\omega,u)\big]
 =-\frac{  \sigma'(u)}{\big(  G^{-1}_{\cc}\!(\omega)+[  \sigma](u)\big)^2} \label{eq:tilde-replica-01} 
 \\
  \left[   \sigma \right] (u)
 = u   \sigma (u) - \int_0^u dv \,   \sigma (v) \label{eq:tilde-replica-02} 
  \\
  \tilde G(\omega) -   G(\omega,u)
 = \frac{1}{u}  \frac{1}{  G^{-1}_{\cc}\!(\omega) + [  \sigma](u)} - \int^1_u \frac{dv}{v^2}  \frac{1}{  G^{-1}_{\cc}\!(\omega) +[  \sigma](v)} \label{eq:tilde-replica-04}
  \\
  \tilde G(\omega)
 = \frac{1}{  G^{-1}_{\cc}\!(\omega)} \left( 1 + \int_0^1 \frac{dv}{v^2} \frac{[  \sigma](v)}{  G^{-1}_{\cc}\!(\omega) + [  \sigma](v)} + \frac{  \sigma (0)}{  G^{-1}_{\cc}\!(\omega)} \right) \label{eq:tilde-replica-03}
\end{eqnarray}
where the connected term writes ${  G^{-1}_{\cc}\!(\omega)=\omega^2}$, as mentioned in \eref{eq-connected-parts-STS}.

Besides, in order to go further in explicit computations, we choose specifically a Gaussian function for the disorder two-point correlator \eref{eq-disorder-correlator-scaling}:
\begin{equation}
  R_{\mathring\xi}(\lambda) = e^{-\mathring\xi^2\lambda^2}
  \label{eq:defRxiGaussian}
\end{equation}
The integral over the transverse continuous Fourier modes $\lambda$ in the variational equation (\ref{eq:coupled-replicae-tilde_u}) can be computed.
This yields the form of the variational equation announced in \sref{sec:GVM-upgraded-finite-length} of the main text:
\begin{eqnarray}
 \sigma (u)
=
\frac{2}{\sqrt\pi}
\hat\beta^{\frac 32}
\bigg\{
\hat\beta
\mathring\xi^2+
  \sum\limits_\omega
  \big[\tilde G(\omega) - G(\omega,u)\big]
\bigg\}^{-\frac 32}
\label{eq:eqvar_iGVM}
\end{eqnarray}
where, by summing explicitly over $\omega\in 2\pi\mathbb Z$ the hierarchical inversion relations~\eref{eq:tilde-replica-04}, one has
  \begin{eqnarray}
\fl\qquad
    \sum_\omega  \big[\tilde G(\omega) - G(\omega,u)\big]
    &= \frac{1}{u} \:
      \frac{\coth \argp{\tfrac 12\sqrt{[ \sigma](u)}}}{2\sqrt{[ \sigma](u)}}
      - \int^1_u \frac{dv}{v^2} \:
      \frac{\coth \argp{\tfrac 12\sqrt{[ \sigma](v)}}}{2\sqrt{[ \sigma](v)}}
\label{eq:sum_omega_inv_iGVM}
\end{eqnarray}
This equation resembles that of the infinite-$\tf$ Hamiltonian GVM in \cite{agoritsas_2010_PhysRevB_82_184207}:
  \begin{eqnarray}
\fl\qquad
    \int_{\mathbb{R}} \dbar q \, \big[\tilde G(q) - G(q,u)\big]
    &= \frac{1}{u} \:
      \frac{1}{2\sqrt{[ \sigma](u)}}
      - \int^1_u \frac{dv}{v^2} \:
      \frac{1}{2\sqrt{[ \sigma](v)}}
\label{eq:sum_q_inv_iGVM-PRB2010}
\end{eqnarray}
since the only difference is that $\coth \argp{\tfrac 12\sqrt{[ \sigma](u)}}$ is replaced by $1$.
Technically, this difference is the precise origin of the distinct scalings presented by the two GVMs (due to the singular behaviour of $\coth x$ as
$x\to 0$). Physically, this is understood by the fact that the small-$[\sigma](u)$ regime, where $\coth \argp{\tfrac 12\sqrt{[ \sigma](u)}}$ is singular, governs the large-scale behaviour of the roughness.

%-------------------------
\subsection{Comparison to the Hamiltonian GVM: a self-consistent equation on $[\sigma](u)$}
\label{sec:compar_hGVM_iGVM}
%%%
%%% all results from this subsection from
%%% ~/recherche/GVM-finite-time_interface_sigmau_iGVM.nb
%%%

In the study of the Hamiltonian GVM of the infinite interface (denoted symbolically by GVM$_{\infty}$), solving the variational equation is done by differentiating the variational equation with respect to $u$ and identifying of a self-consistent equation on $[\sigma](u)$ (valid when $[\sigma](u)$ is non-constant). One then finds that it is solved by $[\sigma](u)\propto u^{10}$.
The complete form of $[\sigma](u)$, recalled in~\eref{eq:sigmau_GVMH}, is then a combination between this power-law behaviour and a plateau.
We follow here a similar procedure for the finite-$\tf$ GVM computation, in order to understand the difference between these two procedures.

Since the GVM variational equations are closely related to those of the  GVM$_{\infty}$ (see the comment after~\eref{eq:sum_omega_inv_iGVM}), we indicate using braces the factors which are equal to $1$ in the GVM$_{\infty}$ and different from $1$ in our finite-$\tf$ settings here.
By direct computation, one directly checks from~\eref{eq:sum_omega_inv_iGVM} that:
\begin{eqnarray}
\fl\qquad
\partial_u
    \sum_\omega  \big[\tilde G(\omega) - G(\omega,u)\big]
    &= 
 - \frac{\sigma'(u)}{8\{[\sigma](u)\}^{\frac 32}}
\overbrace{\frac{\sinh \argp{\sqrt{[ \sigma](u)}}+\sqrt{[ \sigma](u)}}{\sinh^2 \argp{\tfrac 12\sqrt{[ \sigma](u)}}}}^{=1~\text{for~GVM}_{\infty}}
\label{eq:sum_omega_inv_iGVM_deru}
\end{eqnarray}
Differentiating~\eref{eq:eqvar_iGVM} with respect to~$u$ yields
\vspace*{-3mm}
\begin{eqnarray}
\fl\qquad
\sigma'(u)
=
\frac{3}{8\sqrt{\pi}}
  \hat\beta^{\frac 32}
\bigg\{
\hat\beta
\mathring\xi^2+
  \sum\limits_\omega
  \big[\tilde G(\omega) - G(\omega,u)\big]
\bigg\}^{-\frac 52}
\nonumber
\\
\fl\qquad\qquad\qquad\qquad\qquad
\times
\frac{\sigma'(u)}{\{[\sigma](u)\}^{\frac 32}}
\underbrace{\frac{\sinh \argp{\sqrt{[ \sigma](u)}}+\sqrt{[ \sigma](u)}}{\sinh^2 \argp{\tfrac 12\sqrt{[ \sigma](u)}}}}_{=1~\text{for~GVM}_{\infty}}
\end{eqnarray}
Hence, if $\sigma'(u)\neq 0$
\begin{eqnarray}
1
&=
\frac{3\pi^{\frac 13}}{2^{14/3}}
\hat\beta^{-1} \sigma(u)^{\frac 53}\: \{[\sigma](u)\}^{-\frac 32}\;\: \frac{\sinh \argp{\sqrt{[ \sigma](u)}}+\sqrt{[ \sigma](u)}}{\sinh^2 \argp{\tfrac 12\sqrt{[ \sigma](u)}}}
\\
\sigma(u)
&=
\Big(\frac{2^{14}}{27\pi}\Big)^{\frac 15}
\:
\hat\beta^{\frac 35}\: \{[\sigma](u)\}^{\frac{9}{10}}
\;
\underbrace{
\bigg[\frac
  {\sinh^2 \argp{\tfrac 12\sqrt{[ \sigma](u)}}}
  {\sinh \argp{\sqrt{[ \sigma](u)}+\sqrt{[ \sigma](u)}}}
\bigg]^{\frac 35}
}_{ =1~\text{for~GVM}_{\infty}}
\end{eqnarray}
Differentiating again with respect to~$u$ and assuming $\sigma'(u)\neq 0$
\begin{eqnarray}
\fl
1
=
\frac 15
\Big(\frac{27\times 2^9}{\pi}\Big)^{\frac 15}
\hat\beta^{\frac 35} \:u\: \{[\sigma](u)\}^{-\frac{1}{10}}
\left\{
3\argc{
  \frac
  {\sinh^2 \argp{\tfrac 12\sqrt{[ \sigma](u)}}}
  {\sinh \argp{\sqrt{[ \sigma](u)}}+\sqrt{[ \sigma](u)}}
  }^{\frac 35}
\right.
+ \nonumber
\\
\quad\qquad
\underbrace{\;
+
\left.
[\sigma](u)
\argc{
  \frac
  {\sinh^5 \argp{\sqrt{[ \sigma](u)}}}
  {\sinh^4 \argp{\tfrac 12\sqrt{[ \sigma](u)}}\;\argp{ \sinh \argp{\sqrt{[ \sigma](u)}}+\sqrt{[ \sigma](u)}}^8}
  }^{\frac 15}
\right\}
}_{ =1~\text{for~GVM}_{\infty}}
\nonumber
\end{eqnarray}
It can also be rewritten
\begin{eqnarray}
\fl
1
=
\frac 35
\Big(\frac{27\times 2^9}{\pi}\Big)^{\frac 15}
\hat\beta^{\frac 35} \:u\: \{[\sigma](u)\}^{-\frac{1}{10}}
\nonumber
\\
\fl\qquad
\times
\underbrace{
\bigg[
  \frac
  {\sinh^2 \argp{ \tfrac 12\sqrt{[ \sigma](u)}}}
  {\sinh \argp{ \sqrt{[ \sigma](u)}}+\sqrt{[ \sigma](u)}}
  \bigg]^{\frac 35}
% \nonumber
% \\
% &\quad\qquad\qquad\qquad\;\quad\;
%\times
\;
\left\{
1+
\frac 13
  \frac
  {[\sigma](u)\coth \argp{ \tfrac 12\sqrt{[ \sigma](u)}}}
  {\sinh \argp{\sqrt{[ \sigma](u)}}+\sqrt{[ \sigma](u)}}
\right\}
}_{ =1~\text{for~GVM}_{\infty}}
\label{eq:sigmaCimplicit}
\end{eqnarray}
as the result announced in the main text at~\eref{eq:sigmaCimplicit_txt}.
This is the equation one has to solve in order to determine $[ \sigma](u)$ in the segment(s) where $\sigma'(u)\neq 0$.
%To understand the behaviour at $u\to 0$, one might assume  $[ \sigma](u)\sim u^\nu$, but this yields $\nu=-5$.
%

%\newpage

% \subsection{Numerical study through the iterative procedure}
% \label{sec:numerical-study_iGVM}

%-------------------------
\subsection{Study of the finite-$\tf$ Hamiltonian GVM solution with a 1-step RSB Ansatz}
\label{sec:study-interm-gvm}

Supported by the numerical results presented in \sref{sec:GVM-upgraded-finite-length_iii}, we study in this subsection a 1-step RSB form of the solution to the variational equations~(\ref{eq:eqvar_iGVM}-\ref{eq:sum_omega_inv_iGVM}) for the finite-$\tf$ Hamiltonian GVM.

%---------
\subsubsection{Form of the 1-step RSB Ansatz}
\label{sec:form-1-step}
~\\[-2mm]

The 1-step RSB Ansatz takes the following form (taking notations similar to those of Ref.~\cite{giamarchi_ledoussal_1995_PhysRevB52_1242}):
\begin{equation}
\sigma(u)
=
\cases{
 0 & $u<u_{\cc} $   
 \\
 \Sigma_1/u_{\cc} & $u>u_{\cc}$
}
\end{equation}
which implies
\begin{equation}
[\sigma](u)
=
\cases{
0 & $u<u_{\cc}    $
\\
\Sigma_1 & $u>u_{\cc}    $
}
\label{eq:1RSB_iGVM}
\end{equation}
From \eref{eq:sum_omega_inv_iGVM}:
\begin{equation}
\sum_\omega \big[\tilde G(\omega) - G(\omega,u)\big]
=
\cases{
+\infty
 & $u<u_{\cc}$
\\%[2mm]
\displaystyle
\frac{\coth \argp{\frac 12 \sqrt{\Sigma_1}}}{2\sqrt{\Sigma_1}}
& $u>u_{\cc}    $
}
\label{eq:gt_1RS1_1_iGVM}
\end{equation}
The GVM equations~(\ref{eq:eqvar_iGVM}-\ref{eq:sum_omega_inv_iGVM}) become
\begin{eqnarray}
 \Sigma_1
&=
\frac{2}{\sqrt\pi}
\hat\beta^{\frac 32}
u_{\cc}
\bigg\{
\hat\beta
\mathring\xi^2+
\frac{\coth \argp{\frac 12 \sqrt{\Sigma_1}}}{2\sqrt{\Sigma_1}}
\bigg\}^{-\frac 32}
\label{eq:eqvar_iGVM_1step}
\end{eqnarray}

%---------
\subsubsection{Variational free energy in the 1-step RSB Ansatz}
\label{sec:vari-free-energy_iGVM}
~\\[-2mm]

In the full-RSB cases recalled in \sref{sec:GVM-previous} from \cite{agoritsas_2010_PhysRevB_82_184207}, the variational equation for ${\sigma(u)}$ contained all the information for solving both the power-law behaviour and the plateau, along with the value of the full-RSB cutoff $u_{\cc}$.
On the contrary, in the $1$-step case,
%In fact, contrarily to the full-RSB case,
the value of the cut-off $u_{\cc}$ is determined by differentiating the variational free energy 
$\mathcal F_{\text{var}}$ with respect to $u_{\cc}$ and one actually needs to explicitly compute $\lim_{n\to 0} \frac 1 n \mathcal F_{\text{var}}$.
The computation of three contributions to the variational free energy
${
\mathcal F_{\text{var}}=
\langle \HHt_\text{el} - \HHt_0 \rangle_0+
\langle \HHt_\text{dis}\rangle_0
+
\mathcal F_0
}$,
defined in~(\ref{eq:Fvar_iGVM}-\ref{eq:Hdis-for-Fvar_iGVM}) is the longest computation %the most involved part
of the solution of the 1-step RSB Ansatz.
One has first:
\begin{equation}
\fl\qquad\qquad
\lim_{n\to 0}\frac 1n
\langle \HHt_\text{el} - \HHt_0 \rangle_0
=
\frac{\hat\beta^{-1}}{2} 
 \sum_\omega \omega^2 \tilde G(\omega)
\end{equation}
secondly:
\begin{eqnarray}
\fl\qquad\qquad
\lim_{n\to 0}\frac 1n
\big\langle
  \HHt_\text{dis}%\big[\tilde {\mathbf y}(\omega),\mathring {\mathbf y},\mathring\xi(\tf)\big]
\big\rangle_0
&
%\stackrel{\eref{eq:Hdis_ave0}}
=
\int \dbar \lambda
R_{\mathring \xi}(\lambda) 
%\int_0^1 \!\! d\hat t
\int_0^1 \!\!du\:
  e^{-\frac{\lambda^2}{\hat\beta}
  \sum\limits_\omega
  \big[\tilde G(\omega)-G(\omega,u)\big]
}
\\
\fl\qquad\qquad
&
%\stackrel{\eref{eq:defRxiGaussian}}
=
\frac{1}{2\sqrt\pi}
%\int_0^1 \!\!\! d\hat t\!
\int_0^1 \!\!\!du\:
\Big\{
\mathring\xi^2+
\hat\beta^{-1}
  \sum\limits_\omega
  \big[\tilde G(\omega)- G(\omega,u)\big]
\Big\}^{-\frac 12}
\end{eqnarray}
and thirdly, up to an additive constant:
\begin{eqnarray}
\fl\quad
\lim_{n\to 0}\frac 1n
\mathcal F_0 
&
%\stackrel{\eref{eq:expression_Fvar0_log-det}}
=
-\frac 12\hat\beta^{-1} 
\lim_{n\to 0}\frac 1n
\sum_\omega \log \det \tilde G(\omega)
%+ \text{const.}
\\
&
\stackrel{\phantom{\eref{eq:expression_Fvar0_log-det}}}
=
\frac 12\hat\beta^{-1} 
\:
\sum_\omega
\bigg[
- \log  G_{\cc}(\omega)
+\int^{1}_{0} \frac{du}{u^2}\: 
 \log \frac{G_{\cc}(\omega) - [ G](\omega, u)}{ G_{\cc}(\omega)}
\bigg] 
\\
&
\stackrel{\phantom{\eref{eq:expression_Fvar0_log-det}}}
=
\frac 12\hat\beta^{-1} 
\:
\sum_\omega
\bigg[
+ \log  \omega^2
+\int^{1}_{0} \frac{du}{u^2}\: 
 \log \Big(1 - \omega^2\: [ G](\omega, u)\Big)
\bigg] 
\end{eqnarray}
with ${G_c (\omega) = \argp{G_c^{-1}(\omega)}^{-1}=1/\omega^2}$
and ${[G](\omega) \equiv \int_0^u dv \, G(\omega,v)}$.
One uses then
\begin{eqnarray}
\fl\quad
  \tilde G(\omega)
%
%\stackrel{\phantom{\eref{eq:tilde-replica-04}}}
\stackrel{\eref{eq:tilde-replica-03}}
=
\frac{1}{\omega^2}\:\frac{\omega^2+\Sigma_1/u_{\cc}}{\omega^2+\Sigma_1}
\label{eq:Gtitildesigma1_iGVM}
\\
\fl\quad
  \tilde G(\omega) - G(\omega,u)
\stackrel{\eref{eq:tilde-replica-04}}
= 
\cases{
\frac{1}{u_{\cc} \omega^2}
+
\Big(1-\frac{1}{u_{\cc}}\Big)
\frac{1}{\omega^2+\Sigma_1}
 & if $u<u_{\cc} $
\\
\frac{1}{\omega^2+\Sigma_1}
 & if $u>u_{\cc}$
}
\label{eq:GtitimGti_1step_iGVM}
\\[3mm]
\fl\quad
\sum_\omega
\Big[ \tilde G(\omega) -  G(\omega,u)\Big]
%
%\stackrel{\phantom{\eref{eq:tilde-replica-04}}}
\stackrel{\eref{eq:GtitimGti_1step_iGVM}}
= 
\sum_\omega
\cases{
\frac{1}{\omega^2}\:\frac{\omega^2+\Sigma_1/u_{\cc}}{\omega^2+\Sigma_1}
 & if $u<u_{\cc}$
\\
\frac{1}{\omega^2+\Sigma_1}
 & if  $u>u_{\cc}$
}
\label{eq:gt_1RSB_iGVM}
\\
\fl\qquad\qquad\qquad\qquad\;\:
\stackrel{\phantom{\eref{eq:tilde-replica-04}}}
= 
\cases{
\mathfrak S(\Sigma_1,u_{\cc})
 & if $u<u_{\cc} $
\\%[2mm]
\frac{\coth \argp{\frac 12 \sqrt{\Sigma_1}}}{2\sqrt{\Sigma_1}}
 & if $u>u_{\cc}$
}
\label{eq:gt_1RSB_iGVM_2}
\\[3mm]
\fl\quad
[ G](\omega,u)
%
%\stackrel{\eref{eq:ronron-replica-04}}
\stackrel{\phantom{\eref{eq:tilde-replica-04}}}
= 
\cases{
0
 & if $u<u_{\cc} $
\\
\frac{1}{\omega^2}\:\frac{\Sigma_1}{\omega^2+\Sigma_1}
 & if  $u>u_{\cc}$
}
\end{eqnarray}
Here the first sum (for $u<u_{\cc}$) in~\eref{eq:gt_1RSB_iGVM} is singular because of the term $\omega=0$, and decomposed as follows, with $\omega_0=0^+$:
\begin{equation}
\fl
\mathfrak S(\Sigma_1,u_{\cc})\equiv
\sum_\omega \frac{1}{\omega^2}\:\frac{\omega^2+\Sigma_1/u_{\cc}}{\omega^2+\Sigma_1}
=
\frac {1}{u_{\cc}\omega_0^2}
+
\frac{1}{12u_{\cc}}
+
\Big(\frac{1}{u_{\cc}}-1\Big)
\Big(\frac{1}{\Sigma_1}-\frac{\coth \argp{\frac 12 \sqrt{\Sigma_1}}}{2\sqrt{\Sigma_1}}\Big)
\label{eq:regulSigmasum}
\end{equation}
%%%
%%% from ~/recherche/GVM-finite-time_interface_intermediate-GVM_singularsums-regul.nb
%%%
Gathering the results, one has
\begin{eqnarray}
\fl
\mathcal F_\text{var}^\text{1RSB}  
%\stackrel{\phantom{\eref{eq:regulSigmasum}}}
=
\:
\frac{\hat\beta^{-1}}{2}
\sum_\omega
\bigg[
\frac{\omega^2+\Sigma_1/u_{\cc}}{\omega^2+\Sigma_1}
\
+
\
\log \omega^2
+
\Big(\frac{1}{u_{\cc}}-1\Big)
\log\frac{\omega^2}{\omega^2+\Sigma_1}
\bigg]
\label{eq:Fvar1RSB_iGVM}
\\
\fl\qquad\qquad
+
\frac{\hat\beta^{\frac 12}}{2\sqrt\pi}
\bigg[
u_{\cc}
\Big\{
\hat\beta \mathring\xi^2
+
  \sum\limits_\omega
\frac{1}{\omega^2}\:\frac{\omega^2+\Sigma_1/u_{\cc}}{\omega^2+\Sigma_1}
\Big\}^{-\frac 12}
+
(1-u_{\cc})
\Big\{
\hat\beta\mathring\xi^2
+
\frac{\coth \argp{\frac 12 \sqrt{\Sigma_1}}}{2\sqrt{\Sigma_1}}
\Big\}^{-\frac 12}
\bigg]
\nonumber
\\[2mm]
%\stackrel{\eref{eq:regulSigmasum}}
\fl\qquad\;\:
=\:
\frac{\hat\beta^{-1}}{2}
\bigg[
\Big(\frac{1}{u_{\cc}}-1\Big)\:
\frac 12 \sqrt{ \Sigma_1}
\coth \argp{\frac 12 \sqrt{ \Sigma_1}}
\
-
\Big(\frac{1}{u_{\cc}}-1\Big)
\sum_\omega\log \Big(1+\frac{\Sigma_1}{\omega^2}\Big)
%\sum_\omega\frac{1}{\omega^2+\Sigma_1}
%
\bigg]
%
%\nonumber
\label{eq:Fvar1RSB_iGVM_2}
\\
+
\frac{\hat\beta^{\frac 12}}{2\sqrt\pi}
\bigg[
u_{\cc}
\Big\{
\hat\beta \mathring\xi^2+
\mathfrak S(\Sigma_1,u_{\cc})
\Big\}^{-\frac 12}
+
(1-u_{\cc})
\Big\{
\hat\beta\mathring\xi^2+
\frac{\coth \argp{\frac 12 \sqrt{\Sigma_1}}}{2\sqrt{\Sigma_1}}
\Big\}^{-\frac 12}
\bigg]
+``\infty"
\nonumber
\end{eqnarray}
where we have used that
\begin{eqnarray}
\sum_\omega
\frac{\omega^2+ \Sigma_1/u_{\cc}}{\omega^2+ \Sigma_1}
&=
\sum_\omega
\bigg\{
\Big(\frac{1}{u_{\cc}}-1\Big)
\:\frac{ \Sigma_1}{\omega^2+ \Sigma_1}
+1
\bigg\}
\\
&=
\Big(\frac{1}{u_{\cc}}-1\Big)\:
\frac 12 \sqrt{ \Sigma_1}
\coth \argp{\tfrac 12 \sqrt{ \Sigma_1}}
+
``\infty"
\label{eq:eqvariation_uc_2}
\end{eqnarray}
where $+\infty$ depends on no parameters.
Using Euler's formula for infinite products, one has
\begin{equation}
\sum_\omega
\log\big(1+\tfrac{\tilde\Sigma_1}{\omega^2}\big)
=
\log  {\frac {\sinh \argp{\tfrac 12 \sqrt{\tilde\Sigma_1}}}{\tfrac 12 \sqrt{\tilde\Sigma_1}}}
\end{equation}
and finally
\begin{eqnarray}
\fl
\mathcal F_\text{var}^\text{1RSB}  
%\stackrel{\phantom{\eref{eq:regulSigmasum}}}
=
&\:
\frac{\hat\beta^{-1}}{2}
\bigg[
\Big(\frac{1}{u_{\cc}}-1\Big)\:
\frac 12 \sqrt{ \Sigma_1}
\coth \argp{ \frac 12 \sqrt{ \Sigma_1}}
\
-
\Big(\frac{1}{u_{\cc}}-1\Big)
\log \frac {\sinh \argp{ \tfrac 12 \sqrt{\tilde\Sigma_1}}}{\tfrac 12 \sqrt{\tilde\Sigma_1}}
%\sum_\omega\log \Big(1+\frac{\Sigma_1}{\omega^2}\Big)
%\sum_\omega\frac{1}{\omega^2+\Sigma_1}
%
\bigg]
\label{eq:Fvar1RSB_iGVM_3}
%\nonumber
\\
\fl\quad
&
+
\frac{\hat\beta^{\frac 12}}{2\sqrt\pi}
\bigg[
u_{\cc}
\Big\{
\hat\beta \mathring\xi^2+
\mathfrak S(\Sigma_1,u_{\cc})
\Big\}^{-\frac 12}
+
(1-u_{\cc})
\Big\{
\hat\beta\mathring\xi^2+
\frac{\coth \argp{\frac 12 \sqrt{\Sigma_1}}}{2\sqrt{\Sigma_1}}
\Big\}^{-\frac 12}
\bigg]+``\infty"
\nonumber
\end{eqnarray}
Differentiating the free energy w.r.t.~$1/u_{\cc}$, one gets, sending $\omega_0$ to $0$ afterwards:
\begin{eqnarray}
\fl \qquad\qquad
\frac{\hat\beta^{-1}}{2}
\bigg[
\frac 12 \sqrt{ \Sigma_1}
\coth \argp{\frac 12 \sqrt{ \Sigma_1}}
\
&-
\
\log \frac{\sinh \argp{\frac 12 \sqrt{\Sigma_1}}}{\frac 12\sqrt{\Sigma_1}}
\bigg]
 \nonumber
 \\
 &
=
{u_{\cc}^2}
\frac{\hat\beta^{\frac 12}}{2\sqrt\pi}
\Big\{
\hat\beta\mathring\xi^2+
\frac{\coth \argp{\frac 12 \sqrt{\Sigma_1}}}{2\sqrt{\Sigma_1}}
\Big\}^{-\frac 12}
\label{eq:varFreplicated_iGVM}
\end{eqnarray}
This equation will allow to determine the cutoff $u_{\cc}$.

%-------------------------
\subsection{KPZ Scaling of the 1-step RSB solution at large $\tf$}
\label{sec:scaling-1-step_iGVM}

Assuming that at $\xi=0$ the solution of the GVM variational equation is 1-step RSB with self-consistently, 
\begin{equation}
 \Sigma_1\to\infty  \qquad \text{as} \qquad \tf\to\infty
\end{equation}
one obtains from the variational equation~\eref{eq:eqvar_iGVM_1step} that
\begin{equation}
 \Sigma_1\sim u_{\cc} \:\hat\beta^{\frac 32}\:\Sigma_1^{\frac 34}
\qquad
\Longleftrightarrow
\qquad
 \Sigma_1\sim \big( u_{\cc}  \:\hat\beta^{\frac 32}\big)^4
\label{eq:eqvariGVMSigma1}
\end{equation}
We now use the equation \eref{eq:varFreplicated_iGVM} in ${\Sigma_{\cc},u_{\cc}}$,
which depends on $\{\hat{\beta},\mathring\xi\}$,
to determine a relation on $u_{\cc}$: remarking that the dominant terms of the first bracket compensate at large $\Sigma_1$, and that the rest yields a constant, one has
\begin{equation}
  \hat\beta^{-1}
\ \sim \
u_{\cc}^2 \:\hat\beta^{\frac 12}\:\Sigma_1^{\frac 14}
\stackrel{\eref{eq:eqvariGVMSigma1}}
\sim
u_{\cc}^2 \:\hat\beta^{\frac 12}\: u_{\cc}  \:\hat\beta^{\frac 32} 
\ \sim \
u_{\cc}^3\hat\beta^2
\end{equation}
from which we deduce
\begin{equation}
\hat\beta u_{\cc}\sim 1
\label{eq:betahatuc_iGVM}
\end{equation}
This implies that at large $\tf$
\begin{equation}
  u_{\cc}\sim\tf^{-\frac 13}
\qquad
\qquad
\Sigma_1
\stackrel{\eref{eq:eqvariGVMSigma1}}
\sim
\tf^{\frac 23}
\label{eq:uc_Sigma1_largetf_iGVM} 
\end{equation}
which self-consistently justifies the divergence of $\Sigma_1$ as $\tf\to\infty$.

%_____________________________________________________________________________________________________
\renewcommand{\thesection}{B}
\section{Numerical solution to the GVM variational equations}
\label{sec:app-iter-num_GVM}

Solving analytically the variational equations of the GVM approach is not an obvious task. 
To get a hint of the solutions, we present a numerical scheme to solve the variational equations iteratively.
We first test it on the well-studied cases of the infinite-$\tf$ Hamiltonian GVM in \sref{sec:interGVMH}, and the free-energy GVM in \sref{sec:numericGVMtm},
before applying it to the finite-$\tf$ Hamiltonian GVM in~\sref{sec:iternum_iGVM}.

%{Iterative procedure ($i$): the infinite-$\tf$ Hamiltonian GVM}
%	\sref{sec:interGVMH}
%{Iterative procedure ($ii$): the free-energy GVM}
%	\sref{sec:numericGVMtm}
%{Iterative procedure ($iii$): the finite-time Hamiltonian GVM}
%	\sref{sec:iternum_iGVM}

%-------------------------
\subsection{Iterative procedure: ($i$)~the infinite-$\tf$ Hamiltonian GVM}
\label{sec:interGVMH}

\begin{figure}
  \centering
  \hspace*{-28mm}
  \includegraphics[width=.38\columnwidth]{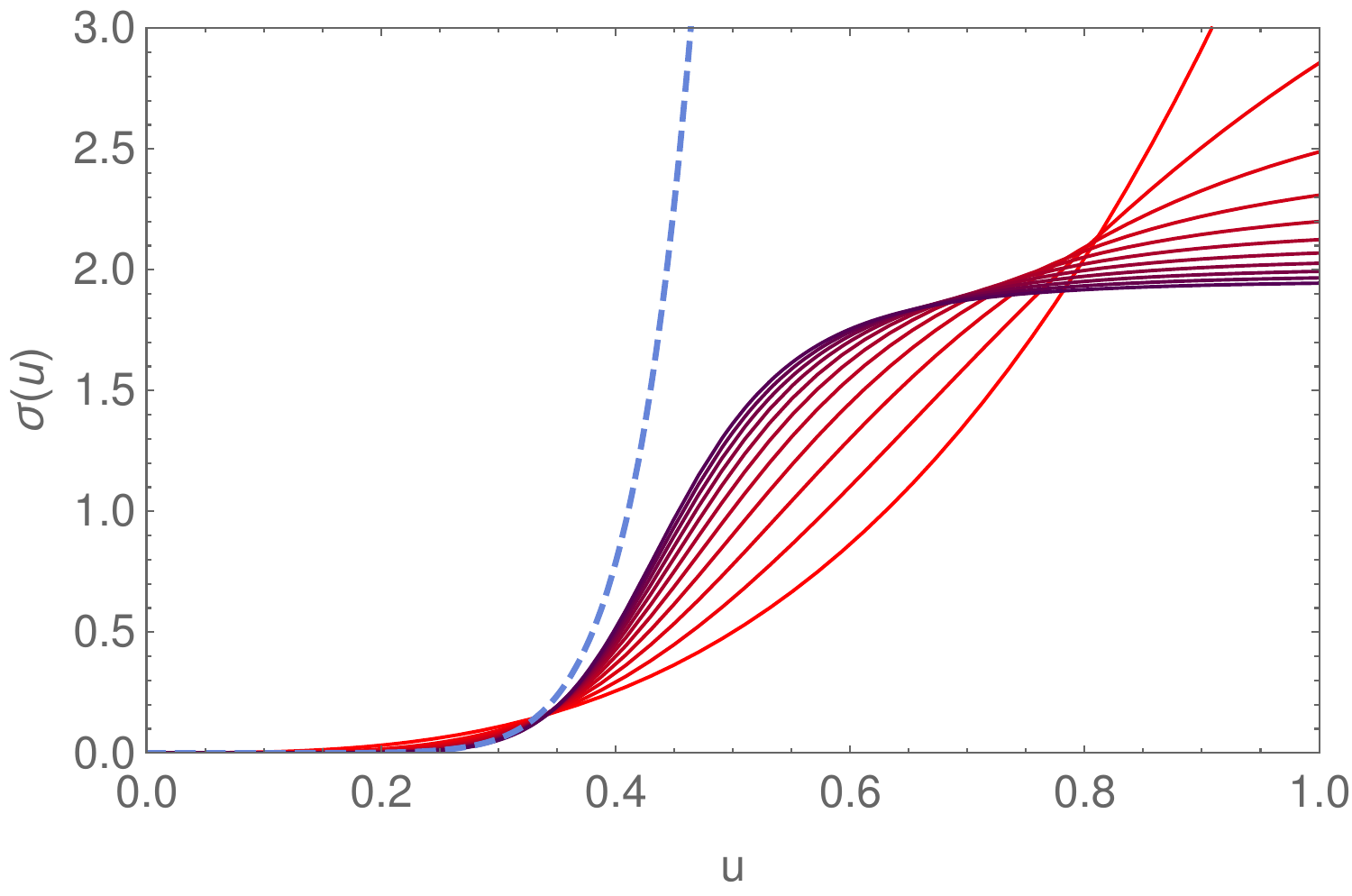}
  \includegraphics[width=.38\columnwidth]{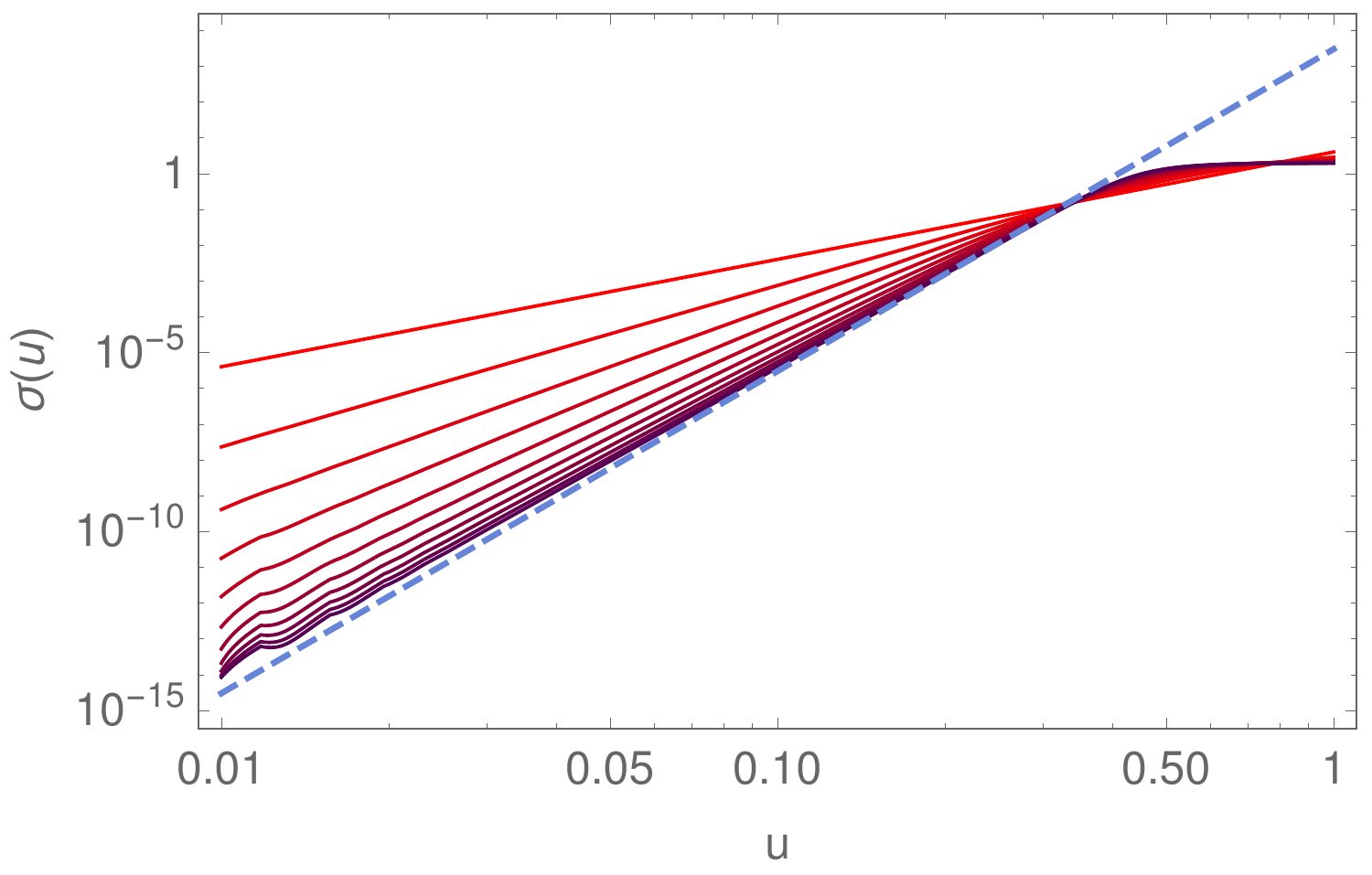}
  \includegraphics[width=.38\columnwidth]{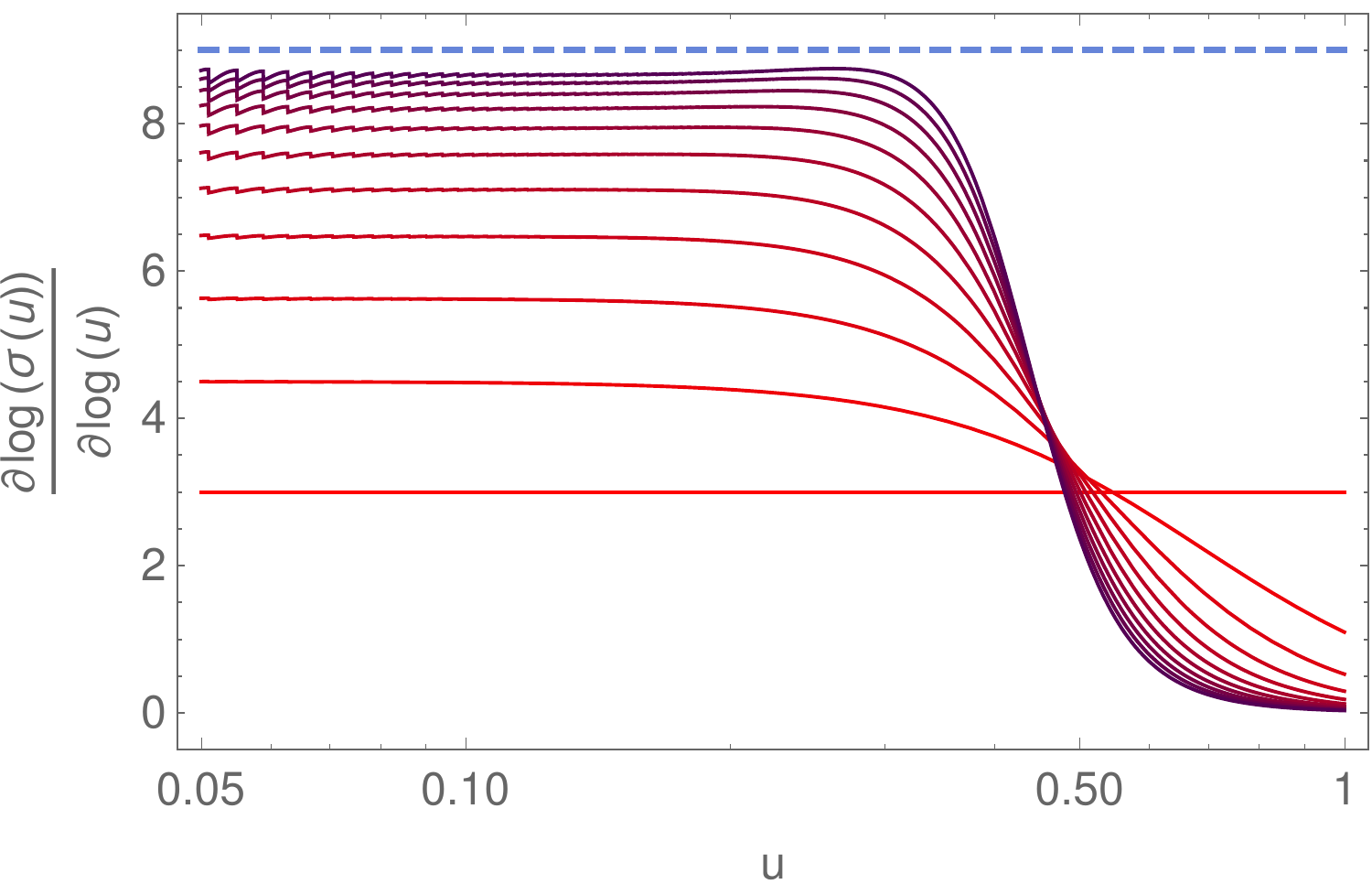}
  \hspace*{-20mm}
  \caption{
10 iterations of the iterative procedure described in \sref{sec:interGVMH} for the infinite-$\tf$ Hamiltonian GVM, showing the convergence of $\sigma_k(u)$ to a form $\sigma(u<u_{\cc})\propto u^9 $ and $\sigma(u>u_{\cc}) =\sigma(u_{\cc})$.
(\textbf{Left})~$\sigma_k(u)$ for $1\leq k\leq 10$. 
(\textbf{Center})~$\sigma_k(u)$ in log-log scale. 
(\textbf{Right})~the logarithmic derivative of $\sigma_k(u)$.
In all figures, $k$ increases from $1$ to $10$ from red to blue.
The dashed lines correspond to the full RSB regime $\sigma(u<u_{\cc})\propto u^9 $.
Parameters are $\hat\beta=2$ and $\mathring\xi=\frac12$.
\label{fig:GVM_ham_iterations}%
}
\end{figure}
%%
%% from ~/recherche/GVM-finite-time_interface_iteration_hamiltonian.nb
%%

Here the variational equation, similar to~\eref{eq:eqvar_iGVM}, reads
\begin{eqnarray}
 \sigma(u)
&=
\frac{2}{\sqrt\pi}
{\hat\beta}^{\frac 32}
\left[
\mathring\xi^2+
\hat\beta^{-1}
  \int_{\mathbb{R}} \! \dbar q\:
  \big[\tilde G(q) - G(q,u)\big]
\right]^{-\frac 32}
\label{eq:varGVMham}
\end{eqnarray}
and the integral over $q$ of the equivalent of~\eref{eq:tilde-replica-04} yields
\begin{eqnarray}
  \int_{\mathbb{R}}\! \dbar q\:
  \big[\tilde G(q) - \ G(q,u)  \big]
 &= \frac{1}{u}  \frac{1}{\sqrt{[ \sigma](u)}} - \int^1_u \frac{dv}{v^2}  \frac{1}{\sqrt{[\sigma](v)}} \label{eq:ronron-replica-04_ham}
\end{eqnarray}
This pair of equations can be seen as a fixed-point equation for $\sigma(u)$.
One can thus solve this pair using an iterative procedure on $\sigma(u)$, which, if converging, gives a stable solution to~(\ref{eq:varGVMham}-\ref{eq:ronron-replica-04_ham}).

Starting from an `initial'  $\sigma_0(u)$, one iterates numerically for $k\geq 0$ the following procedure to evaluate $\sigma_{k+1}(u)$ from $\sigma_{k}(u)$:
\begin{itemize}
\item determine $[\sigma_k](u)$ from $\sigma_k(u)$
\item determine the corresponding $\int_{\mathbb{R}}\! \dbar q\:\big[\tilde G_k(q) - \ G_k(q,u)  \big]$ from~\eref{eq:ronron-replica-04_ham}
\item determine the next iteration $\sigma_{k+1}(u)$ from~\eref{eq:varGVMham} as
\begin{eqnarray}
 \sigma_{k+1}(u)
&=
\frac{2}{\sqrt\pi}
{\hat\beta}^{\frac 32}
\left[
\mathring\xi^2+
\hat\beta^{-1}
  \int_{\mathbb{R}} \! \dbar q\:
  \big[\tilde G_k(q) - G_k(q,u)\big]
\right]^{-\frac 32}
\label{eq:varGVMham_iter}
\end{eqnarray}
\end{itemize}

If the procedure converges as the number of iteration steps increases (\emph{i.e.}~if $\sigma_{k}(u)\to \sigma_{\infty}(u)$ as $k\to\infty$), one expects to obtain a fixed point $\sigma_{\infty}(u)$ which is a stable solution of the variational equation~\eref{eq:varGVMham}.
Figure~\ref{fig:GVM_ham_iterations} confirms the convergence of the procedure to the expected form of the solution that can be obtained analytically by the full solution of the problem, as discussed in \sref{sec:GVM-previous},
and that $\sigma(u<u_{\cc})\propto u^9 $ and $\sigma(u>u_{\cc}) =\sigma(u_{\cc})$, as discussed in \sref{sec:GVM-previous} and in Refs.~\cite{agoritsas_2010_PhysRevB_82_184207,phdthesis_Agoritsas2013}.
%

%-------------------------
\subsection{Iterative procedure: ($ii$)~the free-energy GVM}
\label{sec:numericGVMtm}

\begin{figure}
  \centering
  \hspace*{-28mm}
  \includegraphics[width=.38\columnwidth]{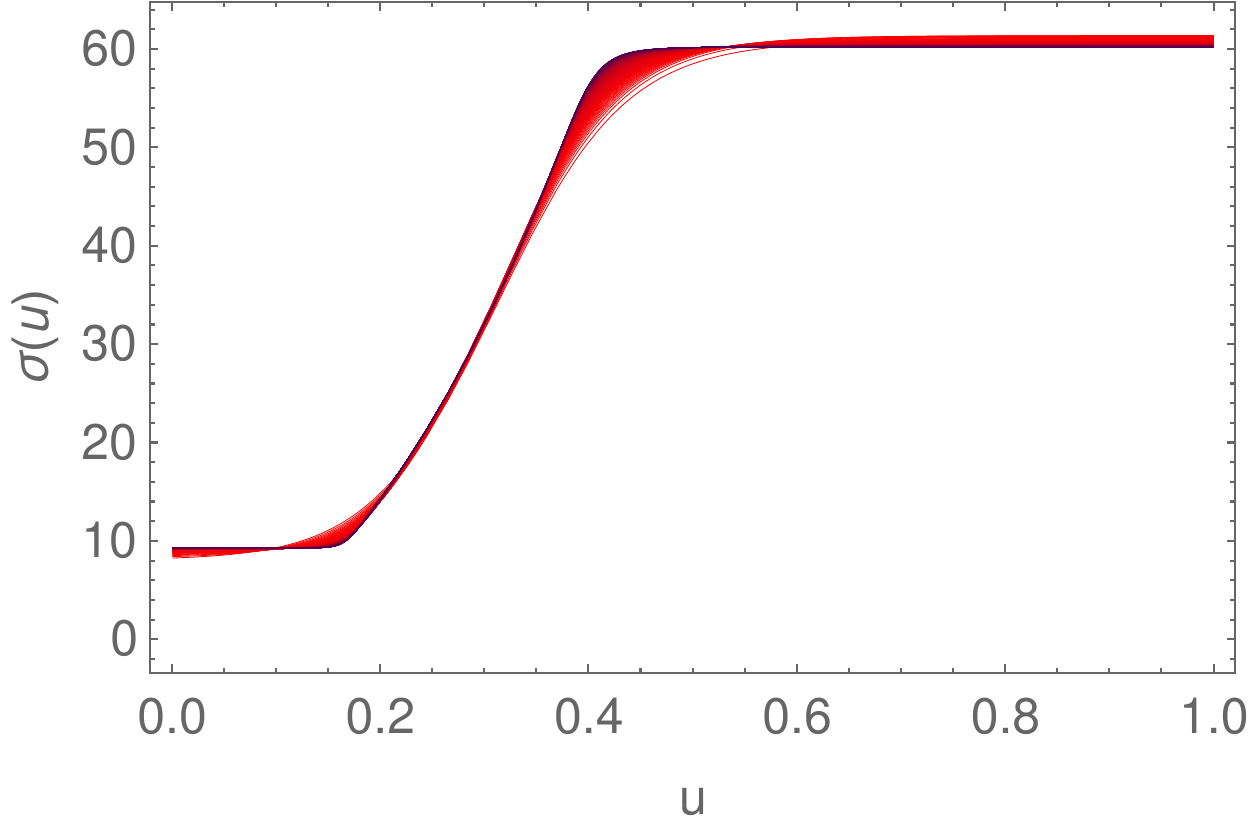}
  \includegraphics[width=.38\columnwidth]{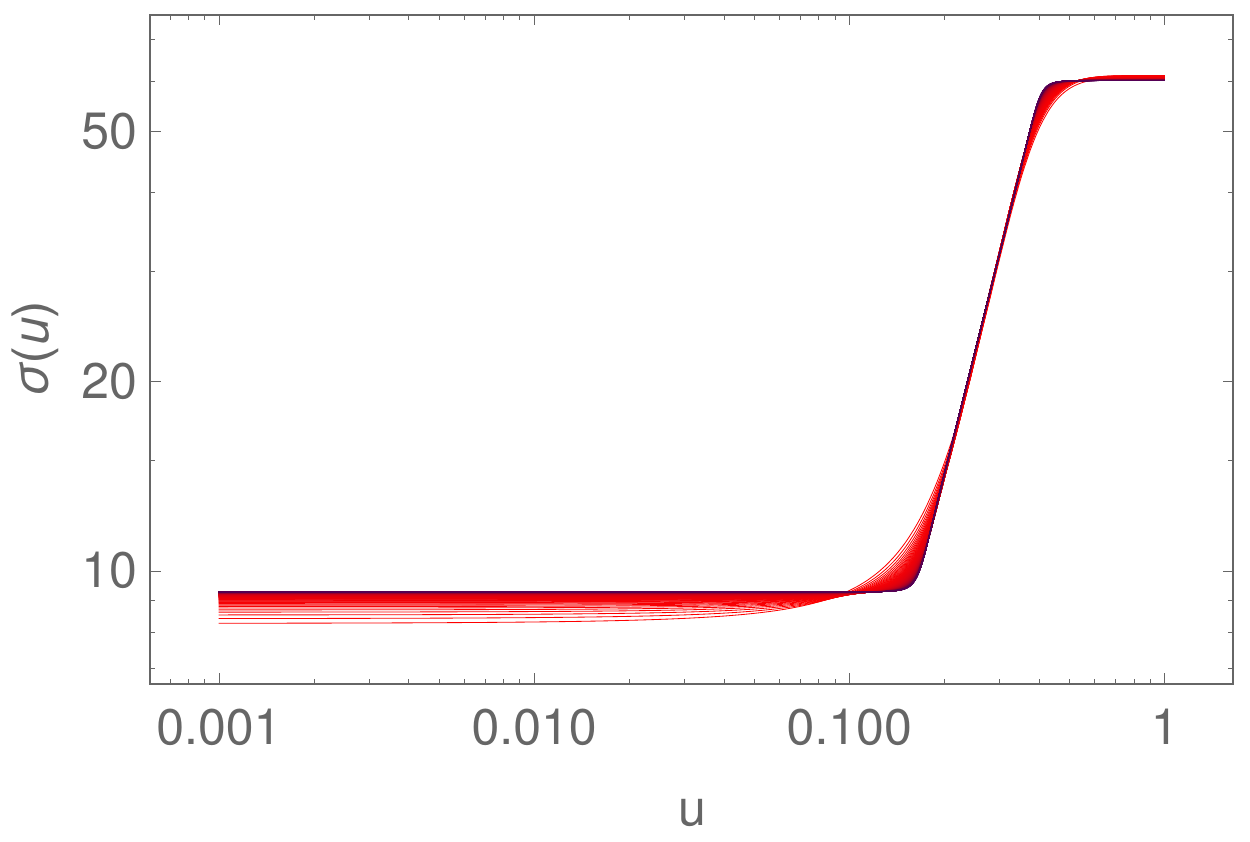}
  \includegraphics[width=.38\columnwidth]{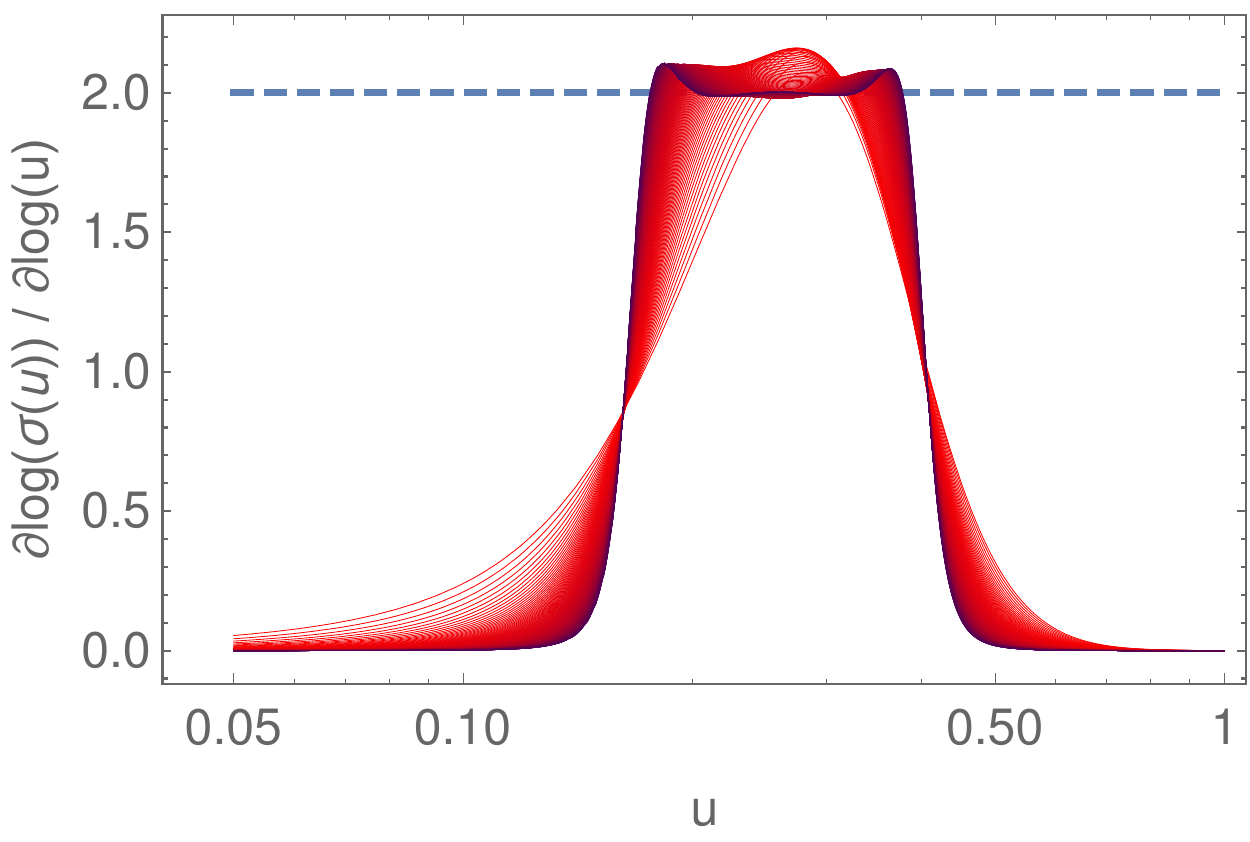}
  \hspace*{-20mm}
  \caption{
$200$ iterations of the fixed-point procedure for the free-energy GVM, described in \sref{sec:numericGVMtm}, showing the convergence of $\sigma_k(u)$ to a form $\sigma(u<u_*)=\sigma(u_*)$, $\sigma(u_*<u<u_{\cc})\propto u^2 $ and $\sigma(u>u_{\cc}) =\sigma(u_{\cc})$.
The index $k$ grows from $1$ to $200$ from red to dark purple.
(\textbf{Left})~$\sigma_k(u)$ for $1\leq k\leq 200$
(\textbf{Center})~$\sigma_k(u)$ in log-log scale. 
(\textbf{Right})~the logarithmic derivative of $\sigma_k(u)$.
The dashed lines correspond to the full RSB regime $\sigma(u_*<u<u_{\cc})\propto u^2 $.
Parameters are $\hat\beta = 20$ and $\mathring \xi=\frac 1{20}$.
\label{fig:GVM_toy_iterations}%
}
\end{figure}

Here the variational equation reads~\cite{agoritsas_2010_PhysRevB_82_184207,phdthesis_Agoritsas2013}:
\begin{eqnarray}
 \sigma(u)
&=
\frac{2}{\sqrt\pi}
{\hat\beta}^{\frac 32}
\left\{
\mathring\xi^2+
\hat\beta^{-1}
  \big[\tilde G - G(u)\big]
\right\}^{-\frac 12}
\label{eq:varGVMtoy}
\end{eqnarray}
and the equivalent of~\eref{eq:tilde-replica-04} yields (with $G^{-1}_{\cc}=1$):
\begin{eqnarray}
 \tilde G -  G(u)
 &= \frac{1}{u}  \frac{1}{ G^{-1}_{\cc} + [\sigma](u)} - \int^1_u \frac{dv}{v^2}  \frac{1}{ G^{-1}_{\cc} +[ \sigma](v)} \label{eq:toy-replica-04}
\end{eqnarray}
We recall that the all the $\tf$ dependence is hidden in the rescaled inverse temperature ${\hat{\beta}}$ and disorder correlation length $\mathring\xi$.
Hence, starting from an `initial'  $\sigma_0(u)$ and iterating numerically the following procedure for $k\geq 0$:
\begin{itemize}
\item determine $[\sigma_k](u)$ from $\sigma_k(u)$,
\item determine the corresponding $\argc{\tilde G_k- \ G_k(u) }$ from~\eref{eq:toy-replica-04},
\item determine the next iteration $\sigma_{k+1}(u)$ from~\eref{eq:varGVMtoy} as
\begin{eqnarray}
 \sigma_{k+1}(u)
&=
\frac{2}{\sqrt\pi}
{\hat\beta}^{\frac 32}
\left\{
\mathring\xi^2+
\hat\beta^{-1}
  \big[\tilde G_k - G_k(u)\big]
\right\}^{-\frac 12} 
\label{eq:varGVMtoy_iter}
\end{eqnarray}
\end{itemize}
One can expect, if the procedure converges with $\sigma_{k}(u)\to \sigma_{\infty}(u)$, to obtain a fixed point $\sigma_{\infty}(u)$ which is a solution of the variational equation~\eref{eq:varGVMtoy}.
Results shown on figure~\ref{fig:GVM_toy_iterations} confirm the convergence to the expected form of the solution obtained analytically by the full solution of the problem  $\sigma(u<u_*)=\sigma(u_*)$, ${\sigma(u_*<u<u_{\cc})\propto u^2}$ and $\sigma(u>u_{\cc}) =\sigma(u_{\cc})$, see \sref{sec:GVM-previous} and Refs.~\cite{agoritsas_2010_PhysRevB_82_184207,phdthesis_Agoritsas2013}.
\begin{figure}[t]
  \centering
  \hspace*{-28mm}
  \includegraphics[width=.38\columnwidth]{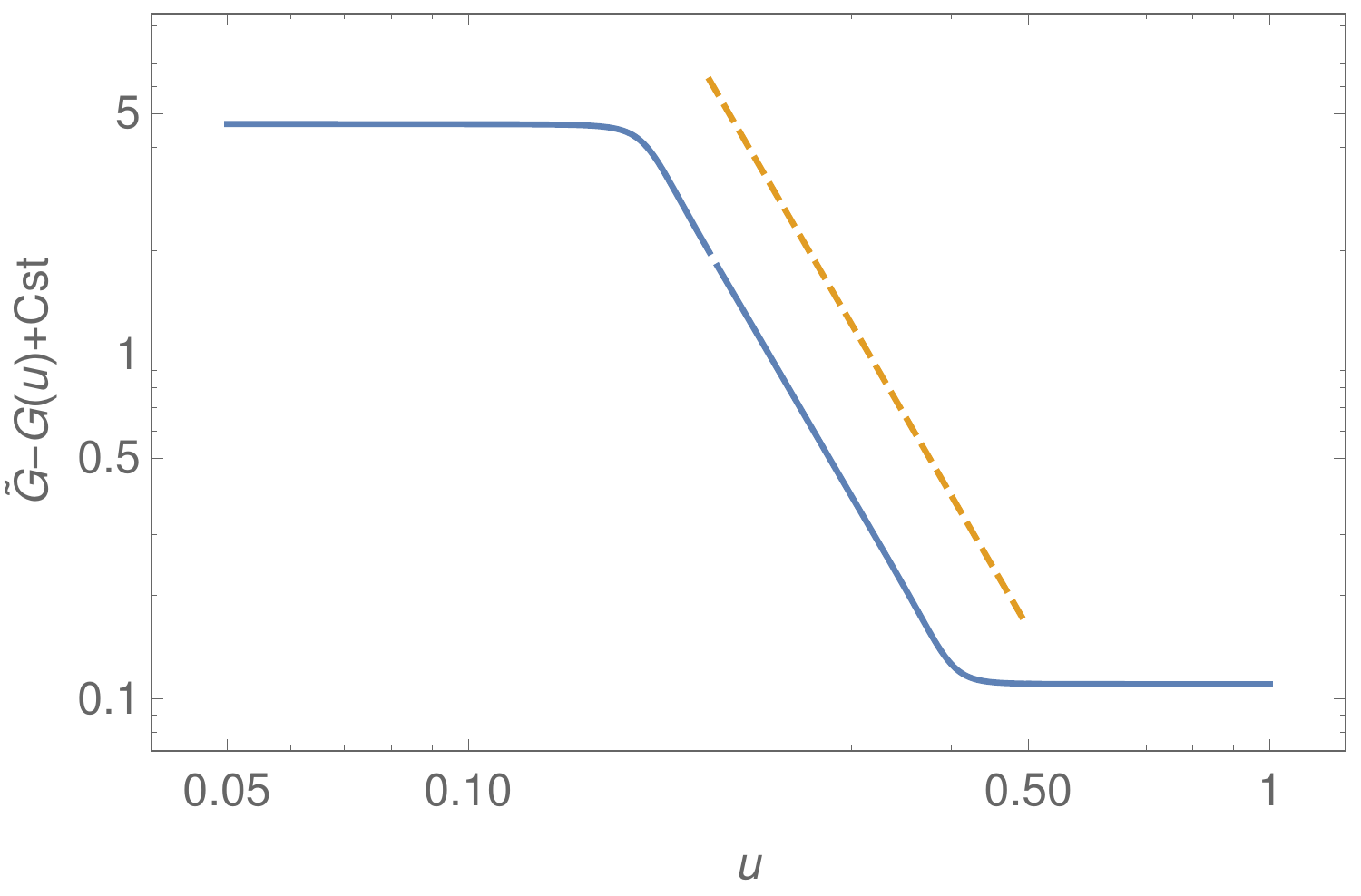}
  \includegraphics[width=.37\columnwidth]{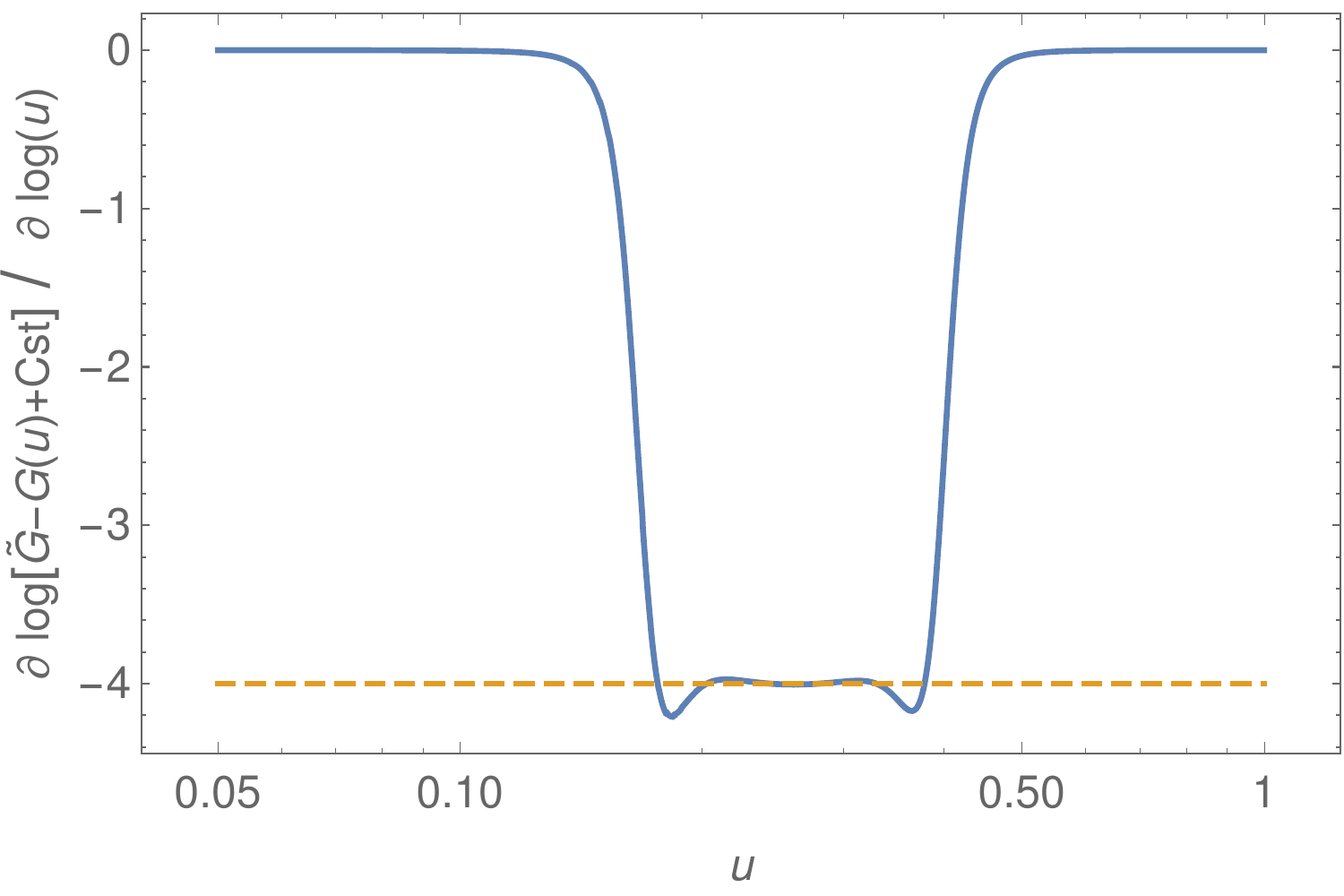}
  \includegraphics[width=.39\columnwidth]{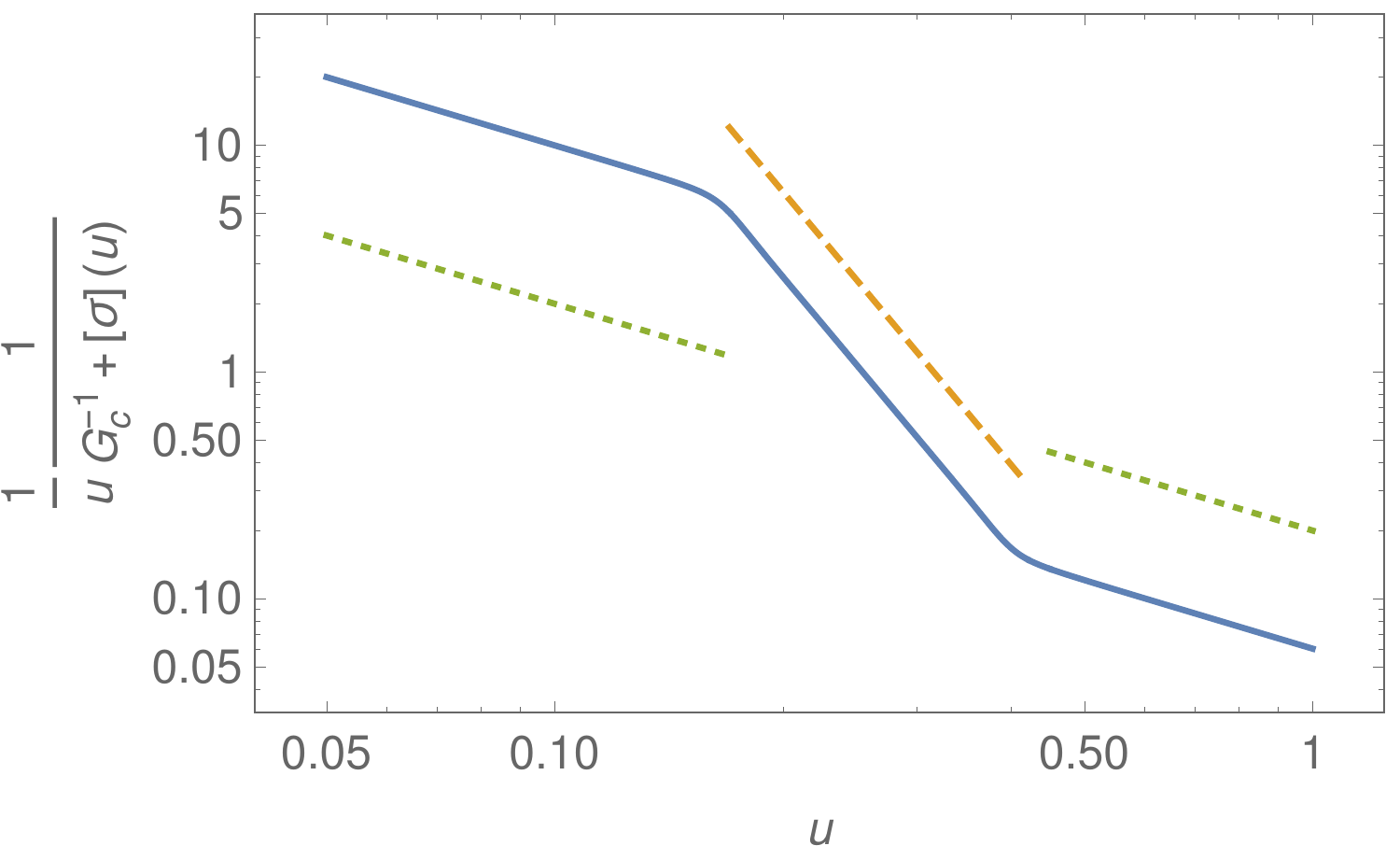}
  \hspace*{-20mm}
  \caption{
$\tilde G-G(u)$ for the free-energy GVM, after 200 iterations. 
(\textbf{Left})~log-log scale. 
(\textbf{Center})~logarithmic derivative. 
(\textbf{Right})~logarithmic derivative of the only first term of~\eref{eq:toy-replica-04}.
The dashed orange lines correspond to the full RSB regime behaviour $\tilde G-G(u_*<u<u_{\cc})\propto u^{-4}$, see~\eref{eq:depGmG_tmGVM}.
The dotted green lines correspond to a $1/u$ behaviour of \eref{eq:depG_tmGVM}.
Parameters are $\hat\beta = 20$ and $\mathring \xi=\frac 1{20}$.
\label{fig:GVM_toy_GGu_iterations}%
}
\end{figure}
%%
%% from ~/recherche/GVM-finite-time_interface_iteration_toy-model.nb
%%

%%
%% from ~/recherche/GVM-finite-time_interface_iteration_toy-model.nb
%%

% Note that the power $\alpha=9$ in $\sigma(u<u_{\cc})\propto u^\alpha $ can be obtained heuristically as follows.
% %
% Assuming that the power-lawe regime $u<u_{\cc}$ is determined self-consistently from the $u\to 0$ asymptotics, one gets from~\eref{eq:ronron-replica-04_ham}
% \begin{eqnarray}
%   \int\! \dbar q\:
%   \big[\tilde G - \ G(q,u)  \big]
%  &\sim u^{-\frac{3+\alpha}{2}}
%  \label{eq:ronron-replica-04_ham_devsmallu}
% \end{eqnarray}
% (where we used $[\sigma](u)\sim u^{1+\alpha}$) and thus from~\eref{eq:varGVMham}
% \begin{equation}
%   u^\alpha \sim u^{\frac{3}{2} \frac{3+\alpha}{2}}
% \end{equation}
% which imposes  $\alpha = {\frac{3}{2} \frac{3+\alpha}{2}}$ and one finds $\alpha=9$.

%\newpage
Note that in~\eref{eq:toy-replica-04} there is a non-trivial interplay between the first and second terms, as seen from figure~\ref{fig:GVM_toy_GGu_iterations}~(right):
\begin{equation}
\frac{1}{u}  \frac{1}{ G^{-1}_{\cc} + [\sigma](u)} 
\sim
\cases{
     \frac 1u & if $0<u<u_*$ \\
     u^{-4}+\text{constant} & if $u_*<u<u_{\cc}$ \\
     \frac 1 u & if $u_{\cc}<u<1$
}
\label{eq:depG_tmGVM}
\end{equation}

The behaviour of $\sigma(u)$ is more complex that in the Hamiltonian GVM.
In fact, one can infer it from the following a heuristic reasoning: if one assumes that
\begin{equation}
\sigma(u)\sim
\cases{
     \sigma(u_*) & if $0<u<u_*$ \\
     u^{\nu-1} & if $u_*<u<u_{\cc}$ \\
     \sigma(u_{\cc}) & if  $u_{\cc}<u<1$
}
\end{equation}
then, in the regime $u_*<u<u_{\cc}$, one has $[\sigma](u)\sim u^{\nu}$.
Thus, as read from~\eref{eq:toy-replica-04} (see figure~\ref{fig:GVM_toy_GGu_iterations}),
\begin{equation}
\tilde G-G(u)\sim
\cases{
     \tilde G-G(u_*) & if               $0<u<u_*$ \\
     u^{-(1+\nu)} & if $u_*<u<u_{\cc}$ \\
     \tilde G-G(u_{\cc}) & if            $u_{\cc}<u<1$
}
\label{eq:depGmG_tmGVM}
\end{equation}
Finally, \eref{eq:varGVMtoy} implies $u^{\nu-1}\sim u^{\frac12(1+\nu)}$, whence $\nu=3$.

%-------------------------
\subsection{Iterative procedure : ($iii$)~the finite-time Hamiltonian GVM}
\label{sec:iternum_iGVM}

The iterative procedure for the GVM equations~\eref{eq:eqvar_iGVM}-\eref{eq:sum_omega_inv_iGVM}, after starting from an initial $\sigma_0(u)$, is thus
\begin{itemize}
\item determine $[\sigma_k](\omega,u)$ from $\sigma_k(u)$,
\item determine the corresponding $\sum_\omega \big[\tilde G_k(\omega) - \ G_k(\omega,u)\big] $, from~\eref{eq:sum_omega_inv_iGVM},
\item determine the next iteration $\sigma_{k+1}(u)$ from~\eref{eq:eqvar_iGVM} as
\begin{eqnarray}
 \sigma_k (u)
=
\frac{2}{\sqrt\pi}
\hat\beta^{\frac 32}
\bigg\{
\hat\beta
\mathring\xi^2+
  \sum\limits_\omega
  \big[\tilde G_k(\omega) - G_k(\omega,u)\big]
\bigg\}^{-\frac 32}
\label{eq:varGVMi_iter}
\end{eqnarray}
\end{itemize}
The numerical results shown on figures~\ref{fig:GVM_toy_iterations_i} and~\ref{fig:GVM_toy_iterations_iGVM_scaling} in the main text, support a \emph{1-step RSB form} for $\sigma(u)$ instead of a full-RSB solution, at least in the regime $\xi\to 0$ that we considered in the numerical study.

%_______________________________________________________________________________________________________
%_______________________________________________________________________________________________________

\section*{References}

\addcontentsline{toc}{section}{References}
\bibliographystyle{plain_url}

\bibliography{zeta23_replicae_article}
%\bibliography{Biblio_DES_UTF-8-postdoc}
%\end{thebibliography}

%_______________________________________________________________________________________________________
%_____________________________________________________________________________________________________

\end{document}